\documentclass[%
 reprint,
 11pt,
nofootinbib,
 amsmath,amssymb,
 aps,
,longbibliography]{revtex4}

\usepackage{graphicx}% Include figure files
\usepackage{dcolumn}% Align table columns on decimal point
\usepackage{bm}% bold math
\usepackage{footmisc}
\usepackage{ulem}
\usepackage{amsmath,amssymb} % easier math formulae, align, subequations \ldots
\usepackage{mathrsfs}
\usepackage{amsfonts}
\usepackage{pgfplots}
\pgfplotsset{compat=1.5}
\usepackage{lscape}
\usepackage{subfig}
\usepackage{floatrow}
\usepackage{caption}
\usepackage{lipsum}

\newcommand\blfootnote[1]{%
	\begingroup
	\renewcommand\thefootnote{}\footnote{#1}%
	\addtocounter{footnote}{-1}%
	\endgroup
}

\def\IB#1{\boldsymbol{#1}} % use \IB{} for italic bold vector
\def\W/!i#1{\Wi} % macro for Wi

\newcommand{\PD}[2]{\frac{\partial #1}{\partial #2}}
\def\bnabla{\boldsymbol{\nabla}}
\newcommand{\sm}[1]{{\color{black} #1}}

\newcommand{\ac}[1]{{\color{black} #1}}
\begin{document}

\preprint{APS/123-QED}

\title{Self-propulsion in 2D Confinement: Phoretic and Hydrodynamic Interactions}

\author{Akash Choudhary$ ^{1,3} $}
\author{K. V. S. Chaithanya$ ^1 $}
\author{Sébastien Michelin$ ^2 \, \ddagger $}
\author{S. Pushpavanam $ ^1 \, \dagger $}

\blfootnote{$ \dagger $ spush@iitm.ac.in}  
 \blfootnote{$ \ddagger $ sebastien.michelin@ladhyx.polytechnique.fr}
%\homepage{}
%\footnotetext{$ \dagger $ sebastien.michelin@ladhyx.polytechnique.fr $ \; \; $ $ \ddagger $ spush@iitm.ac.in}

\affiliation{%
$ {}^{1} $Indian Institute of Technology Madras, Chennai, 600036 TN, India
}%
\affiliation{%
$ {}^{2} $LadHyX, CNRS, Ecole Polytechnique, Institut Polytechnique de Paris, 91120 Palaiseau, France
}%
\affiliation{%
	$ {}^{3} $  Institute of Theoretical Physics, Technische Universit\"{a}t Berlin, 10623 Berlin, Germany
}%

%	In the presence of solute-particle interactions, this chemical gradient results in a flow near the surface, known as the diffusio-osmotic slip, which results in self-propulsion. 

\begin{abstract}
	Chemically active Janus particles generate tangential concentration gradients along their surface for self-propulsion. 
Although this is well studied in unbounded domains, the analysis in biologically relevant environments such as confinements is scarce.
	In this work, we study the motion of a Janus sphere in weak confinement. The particle is placed at an arbitrary location, with arbitrary orientation between the two walls. 
	Using the method of reflections, we study the effect of confining planar boundaries on the phoretic and hydrodynamic interactions, and their consequence on the Janus particle dynamics.
	The dynamical trajectories are analyzed using phase diagrams for different surface coverage of activity and solute-particle interactions.
	In addition to near wall states such as `sliding' and `hovering', we demonstrate that accounting for two planar boundaries reveals two new states: channel-spanning oscillations and damped oscillations around the centerline, which were characterized as `scattering' or `reflection' by earlier analyses on single wall interactions. 
Using phase-diagrams, we highlight the differences in inert-facing and active-facing Janus particles.
	We also compare the dynamics of Janus particles with squirmers for contrasting the chemical interactions with hydrodynamic effects. Insights from the current work suggest that biological and artificial swimmers sense their surroundings through long-ranged interactions, that can be modified by altering the surface properties.
\end{abstract}

\maketitle

%\tableofcontents

	\section{Introduction}
Self-propulsion of biological swimmers occurs through metachronal actuation of ciliary filaments or helical motion of flagella \cite{koch2011collective}.
Janus particles are their artificial analogues, which self-propel by exploiting the catalytic anisotropy on their surface.
In the presence of solute, the region of catalytic coating either consumes or reacts with it to produce chemical gradients, which generate a diffusio-osmoic flow tangential to the particle surface. This results in the particle movement known as self-diffusiophoresis. This self-propulsion generates hydrodynamic and chemical signals that decay as $ \sim O(1/r^{2}) $ and $ O(1/r) $, respectively.
These artificial swimmers offer several potential applications such as drug-delivery micromachines and use in controlled studies of microbial infections \cite{gao2014synthetic}.
Furthermore, these can potentially facilitate payload delivery in confined biological environments such as cerebrospinal-fluid pathways \cite{nelson2014micro}.
Earlier studies on their biological counterparts (such as \textit{E. coli}, sea-urchin and sperm cells) offered useful insights on microswimmer-surface interaction mechanism. We discuss these studies below.

One of the earliest studies by \citet{rothschild1963non} showed surface accumulation of bull spermatozoa. 
Similar observations were reported for \textit{E. coli} by \citet{berke2008hydrodynamic}. Accounting for the far-field hydrodynamics, they showed that the hydrodynamic interaction of force-dipole disturbance (generated by the bacteria to self-propel) causes surface attraction.
This offered a possible explanation of biofilm generation due to large residence time of bacteria near surfaces.
\citet{spagnolie2012hydrodynamics} introduced a general framework that studied the surface-interaction of individual hydrodynamic singularities (in the far-field) and demonstrated that the analytical approaches based on remote interaction offer impressive accuracy even when the particle is a few body lengths away from the wall.
Accounting for higher-order flow singularities and soft-repulsive potential at the boundaries was shown to result in the onset of surface-bound oscillatory states \cite{deGraaf2016May,kuron2019hydrodynamic,Lintuvuori2016Sep}.
In confined spaces, such as narrow microchannels, it has been shown that microswimmers exhibit reduced swimming speed and oscillatory trajectories \cite{jana2012paramecium,zhu2013low}. 
\citet{ahana2019confinement}  analyzed the squirmer dynamics in a channel using 2D lattice Boltzmann simulations. They reported various dynamical states for squirmers namely, (i) sliding, (ii) channel-wide oscillations, and (iii) damped oscillations about the channel centreline. 
It has also been reported that the dynamics are strongly governed by the source-dipole and force-dipole modes and are relatively less sensitive to the higher order squirming modes.
The far-field approach is extended to more complex geometries like tapered channel, and it has been reported that the dynamics are dependent on the initial conditions \cite{dhar2020hydrodynamics}. 
A recent study \cite{daddi2021hydrodynamics} demonstrated that remote hydrodynamic interaction of microswimmers with surfaces can result in strategic trajectories that have a survival advantage.

The self-diffusiophoretic particles interact with confinement differently than their biological counterparts. In addition to $ O  (1/r^{2}) $ hydrodynamic interaction, walls also alter the solute distribution: the slowly decaying chemical signals ($ \sim 1/r $) are altered in the presence of walls, which consequently modifies the net diffusio-osmotic slip on the particle surface, yielding a change in particle kinematics.
The first detailed study on confined Janus particles was performed by \citet{Popescu2009May}. They reported enhanced self-propulsion in spherical shell confinement.
Experiments and Brownian dynamics simulations performed by \citet{Kreuter2013Nov} on magnetically controlled active particles showed the existence of wall-sliding state, accompanied with suppressed rotational diffusion and enhanced directionality.

In the presence of single walls, \citet{crowdy2010two} used complex analysis in conjunction with the reciprocal theorem to find exact expressions for kinematics and dynamical trajectory of a 2D circular Janus particle.
Using boundary element method, \citet{Uspal2014} analyzed a Janus sphere with catalytic coverage ($ \theta_{c} $) ranging from $ \pi/2 $ to $ \pi $ (see Fig. \ref{fig:schematic}-b). They revealed the onset of two near-wall dynamical states: a sliding state at $ \theta_{c}=\cos^{-1} (-0.35) $, and a hovering state at $ \theta_{c} = \cos^{-1} (-0.85) $.
In the sliding state, the particle grazes the wall with its orientation slightly tilted towards the bottom wall; in the hovering state, the particle comes to a halt with its orientation pointing perpendicularly into the wall.
Using a bispherical coordinate system, \citet{Mozaffari2016May} also showed similar observations.
Later, \citet{ibrahim2016walls} used a far-field theory where they employed the method of images to provide analytical expressions for chemical and hydrodynamic wall-effects.
To keep the evaluation tractable, the chemical field was truncated to first four moments of the activity function; the more rapidly decaying disturbances (associated with higher activity moments) were neglected.
A recent study by \citet{Popescu2017Apr} has shed light on the consequence of the truncations (based on far-field approximation) on the dynamics of Janus particles. They note that although the truncations are necessary to facilitate analytical calculations that offer conceptual insights, the near-wall states may modify when considering higher concentration moments. They emphasized on the importance of the second concentration moment, neglected in previous studies \cite{Ibrahim2015Sep}, without which the sliding and hovering states do not appear.
A recent Lattice-Boltzmann study \cite{Shen2018Mar} drew comparisons between squirmers \& Janus particles, in terms of pure hydrodynamic interactions with single wall. They reported that Janus particles with weak to moderate force-dipole field are prone to surface entrapment.

The aforementioned theoretical and numerical studies offer practical insights into designing micromachines that can sense and respond to their immediate surroundings. Findings on near-wall states motivated further studies on exploitation of topographical features and surface chemistry of walls to attain enhanced control over the trajectory \cite{uspal2016guiding,simmchen2016topographical,das2015boundaries,popescu2018effective}. Furthermore, there have been recent studies towards understanding the effect of confinement on active-particle interactions and self-organization \cite{thutupalli2018flow,kanso2019phoretic}.
\ac{\citet{thutupalli2018flow} used experiments and simulations to demonstrate an ordering mechanism of active droplets that can be controlled by the hydrodynamic boundary conditions (such as no-slip and free-slip) at the bounding surfaces.}
In their theoretical study of dilute suspension of Janus spheres in a Hele-Shaw cell, \citet{kanso2019phoretic} highlighted the importance of hydrodynamic and phoretic interactions (between both particle-particle and particle-wall) and reported swirling \& clustering modes.

\begin{figure}
	\centering
	\includegraphics[scale=0.37]{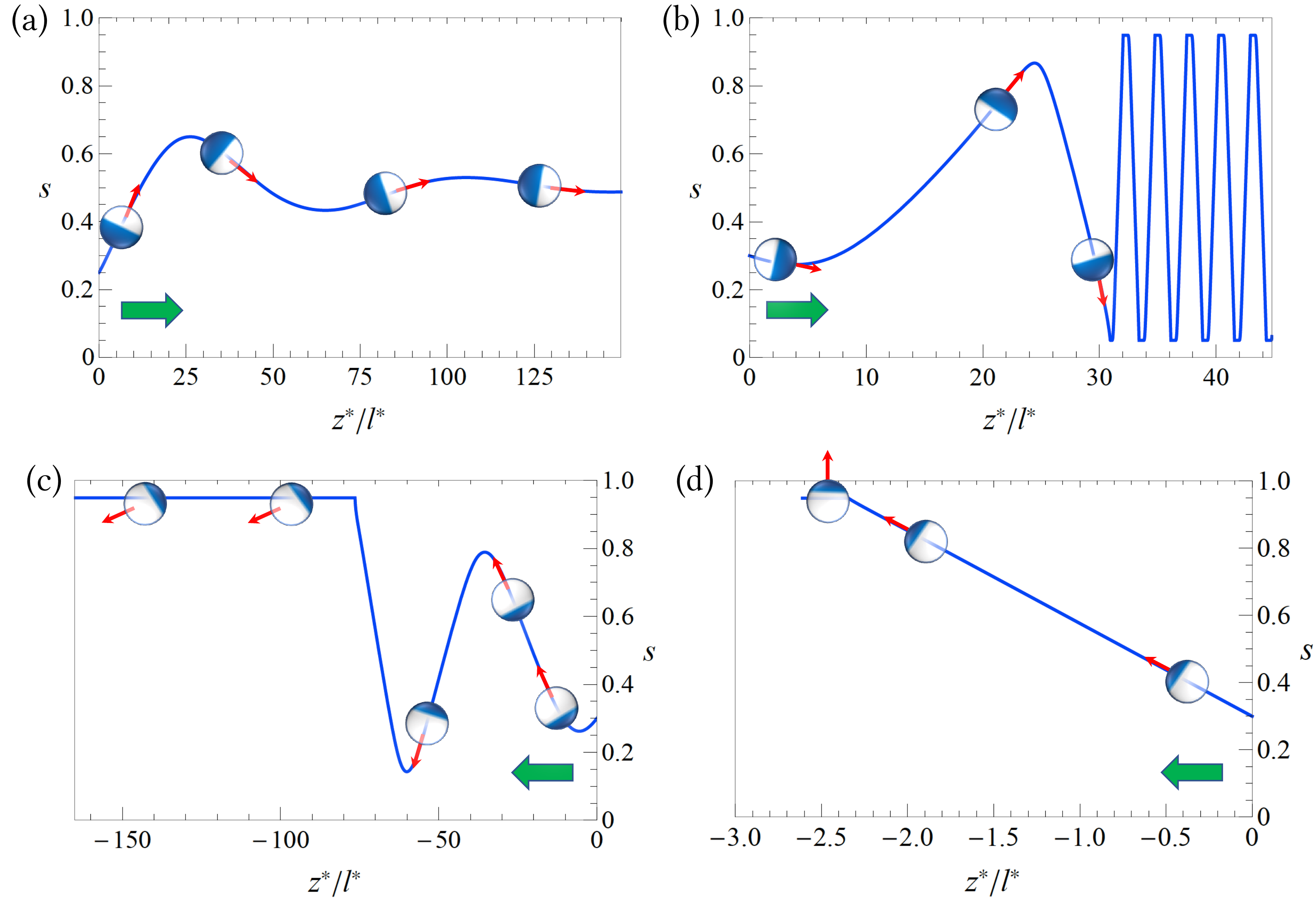}
	\caption{{\small Illustrations of the two new states (a) damped oscillations, (b) channel oscillations, which occur for inert and active facing particles, respectively. (c) Sliding and (d) Hovering states for same activity coverage but for different mode of propulsion: inert and active facing, respectively.}}%
	\label{fig:intro_illustration}%
\end{figure}

Single wall analysis can only provide insights into the near-wall states (hovering and sliding). It remains currently unexplored as to what happens to the \textit{reflected/scattered} particle in a confinement? And how accurate are single-wall corrections \cite{ibrahim2016walls} for determining  dynamics in 2D confinement?
To answer this, we take a far-field approach based on the method of reflections used in conjunction with the assumption of short-ranged repulsion at the walls.
In addition to the near-wall sliding and hovering states (see Fig. \ref{fig:schematic} c-d), we find two new dynamical states in the channel bulk: $ (i) $ \textit{damped oscillations} around the centerline (Fig. \ref{fig:intro_illustration}a), and $ (ii) $ channel-wide periodic \textit{oscillations} (Fig. \ref{fig:intro_illustration}b).
These states are found to depend on the surface characteristics of the particle.
%If solute molecules are attracted (or repelled) to the particle, the particle propels with inert face forward, provided the solute is consumed \cite{golestanian2007designing}. If either of the conditions is reversed (\textit{i.e.} solute repulsion + consumption or solute attraction+release), the particle propels with active face forward.
We find that the wall-induced chemical effects act to repel the inert facing and attract the active facing Janus particle, also reported in single wall studies \cite{Ibrahim2015Sep,ibrahim2016walls}.
%For symmetrically coated Janus particles, 
Due to this, when hydrodynamic effects are weak (for near-half activity coverage), the chemical effects dominate the trajectory, which results in damped and periodic oscillations for inert and active facing particles, respectively.
This dependence of channel-spanning states on the mode of propulsion is also prevalent in near-wall  states. 
Fig.\ref{fig:intro_illustration}(c) shows a sliding state for inert facing Janus particle and fig. \ref{fig:intro_illustration}(d) shows a hovering for the active-face-forward propulsion, while the activity coverage remains constant. 
	The analysis also sheds light on the potential limitation associated with the single wall studies. For instance, in Fig. \ref{fig:intro_illustration} (c) we show that the particle, at first few occasions, reflects from the walls before attaining near-wall sliding state.
%\textcolor{blue}{Do you feel the para. is long?}

The article is organized as follows:
in section II, we outline the problem formulation. 
%The concentration field is evaluated using the method of reflections in section III. Section IV describes the velocity field. 
The concentration and velocity fields are then obtained successively using the method of reflections in Sections III and IV, respectively.
The wall-effects on translation and rotation of the particle are detailed in section V, where
kinematics, trajectories,  and phase diagrams are discussed over a wide range of parameter space (particle size, activity coverage, and orientation). Key conclusions and future scope are discussed in section VI.

%%%%%%%%%%%%%%%%%%%%%%%%%%%%%%%%%%%%%%%
%%%%%%%%%%%%%%%%%%%%%%%%%%%%%%%%%%%%%%%

%\pagebreak

\section{Problem formulation}
\subsection{Phoretic propulsion of Janus particles}
\begin{figure}[H]
	\centering
	\includegraphics[scale=0.55]{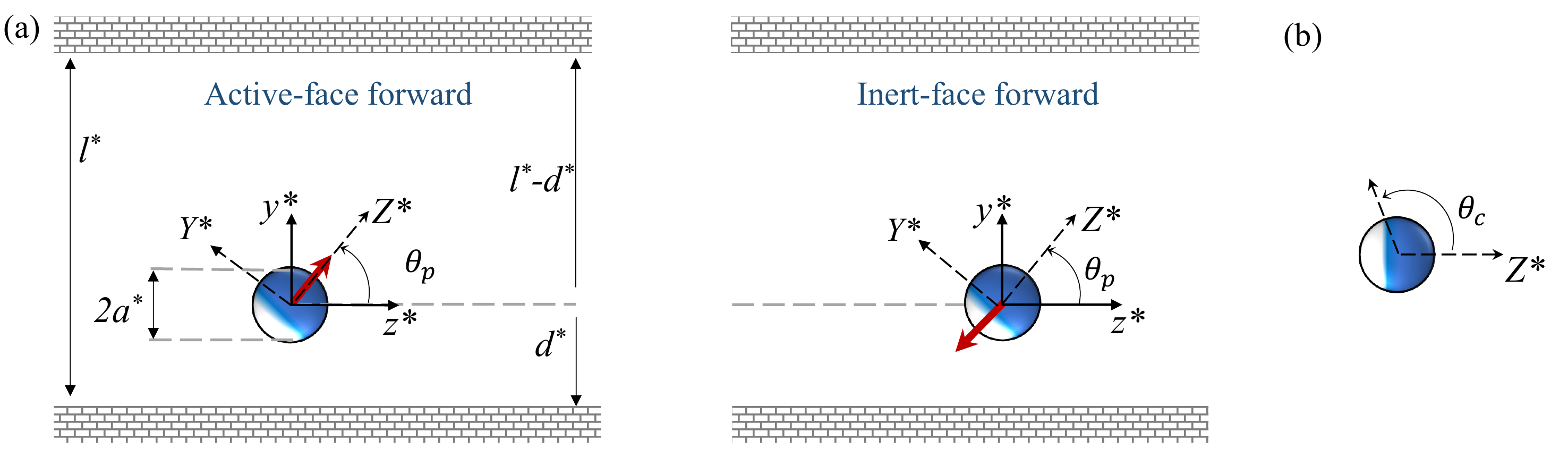}
	\caption{{\small (a) Arbitrarily oriented Janus particle between two parallel walls that extend to infinity in $ x^{*} $ and $ z^{*} $ directions. Here, the red arrow shows the propulsion direction; the blue cap indicates the active part where solute is consumed at a constant rate. The left particle propels with its active face forward, whereas the right one is inert facing.  The two coordinate systems are illustrated: channel frame (solid arrows) and particle frame (dashed arrows).
			(b) Schematic shows the catalytic activity coverage ($ \theta_{c} $).
	}}%
	\label{fig:schematic}%
\end{figure}

We consider a catalytically active particle of radius $ a^{*} $ in a Newtonian fluid. 
	Solute consumption or release at the active site creates a concentration gradient.
	As the solute interacts with the particle surface, this chemical imbalance generates a longitudinal pressure gradient in an infinitesimally thin interaction layer, that drives a net diffusio-osmotic slip along the particle surface. If the particle is freely suspended, it performs a self-diffusiophoretic motion, whose direction is determined by two properties: particle-solute interaction (\textit{i.e.} attractive or repulsive) and consumption/release of solute at the active site.
	In the case of solute consumption at the active site (considered throughout the manuscript unless stated otherwise), attractive (respectively repulsive) particle-solute interaction result in inert-face forward propulsion (respectively active-face-forward). The direction of propulsion would be reversed for particles that release solute from the active site.
%	Inert-face forward movement occurs, provided these two requirements are met: (i) attractive (or  repulsive) particle-solute interaction and (ii) solute consumption (or release) from the active site. If any one of the conditions is reversed, the particle moves with active-face forward.
The schematic in Fig. \ref{fig:schematic} shows a neutrally buoyant Janus particle (both cases of active and inert faced propulsion) in a 2D planar confinement of separation $ l^{*} $, bounded in $ y^{*} $ direction and unbounded in axial $ (z^{*}) $ and in-plane ($ x^{*} $) direction. 
We define two sets of axes centered on the particle at any instant in time $ t $: (i) the first one is ($ x^{*},y^{*},z^{*} $) where $ z^{*} $ axis is parallel to the channel, (ii) the second one is ($ X^{*},Y^{*},Z^{*} $), where $ Z^{*} $ axis is aligned parallel to the symmetry axis of the particle and is directed towards the active face.
%\ac{The propulsion direction of the particle, located at an arbitrary distance from bottom wall ($ d^{*} $), makes an `orientation angle' $ \theta_{p} $ with the $ z^{*} $ axis for an active-face forward particle. }
\ac{$ \theta_{p} $ denotes the `orientation angle' between the $ z^{*} $ and $ Z^{*} $ axes.}
We consider the solute-particle interactions to be uniform over the entire particle surface (\textit{i.e.} uniform mobility coefficient), and surface coverage of activity $ (\theta_{c}) $ to be arbitrary \textit{i.e.} $ 0<\theta_{c}<\pi $. 

The typical self-propulsion speed ($ U^{*} $) of a Janus particle of size ($ a^{*}  $) $ \sim 5 \mu $m is $  \sim10^{-6} $ m/s \cite{golestanian2005propulsion}. For solute diffusion coefficient  $ D^{*} \sim10^{-9} $ m$ ^{2} $/s, the associated  Péclet number ($\mbox{Pe} = U^{*} a^{*}/D^{*} $) is small $ \sim 10^{-2} $.
Therefore, in the current work, we neglect advective effects \textit{i.e.} $ \mbox{Pe} \rightarrow 0 $. Thus, the  disturbance concentration field around the particle is governed by the Laplace equation and boundary conditions are governed by a step flux at the particle surface and no-flux at the channel walls.
\begin{subequations}\label{Conc}
	\begin{gather}
	\nabla^{2} c = 0, \label{GE:conc}\\
	 \left.\PD{c}{r}\right\vert_{r=1}
	 = \mathcal{A}(\theta) = \left\{
	 \begin{array}{ll}
	 	1 \qquad \theta \leq \theta_{c}\\
	 	0 \qquad \theta > \theta_{c}
	 \end{array}
	 \right. , \quad  \left.\PD{c}{y}\right\vert_{walls} = 0, \quad
	 \quad \mbox{and } \quad c\rightarrow0 \;  \; \mbox{as } \lbrace |x^{*}| , |z^{*}| \rbrace \rightarrow \infty.
	 \label{BC:conc}
	\end{gather}
\end{subequations}

The inertia-less hydrodynamics is governed by the Stokes equation. The boundary conditions at the particle surface are that associated with rigid body motion and a diffusio-osmotic slip, whereas the channel walls satisfy the no-slip boundary condition.
\begin{subequations}\label{Hyd}
	\begin{gather}
		\bnabla \cdot \IB{u}=0, \quad -\bnabla p + \nabla^{2}\IB{u}  = \IB{0}, 		\label{GE:vel} \\
		\left.\IB{u}\right\vert_{r=1} = \IB{U}_{s} + \IB{\Omega}_s \times \IB{r} + M \bnabla_{\!s} c \quad\quad \mbox{and } \quad 	\left.\IB{u}\right\vert_{walls} = 0 . \label{BC:hyd}
		\end{gather}
\end{subequations}
The variables in the above equations are non-dimensionalized using $ a^{*} $, $ |\mathcal{A}^{*}|a^{*}/{D^{*}} $, $ |\mathcal{A}^{*}| |M^{*}|/D $, $ \mu^{*}|\mathcal{A}^{*}| |M^{*}|/{a^{*}D^{*}} $ as the characteristic scales for length, concentration ($ c $), velocity ($ \IB{u} $), and pressure ($ p $). Here, $  |\mathcal{A}^{*}| $ and $  |M^{*}| $  are the maximum magnitude of dimensional activity and dimensional mobility, respectively.
In the above boundary conditions (\ref{BC:conc}) and (\ref{BC:hyd}), $ r $ represents the non-dimensionalized radial distance $ (X^{2}+Y^{2}+Z^{2})^{1/2} $. The walls are located at $ y=-s/\kappa $ and $ y=(1-s)/\kappa $, where $ \kappa $ is the particle ($ a^{*} $) to channel size ($ l^{*} $) ratio, and $ s $ is the distance at which particle is located, measured from the bottom wall and scaled by the channel size
\begin{equation}
	\kappa = \frac{a^{*}}{l^{*}} \quad \mbox{and\ } \quad s = \frac{d^{*}}{l^{*}}.
\end{equation} 
$ \IB{U}_{s} $ and $ \IB{\Omega}_{s} $ represent the unknown translational and angular velocities of the spherical Janus particle. The tangential diffusio-osmotic slip velocity is denoted by $ M \bnabla_{\!s} c $, where $M$ is the uniform mobility coefficient and $ \bnabla_{\!s} $ is the surface gradient operator.

The self-propulsion of Janus particle modifies both the fluid flow and solute concentration around it. 
 This disturbance is further altered in the presence of boundaries. Therefore, the current configuration is a three-body interaction problem; namely, the chemo-hydrodynamic interaction between (i) particle and bottom wall, (ii) particle and top wall, (iii) top and bottom wall.
 In the current problem, the hydrodynamic interaction modifies the viscous drag and torque, and the chemical interaction affects the diffusio-osmotic slip at the particle surface $ (M\bnabla_{s}c) $; consequently further modifying the viscous drag and torque.
Thus, the particle interacts both hydrodynamically and chemically with the channel walls.

\ac{\subsection{A note on the diffusion problem in confined environment}\label{note}
	For small $\mbox{Pe}$, the steady diffusion model with Neumann (i.e. fixed-flux) boundary condition, Eq.~\eqref{Conc}, is well-posed in an unbounded 3D-domain as 3D diffusion is able to balance the constant  production or consumption at the particle's surface with a $1/r$-decay of the corresponding concentration field. This is not the case in a confined environment such as a Hele-Shaw cell, where far-field diffusion is two-dimensional ($x^*,z^*\rightarrow \infty$) leading to a $O(\log r)$ divergence of the concentration field if the particle is a net source/sink of solute~\citep{kanso2019phoretic}. This logarithmic far-field divergence arises for $r\gg 1/\kappa$, i.e. at distances much larger than the channel width, as the particle and its image system behave as an effective infinite line of sources. This is problematic when considering the concentration gradient and interactions of multiple particles located at relative distances of $1/r$, and the problem must be regularised by considering the long-time behaviour of the unsteady diffusive dynamics~\citep{sondak2016}, a small bulk reactivity of the solute to relax to its background concentration (effectively replacing the Laplace equation by a Helmholtz one, $\nabla^2 c=k^2 c$ with $k\rightarrow 0$) or a large but finite absorbing boundary~\citep{hu2019}.
	
	Here, however, we are interested in the interaction of a single particle with the confining walls, thereby focusing specifically and exclusively on the behaviour of the concentration field over $O(1/\kappa)$-distances from the particle's center. At such scales, the presence of the wall introduces an infinite yet discrete set of images to reconstruct the leading order effect of the confining walls, leading to a logarithmic divergence of the mean (uniform) concentration level but not of its spatially dependent part. At the scale of the channel width, the steady diffusion problem is therefore regular and well-posed for the concentration gradient, which completely determines the phoretic forcing on the hydrodynamic flow, and hence the propulsion problem: note that the phoretic propulsion problem is completely independent of the mean concentration level from a mathematical point of view. This justifies the well-posedness of the steady state diffusion formulation in Hele-Shaw configuration.
	
	Nevertheless, in particular when the particle is consuming solute, the mean relative concentration is important physically as an implicit constraint in the formulation is that the solute is not entirely consumed. Physically, however, we can see the present steady diffusion formulation as a limit of one of the three regularised formulations introduced above (i.e. long-time unsteady dynamics, small bulk reactivity of the solute or large neutralising side boundaries) with a large enough background concentration of solute so that the total concentration of solute remains positive over the time scale of the dynamics considered here.}

\subsection{Method of reflections}

Assuming that the particle is significantly smaller than the channel width $ (\kappa \ll 1) $, we use the method of reflections to seek solution to (\ref{Conc}) \& (\ref{Hyd}). 
When the small particle is not too close to the walls (\textit{i.e.} $ s \gg \kappa $), the disturbance variables $ \xi $ (representing $ c $, $ \IB{u} $, $ p $, $ \IB{U}_{s} $, and $ \IB{\Omega}_{s} $) can be sought as successive reflections: $ \xi = \xi_{1} + \xi_{2} + \xi_{3} \cdots $ \cite{brenner1962,happel2012low}.
Here, $ \xi_{i} $ represents the $ i^{th} $ reflection, with the odd reflections satisfying boundary conditions at the particle surface and even reflections satisfying boundary conditions at the walls.
Accounting for each successive pair of reflections increases the accuracy by $ O(\kappa) $ \cite{happel2012low}. This iterative process is performed until a desired accuracy is obtained.
%In this work, we solve for first three reflections of concentration and velocity field.

The equations governing the reflections of concentration field are:
\begin{subequations}\label{MOR:Conc}
	\begin{gather}
		\nabla^{2} c_{1} = 0, \qquad 
		\left.\PD{c_1}{r}\right\vert_{r=1}
		=\mathcal{A}(\theta) 	,
		\qquad c_{1} \rightarrow 0 \mbox{ as\ } r \rightarrow \infty;
			\label{MOR:Conc-1}\\ \nonumber \\
				\nabla^{2} c_{2} = 0, \qquad 
				\left.\PD{c_{2}}{y}\right\vert_{walls}
		=-\left.\PD{c_{1}}{y}\right\vert_{walls}
		\mbox{ at\ } y=-s/\kappa \; \& \; (1-s)/\kappa;
		\label{MOR:Conc-2}\\ \nonumber \\
		\nabla^{2} c_{3} = 0, \qquad 
		\left.\PD{c_3}{r}\right\vert_{r=1}
		=-\left.\PD{c_{2}}{r}\right\vert_{r=1} 	,
		\qquad c_{3} \rightarrow 0 \mbox{ as\ } r \rightarrow \infty.
		\label{MOR:Conc-3}
	\end{gather}
\end{subequations}
Similarly, we write the equations governing the reflections of velocity field
\begin{subequations}\label{MOR:Vel}
	\begin{gather}
		\bnabla \cdot \IB{u}_{1}=0, \qquad \nabla^{2} \IB{u}_{1} -\bnabla p_{1}= 0, \qquad 
\left.\IB{u}_{1}\right\vert_{r=1} = \IB{U}_{s}^{\infty} + \IB{\Omega}_s^{\infty} \times \IB{r} + \left.M \bnabla_{\!s} c_{1}\right\vert_{r=1} 		\label{MOR:vel-1}
 \\ \nonumber \\
				\bnabla \cdot \IB{u}_{2}=0, \qquad \nabla^{2} \IB{u}_{2} -\bnabla p_{2}= 0 \qquad
			\left. \IB{u}_{2} \right\vert_{walls} = \left.-\IB{u}_{1} \right\vert_{walls} 	\mbox{ at\ } y=-s/\kappa \; \& \; (1-s)/\kappa
		\label{MOR:vel-2}
%				%
%		%
%		%
%				\bnabla \cdot \IB{u}_{3}=0, \qquad \nabla^{2} \IB{u}_{3} -\bnabla p_{3}= 0, \qquad 
%		\left.\IB{u}_{3}\right\vert_{r=1} = \left.-\IB{u}_{2}\right\vert_{r=1}  + \left.M \bnabla_{\!s} (c_{2} + c_{3}) \right\vert_{r=1} 
%		\label{MOR:vel-3}
	\end{gather}
\end{subequations}
\ac{$ \IB{U}_{s}^{\infty} $ and $ \IB{\Omega}_{s}^{\infty} $ in the above equations represent the unbounded particle velocities, which will be obtained by using force-free and torque-free conditions.}
%The first three reflections of concentration and velocity field capture the leading order wall effects. 
It is important to consider the third reflection of velocity field to obtain the leading order wall-induced correction of the particle translation and rotation velocity (the particle is absent from the second reflection problem). Since we are only interested in evaluating translation and rotational velocities, we can avoid solving for the third reflection by making use of Faxen's laws for evaluating leading order hydrodynamic wall effects:
%It is via the third reflection, the first order wall-effects enter.
\begin{equation}\label{FAXEN}
	\IB{U}_{s}^{hyd} =  \IB{u}_{2} ({\IB{r}}_0) , \, \mbox{and\ } \, \IB{\Omega}_{s}^{hyd} = \frac{1}{2} \IB{\omega}({\IB{r}}_0),
\end{equation}
where $ \IB{\omega} $ represents the curl ($ \nabla \times \IB{u}_{2} $), and $ {{\IB{r}}}_{0} $ represents the position of particle's center.
To evaluate leading order chemical effects, we use the reciprocal theorem \cite{stone1996propulsion}: 
\begin{equation}\label{LRT}
	\IB{U}_{s}^{chem} = \frac{-1}{4\pi} \int_{S_{p}}^{} M \bnabla_{\!s} (c_{2} + c_{3}) \, dS.
\end{equation}
\sm{We note that the previous equation includes only chemical interactions with the walls, by computing the free-space swimming velocity change resulting from the chemical reflections. Hydrodynamic influence of the confining walls on such corrections (i.e. chemo-hydrodynamic interactions) are indeed subdominant -- see \citet{Varma2019}), \ac{and are therefore not accounted in the current work}.}

\section{Concentration field in 2D confinement}

\subsection{First reflection: $ c_{1} $}
Following \citet{golestanian2007designing}, we obtain solution for the first reflection (particle in unbounded domain) of the concentration field (\ref{MOR:Conc-1}) as
\begin{equation}
	c_{1}(r,\theta)=\displaystyle\sum_{n=0}^{\infty} {\frac{-\mathcal{A}_{n}}{(n+1)}}\, \frac{P_{n}(\cos \theta)}{r^{n+1}},
	\label{c1}
\end{equation}
where $ P_{n} $ is the $ n $th order Legendre polynomial and $ \mathcal{A}_{n} $ are the coefficients of activity distribution:
\begin{equation}\label{Kcoeff}
	\mathcal{A}(\theta)=\sum_{n=0}^{\infty} \mathcal{A}_{n} P_{n}(cos\theta) . 
\end{equation}
These coefficients are found by taking an inner product of (\ref{Kcoeff}) with the Legendre polynomials, and are obtained as
\begin{equation}\label{A}
	\mathcal{A}_{0}=\frac{(1-\cos \theta_{c})}{2} \; \mbox{ and\ } \;  \mathcal{A}_{n}=\frac{-1}{2} (P_{n+1}(\cos \theta_{c}) - P_{n-1}(\cos \theta_{c})) \; \mbox{for\ } n \geq 1.
\end{equation}
The concentration field (\ref{c1}) can be transformed into the Cartesian frame and represented at the leading order as:
\begin{equation}
	c_{1} = \mathcal{K}_{0}\frac{1}{r} + \mathcal{K}_{1} \frac{Z}{r^{3}} + \mathcal{K}_{2} \left(  \frac{3Z^{2}}{2r^{5}} -  \frac{1}{2r^{3}} \right) + O\left(\frac{1}{r^{4}}\right),
	\label{c1Trunc}
\end{equation}
where 
\begin{equation}\label{K}
	\mathcal{K}_{n} = -\mathcal{A}_{n}/(n+1)
\end{equation}
The above solution is written in particle axes ($ X,Y,Z $), which makes an angle $ \theta_{p} $ with the channel's set of axes ($ x,y,z $) \textit{i.e.} $ Z = y \sin \theta_{p} + z \cos \theta_{p} $ (see Fig. \ref{fig:schematic}). Thus, in the channel set of axes, we obtain:
\begin{equation}
	c_{1} = \mathcal{K}_{0}\frac{1}{r} + \mathcal{K}_{1} \frac{(y \sin \theta_{p} + z \cos \theta_{p})}{r^{3}} + \mathcal{K}_{2} \, \frac{3}{2r^{5}} \, \left(z^{2} \cos^{2} \theta_{p}  + y^{2} \sin^{2} \theta_{p}   +  y \, z \,  \sin 2\theta_{p}  \right)- \mathcal{K}_{2} \frac{1}{2r^{3}} + O\left(\frac{1}{r^{4}}\right)
	\label{c1Trunc_transformed}
\end{equation}

\subsection{Second reflection: $ c_{2} $} \label{sec:c2}

The first reflection varies on the scale of the particle size $ a^{*} $, whereas the second reflection varies on the channel length scale $ l^{*} $ which is $ O(1/\kappa) $ in comparison \textit{i.e.} the walls are remotely
located. Therefore, the coordinates for second reflection are stretched, and are termed
as ‘outer’ coordinates. These are defined as
\begin{equation}
	 \tilde{r}=\kappa r, \quad \tilde{x}=\kappa x, \quad \tilde{y}=\kappa y, \quad \tilde{z}=\kappa z .
\end{equation}

To evaluate the wall reflection of concentration field, we first rescale the first reflection (\ref{c1Trunc_transformed}) in the outer variables: 
\begin{equation}
	\tilde{c}_{1} = \mathcal{K}_{0} \, \kappa\frac{1}{\tilde{r}} + \mathcal{K}_{1} \, \kappa^{2} \frac{(\tilde{y} \sin \theta_{p} + \tilde{z} \cos \theta_{p})}{\tilde{r}^{3}} + \mathcal{K}_{2} \, \kappa^{3} \, \frac{3}{2\tilde{r}^{5}} \, \left(\tilde{z}^{2} \cos^{2} \theta_{p}  + \tilde{y}^{2} \sin^{2} \theta_{p}   +  \tilde{y} \, \tilde{z} \,  \sin 2\theta_{p}  \right)- \mathcal{K}_{2} \, \kappa^{3} \frac{1}{2\tilde{r}^{3}} + O(\kappa^{4}).
	\label{c1_outer}
\end{equation}
\citet{faxen1922}, in his seminal work, represented the fundamental solution of Laplace’s equation in the form of Fourier integrals. Following Faxén, we write:
\begin{equation}
\sm{	\frac{1}{\tilde{r}} = \frac{1}{{4\pi }}\int\limits_{0}^{ + \infty } {\int\limits_{0}^{2\pi} {\exp\left[ \mathrm{i}\lambda\Theta-\frac{\lambda|\tilde{y}|}{2} \right] \, \mathrm{d}\phi \mathrm{d}\lambda} } \mbox{, and\ }}
	\label{1byR}
\end{equation}
\begin{equation}
							\sm{	\tilde{r} = \frac{1}{{\pi }}\int\limits_{0}^{ + \infty } {\int\limits_{0}^{ 2\pi} {\left( {1 - \exp\left[ \mathrm{i}\lambda\Theta-\frac{\lambda|\tilde{y}|}{2} \right]} \right)\left( {1 + \frac{{\lambda \left| \tilde{y} \right|}}{2}} \right)\frac{{ \mathrm{d}\phi \mathrm{d}\lambda }}{{{{{\lambda ^2}} }}}}} .}							
	\label{R}
\end{equation}
%Here, $ \Theta  = (\xi \tilde{x}+ \eta \tilde{z})/2 $ and $ \lambda=(\xi^{2}+\eta^{2})^{1/2} $. $ \xi $ and $ \eta $ are the variables in Fourier space. 
\sm{Here, $(\lambda,\phi)$ are the polar variables in Fourier space and $ \Theta  = (\tilde{x}\cos\phi+\tilde{z}\sin\phi)/2$.} 

Using (\ref{1byR}) and (\ref{R}), we transform the first reflection (\ref{c1_outer}) by taking derivatives of the above equations \cite{choudhary2019inertial}. 
\ac{For example: $ O(\kappa^{2}) $ term in (\ref{c1_outer}) contains $ \tilde{z}/\tilde{r}^{3} $ and $ \tilde{y}/\tilde{r}^{3} $, which are expressed using the above transformations as:
	\begin{subequations}
		\begin{gather}
%			\frac{\tilde{z}}{\tilde{r}^{3}} =  - \PD{}{\tilde{z}}\left( {\frac{1}{\tilde{r}}} \right) = \frac{1}{{2\pi }}\int\limits_{ - \infty }^{ + \infty } {\int\limits_{ - \infty }^{ + \infty } {\exp\left[ {{\mathrm{i}} \Theta  - {\lambda \left| \tilde{y} \right|}}/{2} \right]
%					\left( { - \frac{{\mathrm{i}\eta }}{2}} \right)\frac{{ \mathrm{d}\xi \mathrm{d}\eta }}{{2\lambda }}} },
	\sm{							\frac{\tilde{z}}{\tilde{r}^{3}} =  - \PD{}{\tilde{z}}\left( {\frac{1}{\tilde{r}}} \right) = \frac{1}{{4\pi }}\int\limits_{ - 0}^{ + \infty } {\int\limits_{0}^{2\pi} {\exp\left[ \mathrm{i}\lambda\Theta-\frac{\lambda|\tilde{y}|}{2} \right]
					\left( { - \frac{{\mathrm{i} \lambda \, \ac{\sin \phi} }}{2}} \right) \mathrm{d}\phi \mathrm{d}\lambda }},} 
			\label{ZbyR3} \\
			%
%			\frac{\tilde{y}}{\tilde{r}^{3}} =  - \PD{}{\tilde{y}}\left( {\frac{1}{\tilde{r}}} \right) = \frac{1}{{2\pi }}\int\limits_{ - \infty }^{ + \infty } {\int\limits_{ - \infty }^{ + \infty } {\exp\left[ {{\mathrm{i}} \Theta  - {\lambda \left| \tilde{y} \right|}}/{2} \right]
%					\left( \frac{\lambda}{2} \frac{\tilde{y}}{|\tilde{y}|} \right)\frac{{ \mathrm{d}\xi \mathrm{d}\eta }}{{2\lambda }}} }.
\sm{\frac{\tilde{y}}{\tilde{r}^{3}} =  - \PD{}{\tilde{y}}\left( {\frac{1}{\tilde{r}}} \right) = \frac{1}{{4\pi }}\int\limits_{0}^{ + \infty } {\int\limits_{0}^{2\pi} {\exp\left[ \mathrm{i}\lambda\Theta-\frac{\lambda|\tilde{y}|}{2} \right]
					\left( \frac{\lambda}{2} \frac{\tilde{y}}{|\tilde{y}|} \right) \mathrm{d}\phi \mathrm{d}\lambda} }.	}									
			\label{YbyR3}
		\end{gather}
\end{subequations}
}
At the leading order, the Fourier integral representation for the first reflection is:
\begin{equation}
	%\tilde{c}_{1} = \frac{1}{{2\pi }}\int\limits_{ - \infty }^{ + \infty } {\int\limits_{ - \infty }^{ + \infty } {\exp\left[ {{\mathrm{i}} \Theta  - {\lambda \left| \tilde{y} \right|}}/{2} \right]
	%	\;  h_{1}  \; \frac{{ \mathrm{d}\xi \mathrm{d}\eta }}{{2\lambda }}} } \; + \;
	%O(\kappa^{3}),
	\sm{\tilde{c}_{1} = \frac{1}{{4\pi }}\int\limits_{ 0 }^{ + \infty } {\int\limits_{ 0}^{ 2\pi } {\exp\left[ \mathrm{i}\lambda\Theta-\frac{\lambda|\tilde{y}|}{2} \right]
		\;  h_{1}  \; \mathrm{d}\phi\mathrm{d}\lambda} } \; + \;
	O(\kappa^{4}),}
	\label{c1_fax}
\end{equation}
with 
%\begin{align}
%&	h_{1}(\kappa,\eta, \xi, {\rm sgn(\tilde{y})},\theta_{p}) = \nonumber \\
%&	\kappa \, \mathcal{K}_{0} 
%	+ \kappa^{2} \, \mathcal{K}_{1} \left[\cos \theta_{p} (\frac{{-\mathrm{i}} \eta}{2}) + \sin \theta_{p} (\frac{\lambda}{2} \frac{\tilde{y}}{|\tilde{y}|}) \right] 
%	+ \kappa^{3} \mathcal{K}_{2} \left[ \cos^{2} \theta_{p} (\frac{-\eta^{2}}{8}) + \sin^{2}\theta_{p} (\frac{\lambda^{2}}{8}) + \sin 2\theta_{p} (\frac{- {\rm i} \eta }{8}  \frac{\lambda \tilde{y}}{|\tilde{y}|}) \right] + O(\kappa^{4}). 
%\end{align}
\sm{\begin{align}
&	h_{1}(\kappa,\lambda,\phi, {\rm sgn(\tilde{y})},\theta_{p}) = \nonumber \\
&	\kappa \, \mathcal{K}_{0} 
	+ \frac{\lambda\kappa^{2}}{2} \, \mathcal{K}_{1} \left[-\mathrm{i}\sin\phi\cos \theta_{p}  + \frac{\tilde{y}}{|\tilde{y}|}\sin \theta_{p} \right] 
	+ \frac{\lambda^2\kappa^{3}}{8} \mathcal{K}_{2} \left[ \sin^{2}\theta_{p} -\cos^{2} \theta_{p}\sin^2\phi  -\frac{\mathrm{i} \tilde{y}}{|\tilde{y}|}\sin 2\theta_{p}\sin\phi\right] + O(\kappa^{4}). 
\end{align}
}

The wall-reflected concentration field can be assumed to take the form of $ \tilde{c}_{1} $
%\begin{equation}
%	\tilde{c}_{2} = \frac{1}{{2\pi }}\int\limits_{ - \infty }^{ + \infty } \int\limits_{ - \infty }^{ + \infty } 
%	\frac{\exp\left[ {{\mathrm{i}} \Theta} \right]}{2\lambda}
%			\;  \left( h_{2} \, \exp[-\lambda \tilde{y}/2]  + h_{3}\, \exp[+\lambda \tilde{y}/2]  \right)  
%			\; \mathrm{d}\xi \mathrm{d}\eta ,
%	\label{c2_fax}
%\end{equation}
\sm{\begin{equation}
	\tilde{c}_{2} = \frac{1}{{4\pi }}\int\limits_{0}^{ + \infty } \int\limits_{0}^{ 2\pi } \exp\left[ {{\mathrm{i}} \lambda \Theta} \right]
	\left( h_{2} \, \exp[-\lambda \tilde{y}/2]  + h_{3}\, \exp[\lambda \tilde{y}/2]  \right)  
			\; \mathrm{d}\phi \mathrm{d}\lambda ,
	\label{c2_fax}
\end{equation}
}
where $ h_{2} $ and $ h_{3} $ are the unknown terms, which will be determined by using the boundary condition (\ref{MOR:Conc-2}). The derivatives involved in the boundary condition (\ref{MOR:Conc-2}) are
\sm{
\begin{subequations}
	\begin{gather}
	\PD{\tilde{c}_{1}}{\tilde{y}} =  \frac{1}{{4\pi }}\int\limits_{0 }^{ + \infty } {\int\limits_{0}^{2\pi}}\exp\left[ {\mathrm{i}}  \lambda \Theta -\frac{\lambda |\tilde{y}|}{2} \right] 
	\left( \frac{-\lambda}{2} \frac{\tilde{y}}{|\tilde{y}|} \right)  h_{1}   \;
	{\rm d} \phi {\rm d} \lambda ,
		\label{dc1/dy} \\
	\PD{\tilde{c}_{2}}{\tilde{y}} =  \frac{1}{{4\pi }}\int\limits_{ 0 }^{ + \infty } {\int\limits_{ 0}^{2\pi}} \exp\left[ {\mathrm{i}}  \lambda  \Theta \right]
\left( \frac{-\lambda}{2} \exp\left[ \frac{-\lambda |\tilde{y}|}{2} \right] h_{2} + \frac{\lambda}{2} \exp\left[ \frac{\lambda |\tilde{y}|}{2} \right] h_{3} \right)    \;
{\rm d} \phi {\rm d} \lambda .
		\label{dc2/dy}
	\end{gather}
\end{subequations}
}
Using the above two equations in (\ref{MOR:Conc-2}) , we obtain 
\begin{subequations}
	\begin{gather}
h_2=\frac{h_{1}^{+} + h_{1}^{-} e^{\lambda(1-s)}}{-1+e^{\lambda}}\quad \mbox{and\ } \quad
h_3=\frac{h_{1}^{-} + h_{1}^{+} e^{s \, \lambda}}{-1+e^{\lambda}}, \mbox{ where\ }
	\label{h2-h3}  \\
	 h_1^{+} = h_1|_{\tilde{y} = 1-s} \mbox{ and\ } h_1^{-}= h_1|_{\tilde{y} = -s} .
	\end{gather}
\end{subequations}
This completes the solution for wall reflected concentration field \eqref{c2_fax}. 

\sm{
\textbf{A note on the reflected concentration}: In \S\ref{note}, we clarified that gradient of the concentration field is regular and well-posed for a source in a Hele-Shaw geometry, although the absolute value of the concentration itself may possess a logarithmic singularity. This appears in the form of the $1/\lambda$-divergence of the integrand in Eq.~\eqref{c2_fax} arising from the contribution of $\mathcal{K}_0$ to $h_2$ and $h_3$. This singular contribution is however independent of $(\tilde{x},\tilde{y},\tilde{z})$ and therefore does not impact the solution of the phoretic problem. Alternatively, one may remove such singularity by defining $\tilde{c}_2$ as
\begin{equation}
	\tilde{c}_{2} = 
	\int\limits_{0}^{ + \infty } \left[  \frac{1}{{4\pi }} \int\limits_{0}^{ 2\pi } 
	 \exp\left[ {{\mathrm{i}}  \lambda \Theta} \right]
	\left( h_{2} \, \exp\left[-\frac{\lambda \tilde{y}}{2}\right]  + h_{3}\, \exp\left[\frac{\lambda \tilde{y}}{2}\right]  \right)  	\; \mathrm{d}\phi 
	\;
	-  \left( \frac{2\kappa\mathcal{K}_0}{\lambda} \;
	\right) 
	\right]  	\mathrm{d}\lambda 
			,
	\label{c2_fax_alt}
\end{equation}
which also satisfies the boundary condition Eq.~\eqref{MOR:Conc-2}, as the added correction is spatially uniform.
}

\subsection{Third reflection: $ c_{3} $} \label{sec:c3}
The third reflection of concentration is governed by (\ref{MOR:Conc-3}); its boundary condition relates $ c_{3} $ with second reflection $ c_{2} $. Since $ c_{2} $ is represented in Fourier integrals, it is convenient to use the following Taylor series approximation:
\begin{equation}\label{c3_BC_apx}
 	\left.\PD{c_2}{r}\right\vert_{r=1}  =  \kappa \left( \lim_{\tilde{r} \rightarrow 0}  \PD{\tilde{c}_{2}}{\tilde{r}}  \right)
 	+  \kappa^{2}   \left(\frac{\tilde{r}}{1!} \lim_{\tilde{r} \rightarrow 0} \frac{\partial^{2} \tilde{c}_{2} }{\partial \tilde{r}^{2}}\right) + \cdots.
\end{equation}
Substituting (\ref{c2_fax}) in (\ref{c3_BC_apx}) and simplifying, we obtain the boundary condition for $ c_{3} $ as:
\begin{equation}
	\left.\PD{c_3}{r}\right\vert_{r=1}  = -\left(	\left.\PD{c_2}{r}\right\vert_{r=1} \right) =
	 \mathcal{I}_{c \, i} \cos\theta - \mathcal{I}_{c \, ii} \sin \theta \sin \phi + O(\kappa^{4}),
\end{equation}
where 
\begin{subequations}
	\begin{gather}
	\mathcal{I}_{c \, i} = \kappa^{3} \int_{0}^{\infty} \left(\frac{- \mathcal{K}_{1} \cos \theta_{p}}{16}\right) \frac{2+e^{s \lambda} + e^{\lambda (1-s)}}{-1 + e^{\lambda}} \lambda^{2} \, d \lambda 
\quad 
\mbox{ and\ }  \nonumber  \\
\mathcal{I}_{c \, ii} = \kappa^{2} \int_{0}^{\infty} \frac{ \mathcal{K}_{0} \left( e^{s \lambda} -e^{\lambda(1-s)} \right)  + \kappa \mathcal{K}_{1} \sin \theta_{p} (\lambda/2) \left( e^{s\lambda} + e^{\lambda (1-s)} -2 \right)  }{4 (-1+e^{\lambda})}  \lambda \, d \lambda
\nonumber
	\end{gather}
\end{subequations}
The solution to Laplace equation governing the third reflection is obtained as
\begin{equation}\label{c3}
	c_{3} = \frac{-\mathcal{I}_{c \, i}}{2r^{2}} \cos \theta + \frac{\mathcal{I}_{c \, ii}}{2r^{2}} \sin \theta \sin \phi + O(\kappa^{4}) .
\end{equation}
The first three reflections of concentrations are thus evaluated and are given by (\ref{c1Trunc}), (\ref{c2_fax}), and (\ref{c3}), respectively. In the next section, we evaluate the reflections of velocity field using a similar approach.

\section{Velocity field in 2D confinement}

\subsection{First reflection: $ \IB{u}_{1} $}

The leading order solution to the first reflection (\ref{MOR:vel-1}) can be evaluated using Lamb's general solution \cite{lamb}. 
	\begin{align}\label{v1-Lamb}
	\IB{u}_{1} &= A_{1}^{z} \left( \IB{e}_{z} + \frac{z \, \IB{r}}{r^{2}} \right) \frac{1}{r} +
	\;  A_{1}^{y} \left( \IB{e}_{y} + \frac{y \, \IB{r}}{r^{2}} \right) \frac{1}{r}
		\;+ \; C_{1}^{x} \left(  \frac{y \IB{e}_{z}}{r^{3}} - \frac{z \IB{e}_{y}}{r^{3}}  \right)
		\nonumber \\
		& \quad +\; D_{1}^{i} \left( \frac{-\IB{r}}{r^{3}} + \frac{3z^{2} \IB{r}}{r^{5}}  \right) 
		\; +\; D_{1}^{ii} \left( \frac{-\IB{r}}{r^{3}} + \frac{3y^{2} \IB{r}}{r^{5}}  \right) 
		\; +\; D_{1}^{iii} \left( \frac{3y\, z \, \IB{r}}{r^{5}} \right) \nonumber \\
	& \quad+ \; 
	 B_{1}^{z} \left(-\IB{e}_{z} +\frac{3z\, \IB{r}}{r^{2}} \right) \frac{1}{r^{3}} \, + \,  B_{1}^{y} \left(-\IB{e}_{y} +\frac{3y\, \IB{r}}{r^{2}} \right) \frac{1}{r^{3}} . 
	\end{align}
Here, the coefficients $ A_{1}^{y} \, \& \, A_{1}^{z} $ are the stokeslet coefficient associated with $ y $ and $ z- $direction motion. Similarly, $ C_{1} $, $ D_{1} $, and $ B_{1} $ are the  rotlet, force-dipole, and source-dipole coefficients, respectively. 
$ C_{1}^{x} $ represents the singularity associated with  rotation along $ x- $axis. Since the force-dipole (originating from the second concentration mode) has a quadratic nature, the coefficient $ D_{1} $ is divided here in 3 parts, denoted by the coefficients $ D^{i}, \, D^{ii}, \, D^{iii} $.

	\begin{subequations}\label{v1-Lamb_coeff}
		\begin{gather}
			A_{1}^{y} = \frac{3}{4} U_{s\,y}^{1} + M \mathcal{K}_{1} \frac{\sin \theta_{p}}{2},
		\; \; \;  \quad
		B_{1}^{y} = \frac{-1}{4} U_{s\,y}^{1} - M  \mathcal{K}_{1} \frac{\sin \theta_{p}}{2}, \\
		A_{1}^{z} = \frac{3}{4} U_{s\,z}^{1} + M \mathcal{K}_{1} \frac{\cos \theta_{p}}{2},
		\; \; \;  \quad
		B_{1}^{z} = \frac{-1}{4} U_{s\,z}^{1} - M \mathcal{K}_{1} \frac{\cos \theta_{p}}{2},
			\; \; \;  \quad
			C_{1}^{x} = \Omega_{s\,x}^{1} , \\
			D_{1}^{i} = \frac{3\, M \mathcal{K}_{2}}{2}  \cos^{2} \theta_{p} ,
				\; \; \;  \quad
				D_{1}^{ii} = \frac{3\, M \mathcal{K}_{2}}{2}  \sin^{2} \theta_{p} ,
					\; \; \;  \quad
				D_{1}^{iii} = \frac{3\, M \mathcal{K}_{2}}{2}  \sin 2\theta_{p} .
		\end{gather}
	\end{subequations}
Here, $ \IB{U}_{s}^{1} $ $ \IB{\Omega}_{s}^{1}  $ represent the translation and angular velocity for the first reflection \textit{i.e.} isolated particle.
A force-free and torque-free swimmer in unbounded domain will have no contribution from stokeslet and rotlet singularities. Therefore, we impose $ A_{1}^{y}=A_{1}^{z}=C_{1}^{x}=0 $ and obtain:
\begin{equation}
U_{s\,y}^{1} =	U_{s\,y}^{\infty} = \frac{-2}{3} M \mathcal{K}_{1} \sin \theta_{p}; \quad 	U_{s\,z}^{1} =U_{s\,z}^{\infty} = \frac{-2}{3} M \mathcal{K}_{1} \cos \theta_{p}; \quad \Omega_{s \, x}^{1} = \Omega_{s \, x}^{\infty} = 0.
\end{equation}
The superscript $ \infty $ represents that the velocities correspond to the unbounded domain.
%We employ Brenner's parallel method of reflections \cite{happel2012low} \footnote{As opposed to series method of reflections where force-free and torque-free conditions are used at every reflection to find translational and angular velocity\cite{keh1985}.}, and find the next two reflections, keeping the $ \IB{U}_{s} $ and $ \IB{\Omega}_{s} $ unknown.
%%\textcolor{blue}{later discuss about the dominant disturbances.} 
%
%The second and third reflection of velocity disturbance are evaluated using approach similar to that outlined in \S III. We provide the details in the Appendix.

\subsection{Second reflection: $ \IB{u}_{2} $}
Similar to the procedure outlined in section \ref{sec:c2} for concentration field, we now evaluate $ \IB{u}_{2} $ using Faxen's transformations.
As in section \ref{sec:c2}, we first rescale the first reflection (\ref{v1-Lamb}) in the channel scale, and transform the first order velocity field $ \tilde{\IB{u}}_{1}= \left\lbrace \tilde{u}_{1}, \tilde{v}_{1}, \tilde{w}_{1} \right\rbrace  $:
\sm{
\begin{subequations}\label{v1_fax}
	\begin{gather}
		\tilde{u}_{1} = 
		\frac{1}{{2\pi }}\int\limits_{ 0 }^{ + \infty } \int\limits_{0}^{2\pi } 
		\exp\left[ {{\mathrm{i}}  \lambda \Theta - \frac{\lambda|\tilde{y}|}{2}} \right]
		\;  \left\lbrace  \frac{\mathrm{i}\tilde{y}}{|\tilde{y}|}\cos\phi \left( g_{2} + \frac{\lambda |\tilde{y}|}{2} \, g_{3} \right)  \right\rbrace
		\; \lambda\mathrm{d}\phi \mathrm{d}\lambda \\
		\tilde{v}_{1} = 
		\frac{1}{{2\pi }}\int\limits_{0 }^{ \infty } \int\limits_{0}^{2\pi} 
		\exp\left[ {{\mathrm{i}} \lambda  \Theta - \frac{\lambda|\tilde{y}|}{2}} \right]
		\;  \left\lbrace  
		\left(  \frac{\mathrm{i}\tilde{y}\sin\phi}{|\tilde{y}|}  \right) g_{1} - g_{2}
		- g_{3} \left( 1 + \frac{\lambda |\tilde{y}|}{2} \right)  \right\rbrace
		\; \lambda\mathrm{d}\phi \mathrm{d}\lambda \\
		\tilde{w}_{1} = 
		\frac{1}{{2\pi }}\int\limits_{ 0 }^{ + \infty } \int\limits_{0}^{2\pi} 
		\exp\left[ {{\mathrm{i}}  \lambda \Theta - \frac{\lambda|\tilde{y}|}{2}} \right]
		\;  \left\lbrace g_{1} + \left(  \frac{\mathrm{i}\tilde{y}\sin\phi}{|\tilde{y}|}  \right) \left( g_{2} +  \frac{\lambda |\tilde{y}|}{2} \, g_{3} \right)  \right\rbrace
		\; \lambda\mathrm{d}\phi \mathrm{d}\lambda
	\end{gather}
\end{subequations}
}
Here, $ g_{1}, \, g_{2}, \& \, g_{3} $ are the following terms:
\sm{
\begin{subequations}\label{g1g2g3}
	\begin{align}
		g_{1} &= \frac{A_{1}^{z} \kappa}{\lambda} + \frac{C_{1}^{z}  \kappa^{2} }{4} \frac{\tilde{y}}{|\tilde{y}|} - \mathrm{i}\sin\phi \frac{D_{1}^{i} \kappa^{2} }{2} + \frac{\tilde{y}}{|\tilde{y}|} \frac{D_{1}^{iii} \kappa^{2} }{4}, \\
		g_{2} &= \frac{-B_{1}^{y} \kappa^{3}  \lambda}{8} +   \frac{\mathrm{i}\tilde{y}\sin\phi}{|\tilde{y}|} \left( \frac{B_{1}^{z} \kappa^{3} \lambda}{8} + \frac{A_{1}^{z} \kappa }{2 \, \lambda}  \right) +  \frac{\tilde{y}\sin^2\phi}{|\tilde{y}|} \left( \frac{D_{1}^{i} \kappa^{2} }{4} \right) + \frac{\tilde{y}}{|\tilde{y}|} \frac{D_{1}^{ii} \kappa^{2} }{4}, \\
		g_{3} &= \frac{-A_{1}^{y} \kappa}{2 \, \lambda} + \frac{\mathrm{i}\tilde{y}\sin\phi}{|\tilde{y}|} \left( \frac{A_{1}^{z} \kappa }{2\, \lambda} \right) +  \frac{\tilde{y}}{|\tilde{y}|}  \left( \sin^2\phi\frac{D_{1}^{i} \kappa^{2} }{4} - \frac{D_{1}^{ii} \kappa^{2} }{4}  \right) + \mathrm{i}\sin\phi  \left( \frac{D_{1}^{iii} \kappa^{2} }{4} \right).
	\end{align}
\end{subequations}
}
The boundary condition in (\ref{MOR:vel-2}) suggests that the second reflection takes the form of (\ref{v1_fax}):
\sm{
\begin{subequations}\label{v2_fax}
	\begin{gather}
		\tilde{u}_{2} = 
		\frac{1}{{2\pi }}\int\limits_0^{ + \infty } \int\limits_{0}^{2\pi} 
		e^{{\mathrm{i}} \lambda  \Theta}
		\;  \left\lbrace  e^{-\lambda \tilde{y}/2} \left( g_{5} + g_{6} \,\lambda \tilde{y}/2  \right) 
		-  e^{+\lambda \tilde{y}/2} \left( g_{8} - g_{9} \,\lambda \tilde{y}/2  \right)   \right\rbrace \mathrm{i}\cos\phi
		\; \lambda\mathrm{d}\phi\mathrm{d}\lambda \\
		\tilde{v}_{2} = 
		\frac{1}{{2\pi }}\int\limits_0^{ + \infty } \int\limits_{0}^{2\pi}
		e^{{\mathrm{i}}  \lambda \Theta}
		\left\lbrace  e^{-\lambda \tilde{y}/2} \left( \mathrm{i}g_{4}\sin\phi - g_{5} - g_{6} (1+\lambda \tilde{y}/2)  \right) 
		-  e^{+\lambda \tilde{y}/2} \left( -\mathrm{i}g_{7} \sin\phi- g_{8} - g_{9} (1 - \lambda \tilde{y}/2)  \right)  \right \rbrace
		\; \lambda\mathrm{d}\phi\mathrm{d}\lambda \\
		\tilde{w}_{2} = 
		\frac{1}{{2\pi }}\int\limits_0^{ + \infty } \int\limits_{0}^{2\pi}
		e^{{\mathrm{i}} \lambda  \Theta}
		\;  \left\lbrace  
		e^{-\lambda \tilde{y}/2} \left(  g_{4} + \mathrm{i}\sin\phi\left( g_{5} + g_{6} \lambda \tilde{y}/2 \right)  \right) 
		+  e^{+\lambda \tilde{y}/2} \left(  g_{7} -  \mathrm{i}\sin\phi \left( g_{8} - g_{9} \lambda \tilde{y}/2 \right)  \right) 
		\right\rbrace 
		\; \lambda\mathrm{d}\phi\mathrm{d}\lambda 
	\end{gather}
\end{subequations}
}
Here, the six unknown terms $ g_{4}, \, g_{5}, \, g_{6}, \, \cdots g_{9} $ are determined using the boundary condition in (\ref{MOR:vel-2}); a system of six equations is formed to evaluate these terms as functions of known $ g_{1}, \, g_{2}, \mbox{ and\ } g_{3} $. These six terms are essentially functions of Fourier variable $ \lambda $ and coefficients of Lamb's solution. 
%The expressions are provided in the supplementary material.
The above equations (\ref{v1_fax}-\ref{v2_fax}) are for a general motion of particle (with non-zero stokelet and rotlet). For the current study $ A_{1} $ and $ C_{1} $ are zero for a freely suspended swimmer.

\subsection{Wall-induced modifications to $ \IB{U}_{s} $ and $ \IB{\Omega}_{s} $}\label{points}

Using (\ref{v2_fax}) in the Faxen's law (\ref{FAXEN}) and reciprocal theorem expression (\ref{LRT}), we obtain the velocity modification induced by the wall effects as:
\begin{subequations}\label{UOmega_detail}
	\begin{align}
		\IB{U}_{s}^{chem} &= \left[ \kappa^{2} \, M \mathcal{K}_{0} \mathbb{F}_{\mathcal{K}_0}^{y}     +     \kappa^{3} \, M \mathcal{K}_{1} \sin \theta_{p} \mathbb{F}_{\mathcal{K}_{1}}^{y} \right] \IB{e}_{y} + \left[  \kappa^{3} \, M \mathcal{K}_{1} \cos \theta_{p} \mathbb{F}_{\mathcal{K}_{1}}^{z}  \right] \IB{e}_{z} +O(\kappa^{4}) \label{U_s_chem}\\
		\IB{U}_{s}^{hyd} &= \left[  \kappa^{2} M \mathcal{K}_{2} \left( \cos^{2} \theta_{p} \mathbb{F}_{D}^{i} + \sin^{2} \theta_{p} \mathbb{F}_{D}^{ii} \right)  +  \kappa^{3}  M \mathcal{K}_{1} \sin \theta_{p} \mathbb{F}_{B}^{y}   \right] \IB{e}_{y} + \nonumber \\
		& \qquad \left[ \kappa^{2} \, M\mathcal{K}_{2} \sin 2\theta_{p} \mathbb{F}_{D}^{iii}  +   \kappa^{3} \, M\mathcal{K}_{1} \cos \theta_{p} \mathbb{F}_{B}^{z}  \right] \IB{e}_{z} +O(\kappa^{4})  \label{U_s_hyd}\\
		\Omega_{s\, x} & = \kappa^{3} \, M \mathcal{K}_{2} \sin 2\theta_{p} \mathbb{T}_{D}^{iii} +    \kappa^{4}   M \mathcal{K}_{1} \cos \theta_{p} \mathbb{T}_{B}^{z} + O(\kappa^{5}). \label{Omega_s}
	\end{align}
\end{subequations}
The total particle velocity is written as
\begin{equation}\label{UOmega}
    \IB{U}_{s} = \IB{U}_{s}^{\infty} + \IB{U}_{s}^{chem} + \IB{U}_{s}^{hyd}, \quad \mbox{and\ } \quad 
    	\IB{\Omega} = \Omega_{s\,x} \IB{e}_{x}.
\end{equation}
In the above equations, $ \mathbb{F}(s) $ and $ \mathbb{T}(s) $, represents the wall-induced drag and torque corrections. The subscripts $ B \mbox{ and\ } D$ represent the hydrodynamic corrections arising from  source-dipole and force-dipole singularities, respectively (notation corresponds to Lamb's coefficients in Eq. \ref{v1-Lamb}); the chemical corrections are represented via subscripts $  \mathcal{K}_{0}, \, \mathcal{K}_{1}  $ . The superscript $ y, \, z $ and $ \left\{ i, ii, iii \right\} $ correspond to the direction of the motion and the three force-dipole coefficients in (\ref{v1-Lamb_coeff}), respectively. The expressions for wall corrections are provided in the Appendix section. 
Therein, we also show a comparison of (\ref{UOmega_detail}) with wall-corrections obtained by \citet{ibrahim2016walls}.
A close match reveals that the interactions between top and bottom wall are weak. 
For this particular problem, this suggests that a superposition of single wall expression can accurately represent particle behavior in confinement; attributed to the $ \sim 1/r^2 $ decaying velocity field. 
Nevertheless, the framework outlined here is general and extends previous works \cite{ho1974,choudhary2019inertial} for arbitrary orientations and can be used for a variety of particle-wall problems involving longer-ranged disturbances. See \S VI for further discussion.
%Thus, for the current problem, results obtained for single walls \cite{ibrahim2016walls} can be alternatively used via superimposition of single-wall effects from top and bottom walls.

\vspace{3mm}

A closer look at (\ref{UOmega_detail}) reveals several insights into the particle movement.\\
\noindent
\textbf{1.} In agreement with earlier studies \cite{michelin2015autophoretic,ibrahim2016walls,yariv2016wall,yariv2017boundary}, we find that
one of the leading order wall corrections arising from chemical interactions (\ref{U_s_chem}) is devoid of directionality; in (\ref{UOmega_detail}a), the $ O(\kappa^{2}) $ term is independent of the orientation ($ \theta_{p} $). The proportionality to $ M $ and $ \mathcal{K}_{0} $ can be understood physically: for particles that consume solute ($ \mathcal{K}_{0} < 0 $), the presence of wall generates an increased depletion locally next to the wall and a gradient of solute oriented away from it.
%if the particle is a net source of solute ($\mathcal{K}_{0}>0$), the presence of the wall will generate an accumulation of solute next to it and a gradient of solute oriented toward it. Depending on the sign of the mobility the particle will experience a phoretic drift toward or away from the wall.
Thus, a fully coated catalyst particle (devoid of any directionality) will either be attracted to walls if the product $ M \mathcal{K}_{0} $ is positive and repelled if negative.
 It will be shown later that this leading order chemical interaction plays a critical role in determining the particle trajectories.

\noindent
\textbf{2.} For hydrodynamic wall correction (\ref{U_s_hyd}), the leading order term proportional to $ \cos^{2} \theta_{p} $ suggests a vertical movement of the swimmer for a horizontally orientation ($ \theta_{p}=0 $) \cite{berke2008hydrodynamic}. 
Proportionality to $ \mathbb{F}_{D}^{i} $ suggests that it originates from wall-reflection of force-dipole disturbance. 
%In (\ref{Usz}), the hydrodynamic and chemical wall-corrections affect the horizontal velocity at $ O(\kappa^{3}) $.

\noindent
\textbf{3.} It should be noted that the wall corrections are proportional to first three concentration modes $ \mathcal{K}_{0}, \, \mathcal{K}_{1}, $ and $ \mathcal{K}_{2} $. These are determined by the extent of surface coverage (see eq.\ref{A} and eq.\ref{K}).
% Because of this, 
\ac{Because of this, Janus-pushers and -pullers will experience different magnitude of chemical effects.
%i.e. the trajectories of Janus-pushers ($ \theta_{c}< \pi/2 $; $ M=-1 $) will differ from the trajectories traced by Janus-pullers ($ \theta_{c} > \pi/2 $; $ M=-1 $).
For example, the first three concentration modes ($ \mathcal{K}_{0}, $ $\mathcal{K}_{1}$, $ \mathcal{K}_{2} $) for $ \theta_{c}=\pi/4 $ (Janus-pusher) are $ \left\{ -0.15, -0.19, -0.15  \right\} $; while for a Janus-puller of identical hydrodynamic strength ($ \theta_{c}=5\pi/4 $), these are $ \left\{ -0.85, -0.19, +0.15  \right\} $. Since the latter is more catalytically coated (and thus has higher consumption $ \mathcal{K}_{0} $), it will experience greater wall-induced chemical effects (see $ O(\kappa^{2}) $ term in Eq. \ref{U_s_chem}).
For the same reason, 
%the trajectories of inert-faced propulsion ($ M \mathcal{K}_{1} > 0 $) will qualitatively differ from the trajectories traced by particles which propel with active face forward ($ M \mathcal{K}_{1} < 0 $). 
%In other words, 
a Janus-pusher propelling itself with inert-face forward will experience lower (wall-induced) chemical effects  than a Janus-pusher with active-face forward. Conversely, for Janus-pullers, particles with inert-face forward will experience greater chemical effects than active-face forward pullers. 
}

\noindent
\textbf{4}. As also reported in earlier studies \cite{Ibrahim2015Sep,ibrahim2016walls,kanso2019phoretic}, the rotational velocity arises solely from hydrodynamic wall interactions, which arrive at $ O(\kappa^{3}) $. 
For uniform mobility coefficient, there is no rotation induced by the diffusio-osmotic slip of any concentration distribution \cite{Uspal2014,ibrahim2016walls}.
For symmetrically coated Janus sphere ($\theta_{c}=\pi/2$), the correction to rotational velocity arrives at $ O(\kappa^{4}) $ because the force-dipole vanishes (\textit{i.e.} $ D=0 $, yielding $ \mathbb{F}_{D}=\mathbb{T}_{D}=0 $).
%As we shall show later, $ O(\kappa^{2}) $ term in (\ref{Usz}) and $ O(\kappa^{3}) $ term in (\ref{Omega}) vanish for half catalyst coverage ($ \theta_{c}=\pi/2 $) because the force-dipole is non-existent (\textit{i.e.} $ D=0 $, which yields $ \mathbb{F}_{D}=\mathbb{T}_{D}=0 $).

%The hydrodynamic interactions, however, are equivalent in the two cases.

\vspace{2mm}

Following the theoretical studies on biological swimmers \cite{lauga2009hydrodynamics}, we characterize the strength of force-dipole field relative to the source-dipole field by defining:
\begin{equation}\label{beta}
	\beta =\frac{D_{1}}{B_{1}} = \frac{-9 \mathcal{K}_{2}}{2 \mathcal{K}_{1}},
\end{equation}
where a negative (positive) value represents Janus-pusher (puller), provided that (\textit{i}.) solute is consumed at the active site and (\textit{ii}.) solute is attracted to the particle \textit{i.e.} $ M<0 $. If either of these conditions reverses the definition of Janus-pusher and -puller reverses.
In this work, we explore the catalytic coating ranging from $ \pi/12 \leq \theta_{c} \leq 11\pi/12 $, which correspond to the ratio $ \beta $ in the range: $ -4.83 \leq \beta \leq 4.83 $.
% \textit{i.e.} the study deals with Janus-pushers and -pullers of weak to moderate strength.
%In the next section, we analyze the instantaneous particle velocity, and explore the parameter space to find various  states.

%\vspace{2mm}

%%%%%%%%%%%%%%%%%%%%%%%%%%%%%%%%%%%%%%%
%%%%%%%%%%%%%%%%%%%%%%%%%%%%%%%%%%%%%%%

\pagebreak

\section{Results and discussion}
In this section, we first discuss the effect of confinement on the instantaneous kinematics characterized in terms of translational and angular velocities followed by the long-time dynamics by tracing the trajectories. Later, we classify these trajectories and represent the dynamics in a phase-diagram over a wide range of parameter space.

\subsection{Half coated Janus particle}
\subsubsection{Instantaneous particle velocity}

\begin{figure}[H]
	\centering
	\sidesubfloat[]{{\includegraphics[scale=0.55]{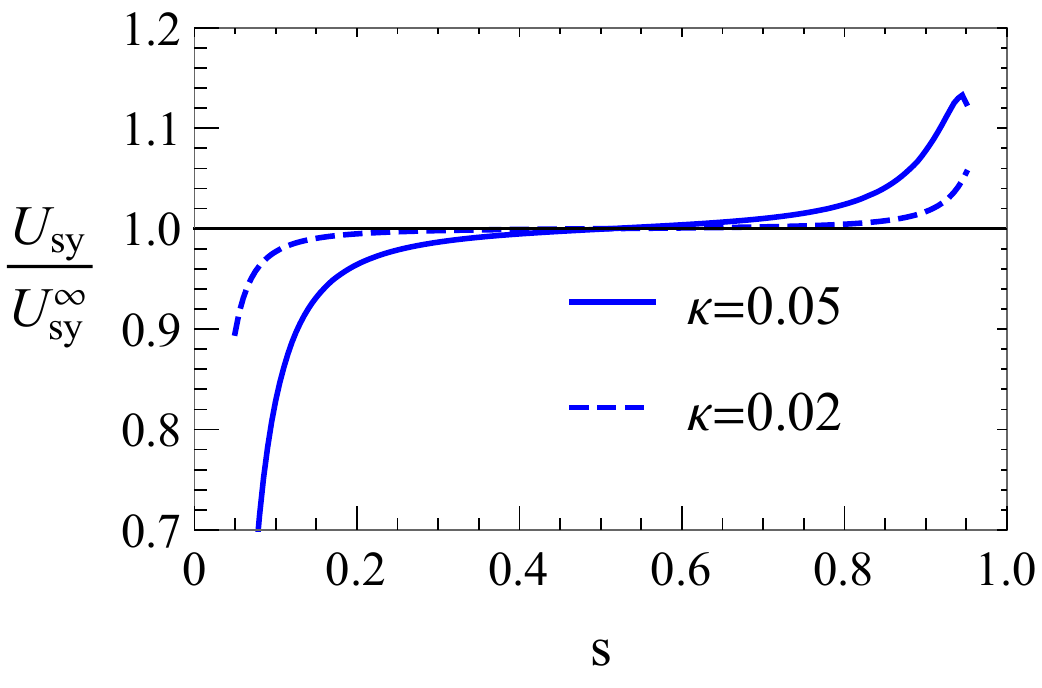} }}%
			\;
	\sidesubfloat[]{{\includegraphics[scale=0.59]{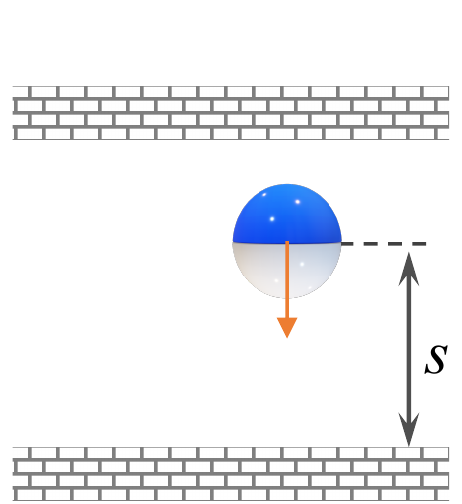} }}
	\;
	\sidesubfloat[]{{\includegraphics[scale=0.5]{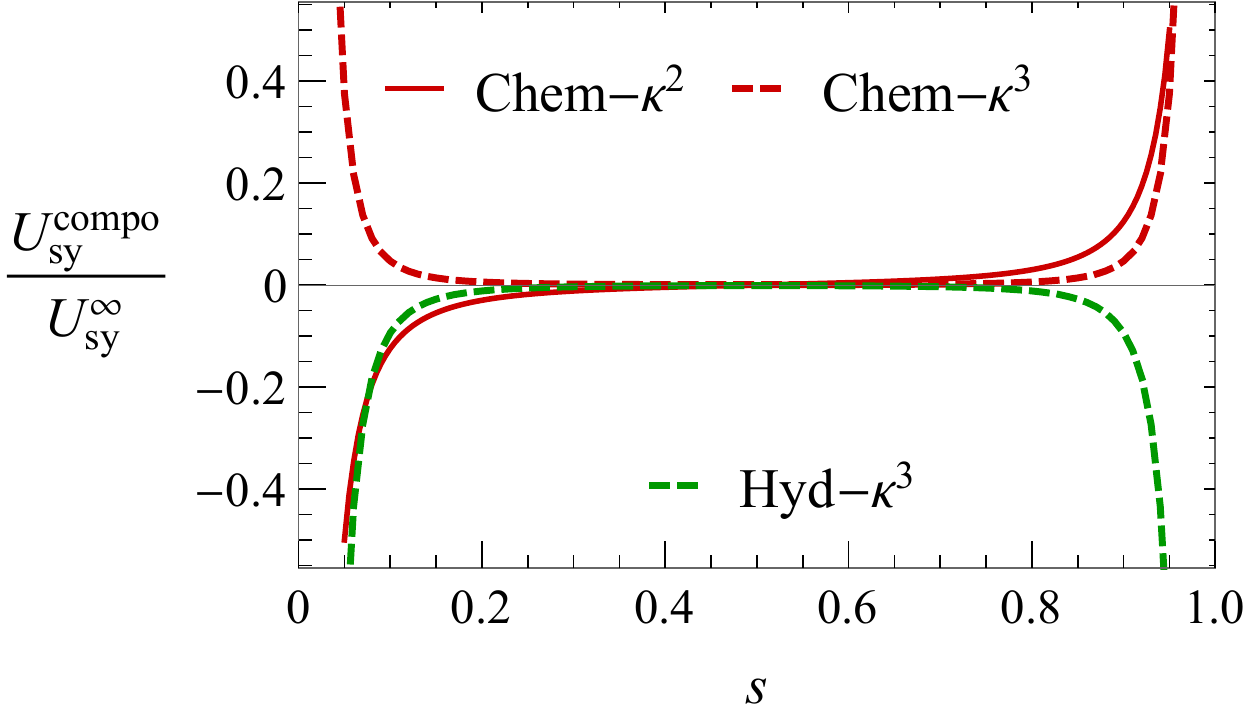} }}
	\caption{{\small (a) Instantaneous perpendicular velocity for a vertically aligned $ (\theta_p=\pi/2) $ Janus particle with half-coating $ (\theta_c=\pi/2) $ and attractive solute-particle interaction $ (M=-1) $. 
		(b) Schematic illustrating the configuration for (a): particle approaches the bottom wall with inert face and leaves the top wall with active face.
				(c) Chemical and hydrodynamic components of the wall-corrections to instantaneous perpendicular velocity for $ \kappa=0.05 $.
	Here the unbounded velocity is $ U_{s\,y}^{\infty} = -0.25. $}}%
	\label{fig:half-inst-perp}%
\end{figure}

We first analyze the instantaneous velocity (\ref{UOmega}) for a relatively simple scenario: half-coated Janus sphere ($ \theta_{c}=\pi/2 $), for which the second concentration mode $ \mathcal{K}_{2} $ is zero. Consequently, the force-dipole disturbance and its associated wall effects vanish \textit{i.e.} $ \mathbb{F}_{D} = \mathbb{T}_{D} = 0 $ in (\ref{UOmega_detail}).
Fig. \ref{fig:half-inst-perp}(a) shows the instantaneous velocity for two size ratios, where the particle is propelled downwards (negative $ y$-axis) and propels with inert face forward ($ M=-1 $) (schematic shown in Fig. \ref{fig:half-inst-perp}-b). 
We find that walls repel the particle: (i) when the particle is leaving the top wall, the velocity is increased ($ U_{s\, y}/U_{s\,y}^{\infty} > 1 $) , (ii) when the particle is approaching the bottom wall, the velocity is decreased ($ U_{s\, y}/U_{s\,y}^{\infty} < 1 $).
The plots show an asymmetric behavior across the centerline ($ s=0.5 $) because the chemical interactions with bottom wall are different from that with the top wall; the bottom wall faces the inert side of the particle, whereas the top wall faces the active side.

As the particle size increases, the curve in Fig. \ref{fig:half-inst-perp} (a) changes its shape because the $ O(\kappa^{3}) $ interactions become comparable to $ O(\kappa^{2}) $ interactions (the different components shown in Fig. \ref{fig:half-inst-perp}(c)). 
The  contribution from the $ O(\kappa^{3}) $ hydrodynamic correction is centerline-symmetric and a negative value represents that the particle slows down when approaching or departing from walls. 
The chemical correction, on the other hand, is composed of two terms. (i) The $ O(\kappa^{2}) $ correction (anti-symmetric about the centerline) repels both the inert and active face. This occurs because the Janus particle consumes solute ($ \mathcal{K}_{0} <0 $) and swims toward regions of high concentration. Thus, excess depletion of solute near the wall leads to repulsion.
(ii) The positive symmetric $ O(\kappa^{3}) $ interaction arises from first concentration mode ($ \mathcal{K}_{1} $) characterized by a potential dipole field ($ \sim y/r^{3} $). For the orientation under consideration (Fig.\ref{fig:half-inst-perp} b), the source ($ + $) (associate with potential dipole) is placed closer to the bottom wall and sink ($ - $) closer to the top wall. A reflection or image of this field at walls would suggest an enhancement in gradient near the walls (along negative $ y- $axis).
Consequently, making the walls attract inert face and repel the active face.
%a positive value below centerline shows that bottom wall speeds up the approaching particle, and for the top wall, a positive value represents that a leaving particle speeds up.
The aforementioned plots would flip around $ s=0.5 $ for a particle moving downward with positive mobility coefficient $ M $ (repulsive particle-solute interactions); \textit{i.e.} the anti-symmetric $ O(\kappa^{2}) $ would qualitatively reverse.

\floatsetup[figure]{style=plain,subcapbesideposition=top}

%
% \floatsetup[figure]{style=plain,subcapbesideposition=top}
%\begin{figure}[H]
%	\centering
%	\includegraphics[scale=0.6]{Omega.pdf}
%	\caption{{\small }}%
%	\label{fig:half-inst-Omega}%
%\end{figure}
%

\begin{figure}
	\centering
	\sidesubfloat[]{{\includegraphics[scale=0.53]{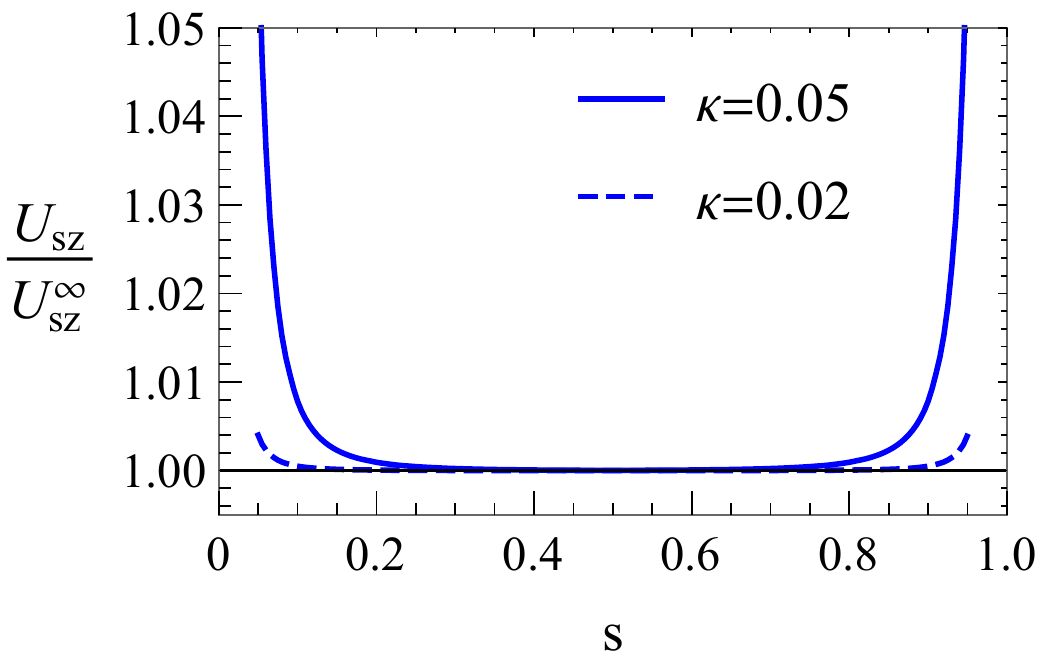} }}% \;
	$ \qquad $
	\sidesubfloat[]{{\includegraphics[scale=0.53]{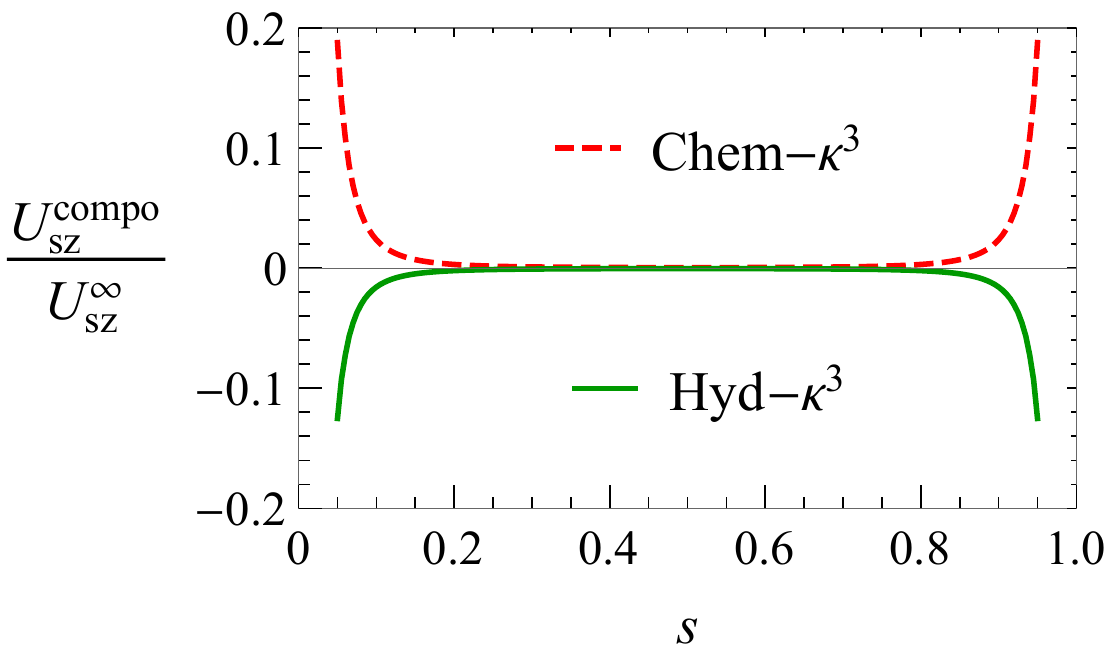} }}%
	\\
	\sidesubfloat[]{{	\includegraphics[scale=0.53]{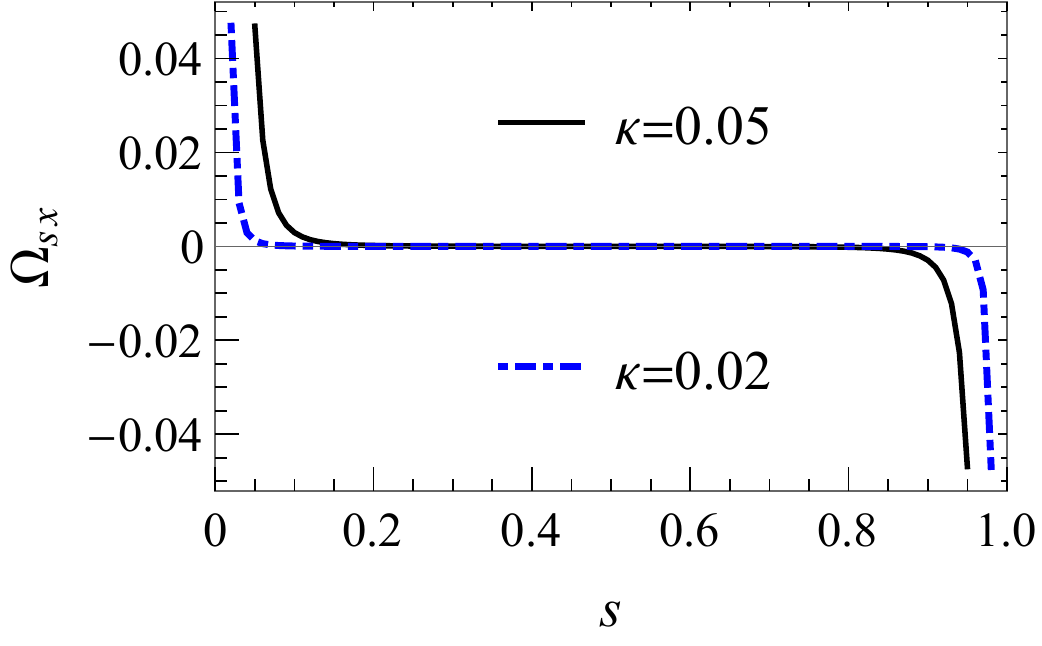}}}
			$ \qquad \qquad \qquad $
			\sidesubfloat[]{{	\includegraphics[scale=0.23]{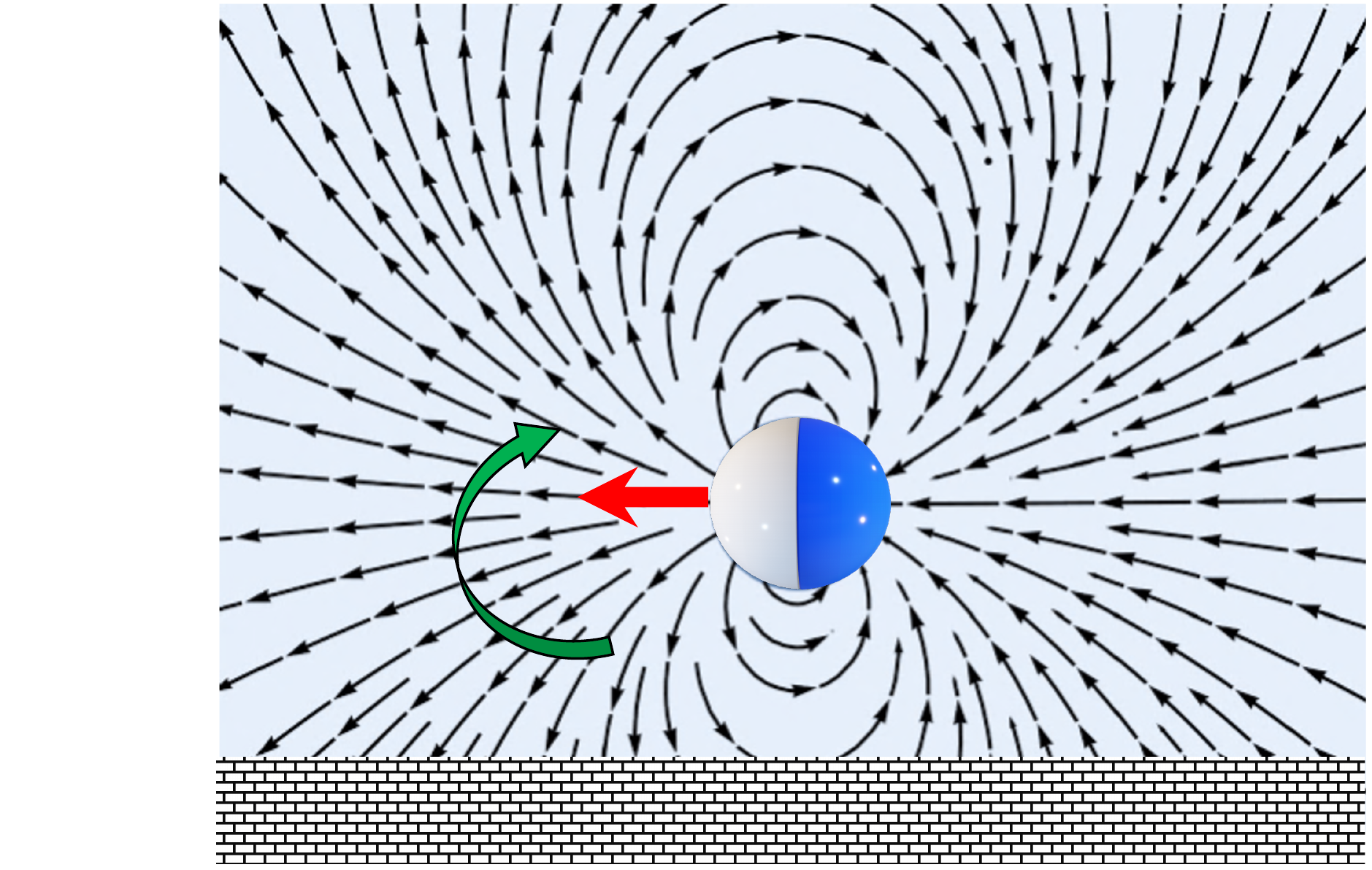}}}
	\caption{{\small (a) Instantaneous axial velocity for horizontally aligned $ (\theta_p=0) $ Janus particle with half-coating $ (\theta_c=\pi/2) $ and attractive solute-particle interaction $ (M=-1) $. (b) Leading order $ O(\kappa^3) $ chemical and hydrodynamic wall- interactions for $ \kappa=0.5 $. (c) Instantaneous rotational velocity for horizontally aligned $ (\theta_p=0) $ Janus particle with half-coating $ (\theta_c=\pi/2) $ and attractive solute-particle interaction $ (M=-1) $. (d) Schematic showing the unbounded flow field around the particle and wall-induced rotation.}}%
	\label{fig:half-inst-horiz}%
\end{figure}

Fig.\ref{fig:half-inst-horiz}(a) shows the instantaneous velocity for horizontal orientation ($ \theta_{p} =0 $). The plot suggests a net enhancement in swimming speed near both walls.
The chemical and hydrodynamic effects compete with each other, as shown in Fig. \ref{fig:half-inst-horiz}(b). The hydrodynamic effects slow down the particle; the more dominant chemical wall-interactions overcome this and result in a net increase in particle velocity near the walls.
This occurs because the effect of wall on chemical distribution is to increase the depletion near the active site (more difficult to refresh the solute distribution), and thus, increasing the concentration gradient for the horizontal propulsion.
Similar observations were also made by \citet{crowdy2010two} in their study of 2D circular Janus swimmers.

The horizontally aligned motion in the presence of walls also imparts a rotational velocity (mathematically represented in eq. \ref{Omega_s}) because the wall-reflected fields have non-zero vorticity, and thus the particle rotates in order to remain torque-free.
This instantaneous rotational velocity is plotted in Fig. \ref{fig:half-inst-horiz}(c) for $ \kappa=0.02 $ $ \& $ $ \kappa=0.05 $, where the entire contribution is from hydrodynamic effects; specifically, from the wall reflection of source-dipole $ \mathbb{T}_{B} $, as the force-dipole and its wall-corrections  are absent here \textit{i.e.} $ \mathbb{T}_{D}=\mathbb{F}_{D}=0 $. The positive value below $ s=0.5 $ suggests that the particle's axis of rotation is in positive x-direction  (into the plane) and vice-versa above the centerline. Thus, the particle rotates away from the wall, as is shown in the schematic Fig. \ref{fig:half-inst-horiz}(d).

\floatsetup[figure]{style=plain,subcapbesideposition=top}

\subsubsection{Trajectories}

\begin{figure}[H]
	\centering
	\sidesubfloat[]{{\includegraphics[scale=0.23]{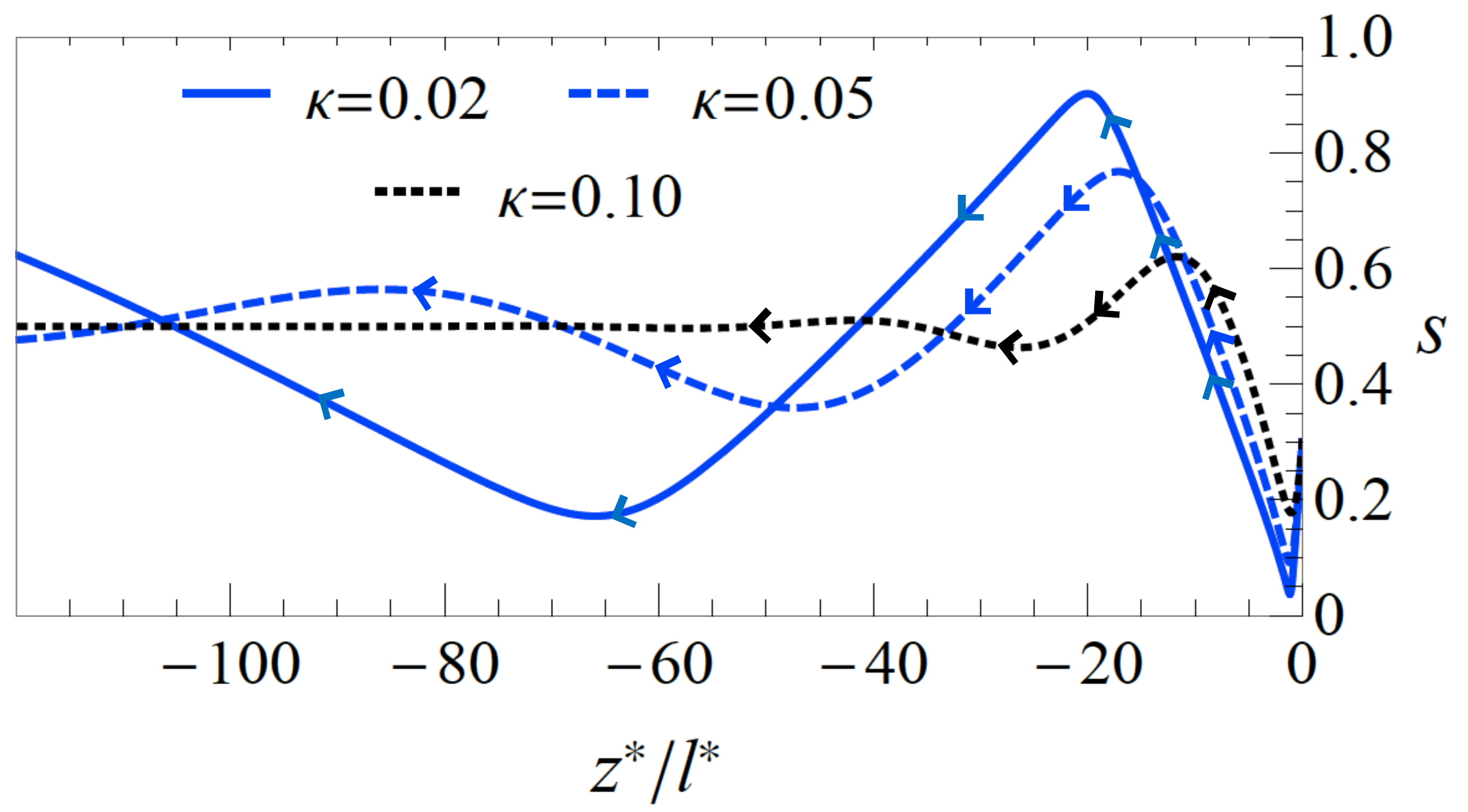} }}%
	\; \qquad
		\sidesubfloat[]{{\includegraphics[scale=0.23]{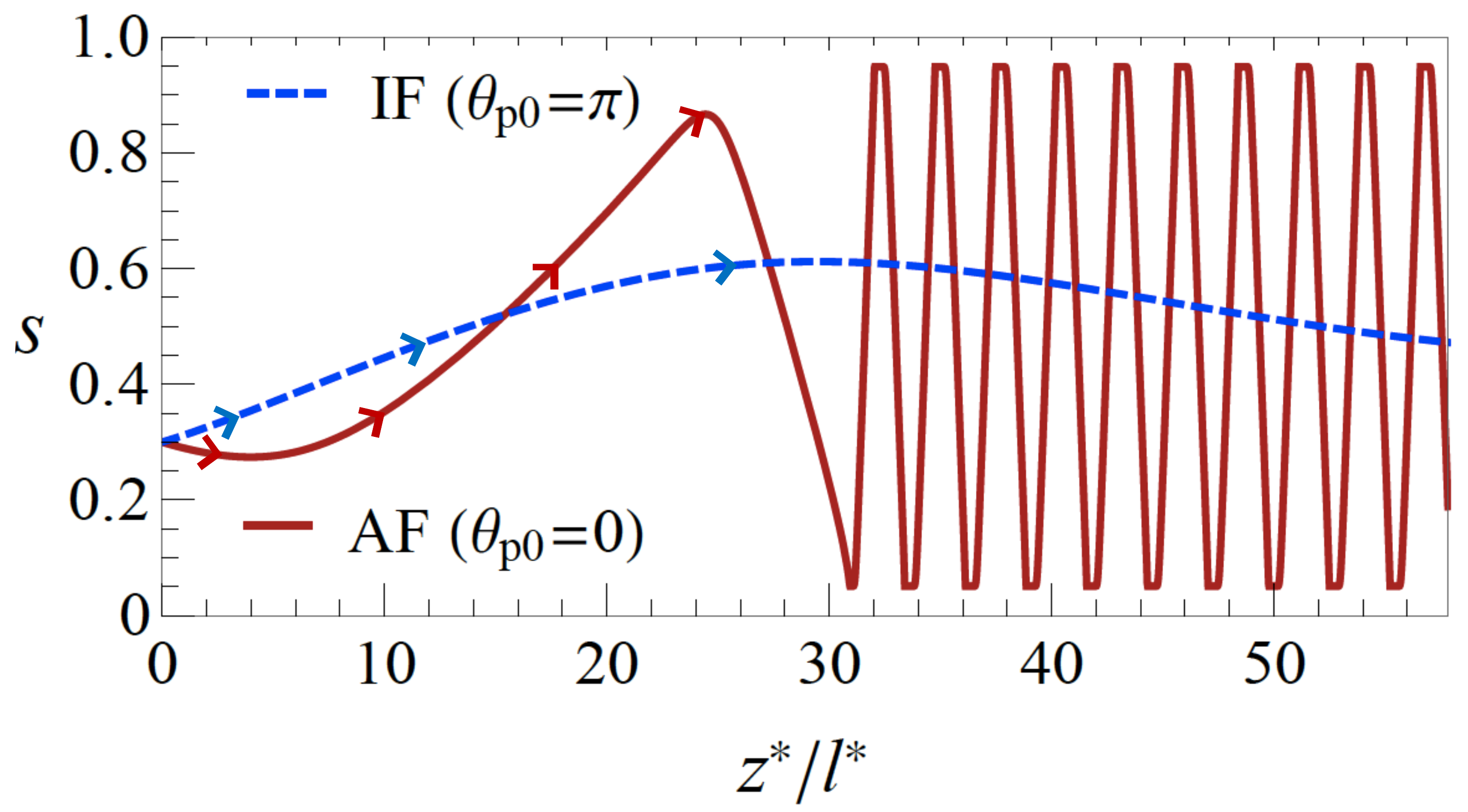} }}%
	\; \;
	\sidesubfloat[]{{\includegraphics[scale=0.5]{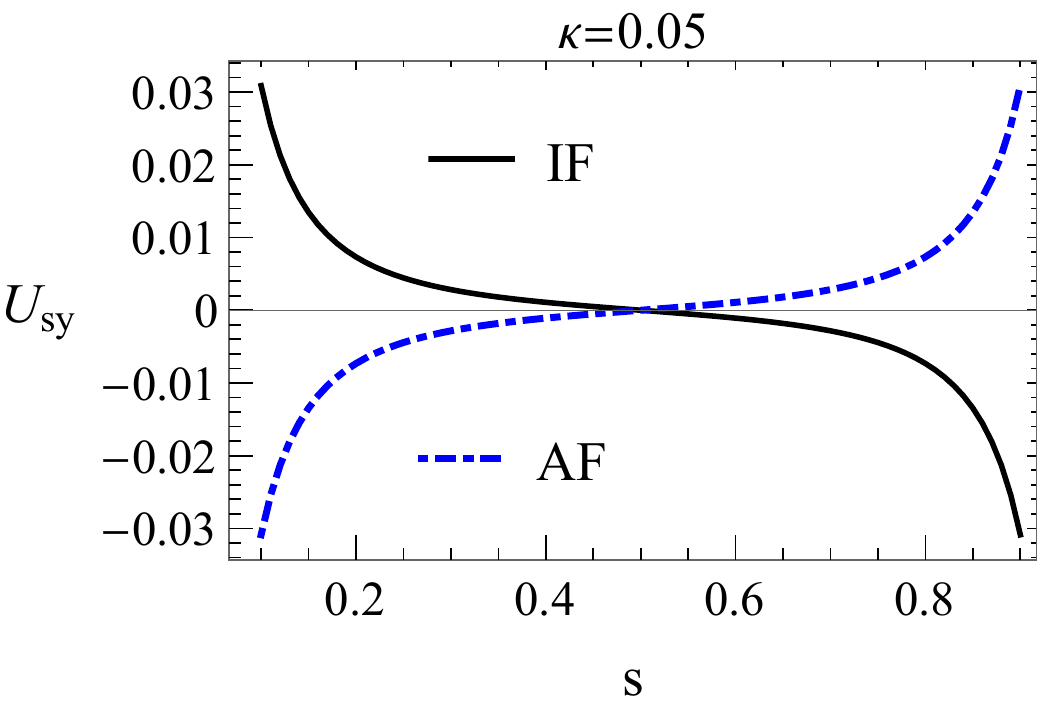} }}%
	\caption{{\small (a) Trajectories of inert facing (IF) Janus particle $ (M=-1) $ for $ \theta_{p_0}=\pi/12 $. (b) Comparison of the trajectories for active facing (AF) Janus particle $ (M=+1) $ with IF Janus particle ($ M=-1 $) for $ \kappa=0.05 $. (c) Comparison of vertical instantaneous velocities for a horizontally aligned particle ($ \theta_{p}=0 $). 
			The arrows in (a) \& (b) depicts the direction in which the particle progresses.}}%
	\label{fig:half-traj}% 
\end{figure}

To understand the dynamical behavior of the Janus particle, we write the equations for temporal evolution of particle's position and orientation as
\begin{equation}\label{dynamic}
		\dot{\tilde{z}} = U_{s\,z}, \quad	\dot{\tilde{y}} = U_{s\,y},  \quad \mbox{and\ }	\; \; \; \;  \dot{\theta} = - \Omega_{s\,x}.
\end{equation}
We also add a repulsion when the particle approaches walls \cite{spagnolie2012hydrodynamics}
\begin{equation}\label{wall-rep}
	U_{s\,y} = \phi_{0} \left[ \frac{ \, e^{-10 s}}{1-e^{-10s}} -  \frac{\, e^{-10 (1-s)}}{1-e^{-10(1-s)}} \right]  , \; \mbox{for\ }  \kappa + s_{cut} \leq s \leq 1- (\kappa  + s_{cut}).
\end{equation}
The cut-off distance is kept as $ s_{cut} = 0.001 $, which corresponds to $ \sim 100 $ nm region around the walls of channel sized $ \sim100 \mu{\rm m} $. \ac{For the results presented here, we impose the potential magnitude ($ \phi_{0} =100 $) high enough to emulate hard repulsion.}
%\footnote{We also used Yukawa-type hard repulsion potential \cite{ibrahim2016walls} and found no change in phase-diagrams.} 
Integrating these equations forward in time provides the trajectories, shown in Fig. \ref{fig:half-traj}\footnote{We used inbuilt `NDSolve' routine in Mathematica 11.3 with `Stiffness-Switching' mode to integrate Eq. \ref{dynamic}.}. For a half-coated Janus sphere, two states are found: (i) damped oscillations and (ii) channel-spanning oscillations, which depends on the nature of solute-particle interaction ($ M $).

Fig.\ref{fig:half-traj} (a) shows the trajectory of an inert facing particle ($ M<0 $) moving in the negative $ z- $direction, with an angle of $ \pi/24 $ with the wall. 
For all particle to channel size ratios ($ \kappa $), the particle exhibits damped oscillations about the centerline, where it eventually finds equilibrium.
An active facing particle, on the contrary, exhibits periodic oscillations across the channel; this comparison is shown in Fig. \ref{fig:half-traj}(b). 
This contrast in trajectory occurs because the leading order wall-induced chemical correction, proportional to $ M \mathcal{K}_{0} $ (see Eq. \ref{U_s_chem}), is opposite for an inert face forward propulsion in comparison to active face forward propulsion. A particle with inert face forward ($ M <0 $) will be repelled from the walls, and attracted if it propels with its active face forward ($ M>0 $), provided the solute is consumed from its active site (\textit{i.e.} $ \mathcal{K}_{0} < 0 $). 
Fig.\ref{fig:half-traj}(c) shows the wall-induced vertical velocity for a horizontally oriented particle; a negative slope of the correction velocity represents wall-repulsion, whereas a positive slope depicts wall-attraction.

 \floatsetup[figure]{style=plain,subcapbesideposition=top}

% \floatsetup[figure]{style=plain,subcapbesideposition=top}
%\begin{figure}[H]
%	\centering
%	\sidesubfloat[]{{\includegraphics[scale=0.25]{RT_Inert_kappa.pdf} }}%
%	\; \qquad
%	\sidesubfloat[]{{\includegraphics[scale=0.25]{RT_Inert_0_02_IC.pdf} }}%
%	\\
%	\sidesubfloat[]{{\includegraphics[scale=0.24]{RT_Active.pdf} }}%
%	\caption{{\small Reduced in-plane trajectories for (a) $ M=-1 $ (b) $ M=-1, \kappa=0.02 $ (c) $ M=+1, \kappa=0.02 $. Green dot indicates the initial condition and red indicates final position. The x-axis is in degrees.}}%
%	\label{fig:half-reduced-traj}%
%\end{figure}

\begin{figure}
	\centering
		\sidesubfloat[]{{\includegraphics[scale=0.25]{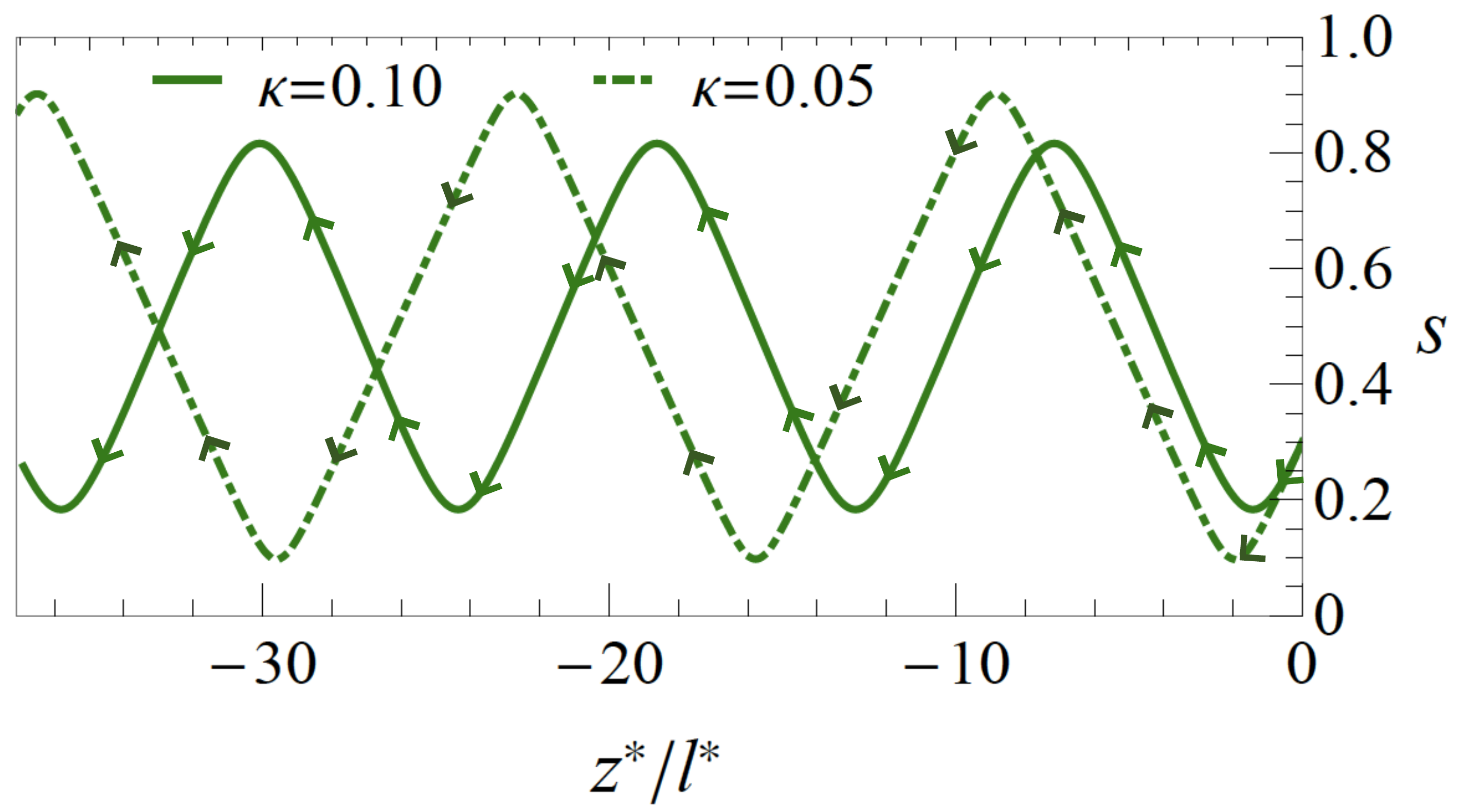} }}%
	\; \;
	\sidesubfloat[]{{\includegraphics[scale=0.54]{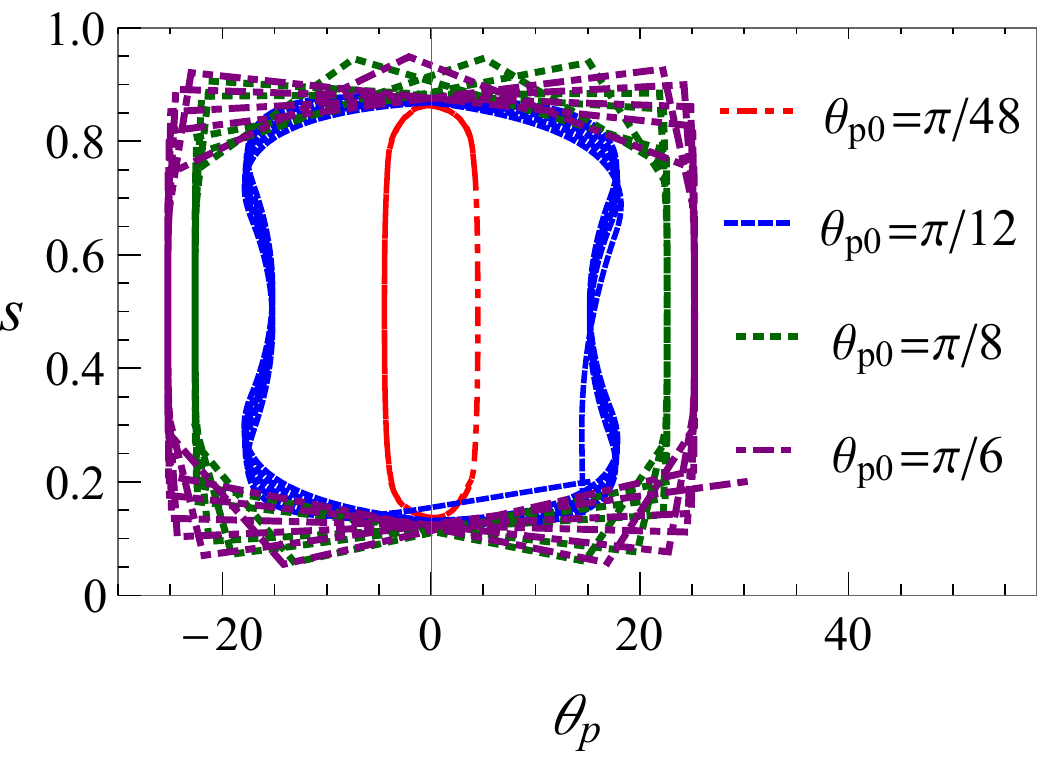} }}%
	\caption{{\small (a) Trajectories arising from pure hydrodynamic interactions for two size ratios for initial conditions $ s=0.3, \, \theta_{p}=\pi/24 $. (b) Reduced in-plane trajectories for four different initial conditions, with fixed $ s=0.2 $ and $ \kappa=0.05 $.}}%
	\label{fig:half-SQ-traj}%
\end{figure}

\textbf{Role of hydrodynamic interactions}:
We now discuss the trajectories of a \textit{squirmer}, which interacts only hydrodynamically with the boundaries. 
It must be noted that the \textit{squirmer} considered here is not arbitrary; it has the slip velocity distribution corresponding to an unbounded Janus particle.
A half-coated Janus sphere corresponds to a neutral squirmer such as \textit{Paramecium}, whose hydrodynamical signature leads with a source-dipole \cite{Ibrahim2015Sep}. Fig.\ref{fig:half-SQ-traj}(a) shows that a neutral squirmer undergoes channel-wide oscillations, with its amplitude dependent on the particle size. However, some of these trajectories are found to be sensitive to initial conditions. 
The reduced trajectory plot in Fig. \ref{fig:half-SQ-traj}(b) shows that for different initial angles of orientation, the particle traces different periodic orbits around the centerline\footnote{ A linear stability analysis of (\ref{dynamic}) reveals that, around the steady state $ s^{*}=0.5,\, \theta_{p}^{*}=0 $ the eigenvalues are purely imaginary, reaffirming the initial condition dependent behavior.}. For a larger initial orientation (for $ \theta_{p_0}=\pi/8 $ or greater), the trajectories are no longer qualitatively dependent on the initial conditions. For such cases, the particle reorients upon impact with the wall and exhibits channel-wide oscillations, identical to those observed for active facing particle (Fig. \ref{fig:half-traj}-b).

\floatsetup[figure]{style=plain,subcapbesideposition=top}

%To summarize, the hydrodynamic interactions results in oscillatory trajectories, whose amplitude depends on the size ratio.
%However, the inclusion of wall-induced chemical effects results in damped oscillations for inert-face-forward Janus particles.

\subsection{Asymmetrically coated Janus particle}

A Janus particle with asymmetric active surface leaves a hydrodynamic signature that leads with $ O(1/r^{2}) $ force-dipole field; the second concentration mode determines its magnitude \cite{ibrahim2016walls}.
For consumption of solute at the active site ($ \mathcal{K}_{0} < 0 $) and attractive solute-particle interaction ($ M<0 $), a particle with activity coverage of less than half (\textit{i.e.} $ \theta_{c} < \pi/2 $) resembles a pusher-type swimmer, whereas a coverage larger than half resembles a puller-type swimmer in unbounded domains \cite{michelin2014phoretic}.
For the case of either net solute release ($ \mathcal{K}_{0}>0 $) or repulsive solute-particle interaction ($ M>0 $), the aforementioned definitions are reversed.
In confinement, the Janus-pushers/pullers differ from the classical pushers/pullers because the former interacts both hydrodynamically and chemically with the walls, whereas the latter interacts only hydrodynamically.

\subsubsection{Instantaneous particle velocity}

\begin{figure}
	\centering
	\sidesubfloat[]{{\includegraphics[scale=0.60]{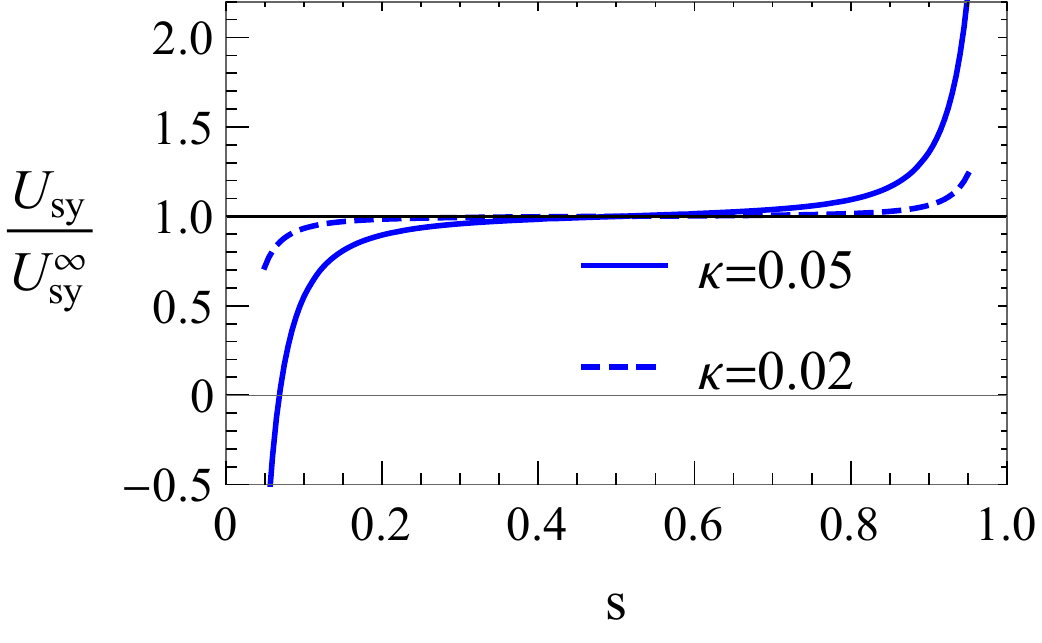} }}%
	\quad
	\sidesubfloat[]{{\includegraphics[scale=0.55]{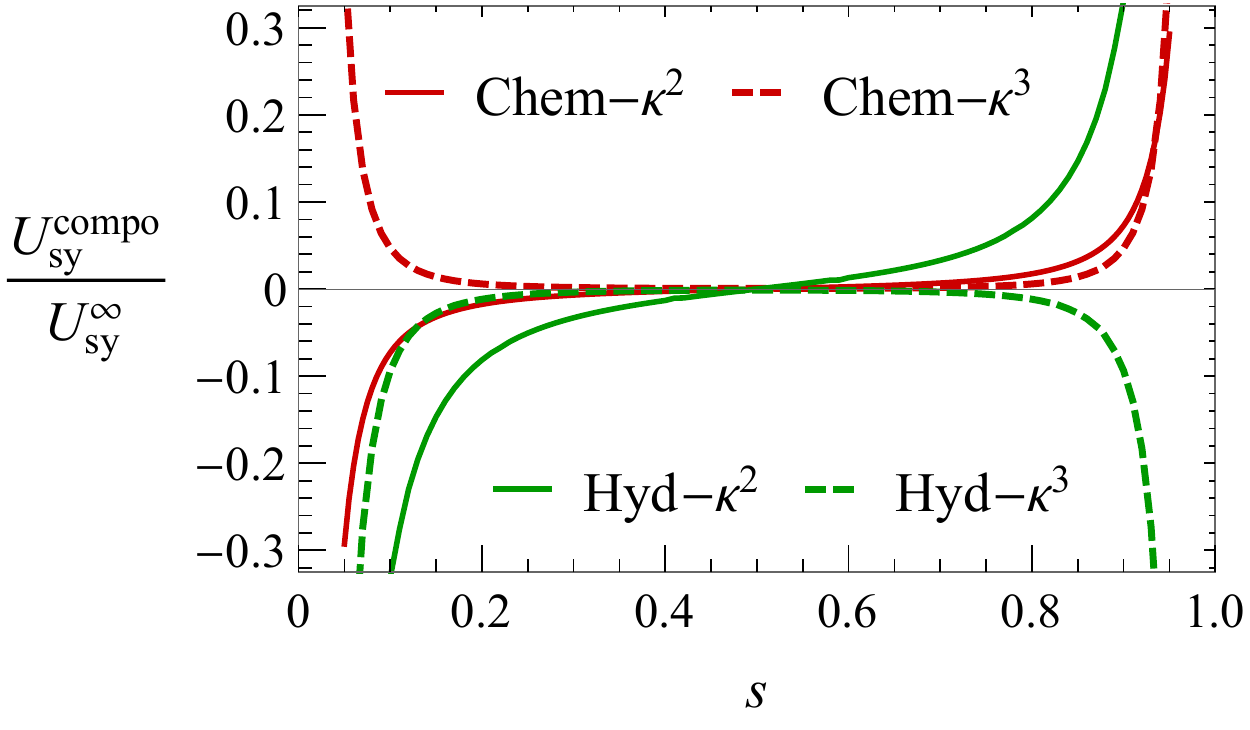} }}
	\caption{{\small (a) Instantaneous perpendicular velocity for vertically aligned $ (\theta_p=\pi/2) $ Janus particle with non-half coating $ \theta_c=\pi/4; $  and attractive solute-particle interaction $ (M=-1) $ \textit{i.e.} Janus-pusher. (b) Chemical and hydrodynamic components of the wall-corrections for $ \kappa=0.05 $. Here $ U_{s\,y}^{\infty} $ is $ -0.125 $.}}%
	\label{fig:non-half-inst-perp}%
\end{figure}

Fig.\ref{fig:non-half-inst-perp}(a) depicts the instantaneous velocity of Janus-pushers oriented perpendicular to walls.
The walls have an effect similar to the case of half-coated particles; as particle size increases the curve becomes asymmetric about centerline because the $ O(\kappa^{3}) $ and $ O(\kappa^{2}) $ effects become comparable. Fig.\ref{fig:non-half-inst-perp}(b) shows various components of wall-corrections for $ \kappa=0.05 $. 
These corrections are qualitatively similar to that observed for half-coated Janus particle, except that there is additional $ O(\kappa^{2}) $ hydrodynamic wall-interaction, originating from the additional force-dipole field.

\floatsetup[figure]{style=plain,subcapbesideposition=top}

\floatsetup[figure]{style=plain,subcapbesideposition=top}
\begin{figure}
	\centering
	\sidesubfloat[]{{\includegraphics[scale=0.55]{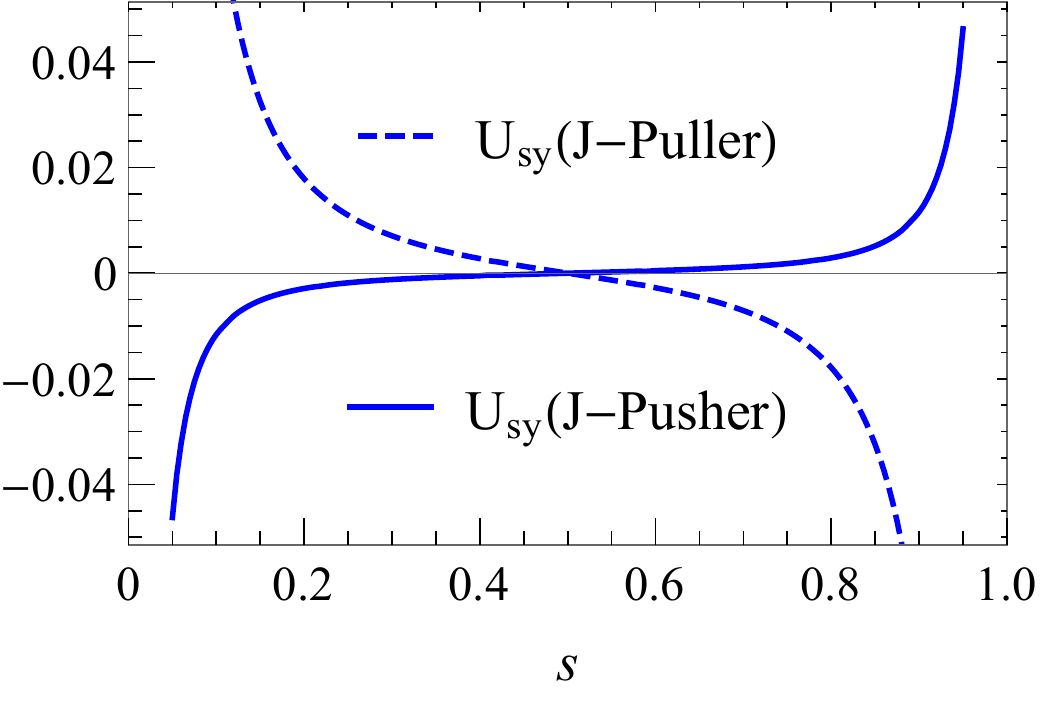} }}%
	\quad
	\sidesubfloat[]{{\includegraphics[scale=0.24]{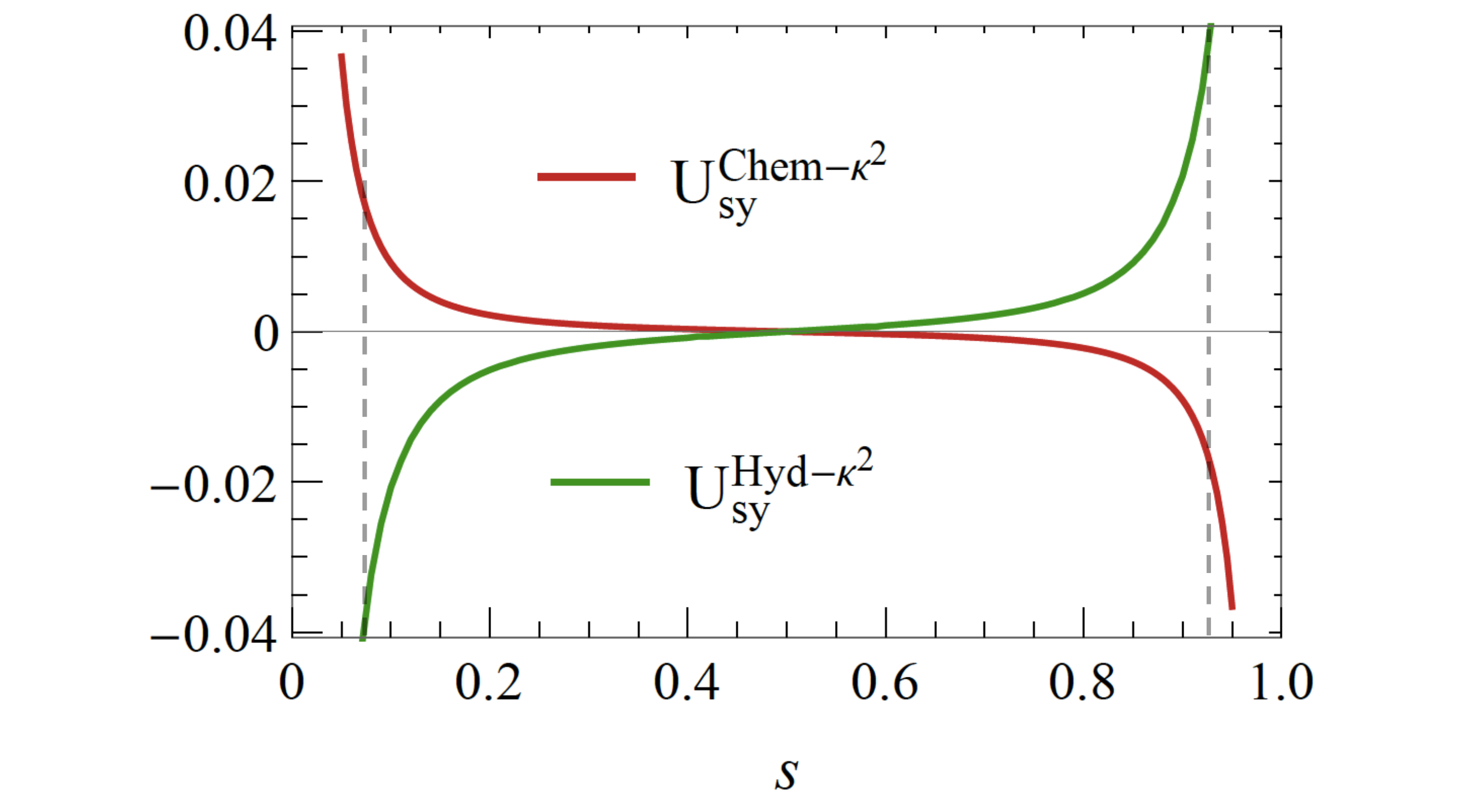} }}%
	\\
	\sidesubfloat[]{{\includegraphics[scale=0.26]{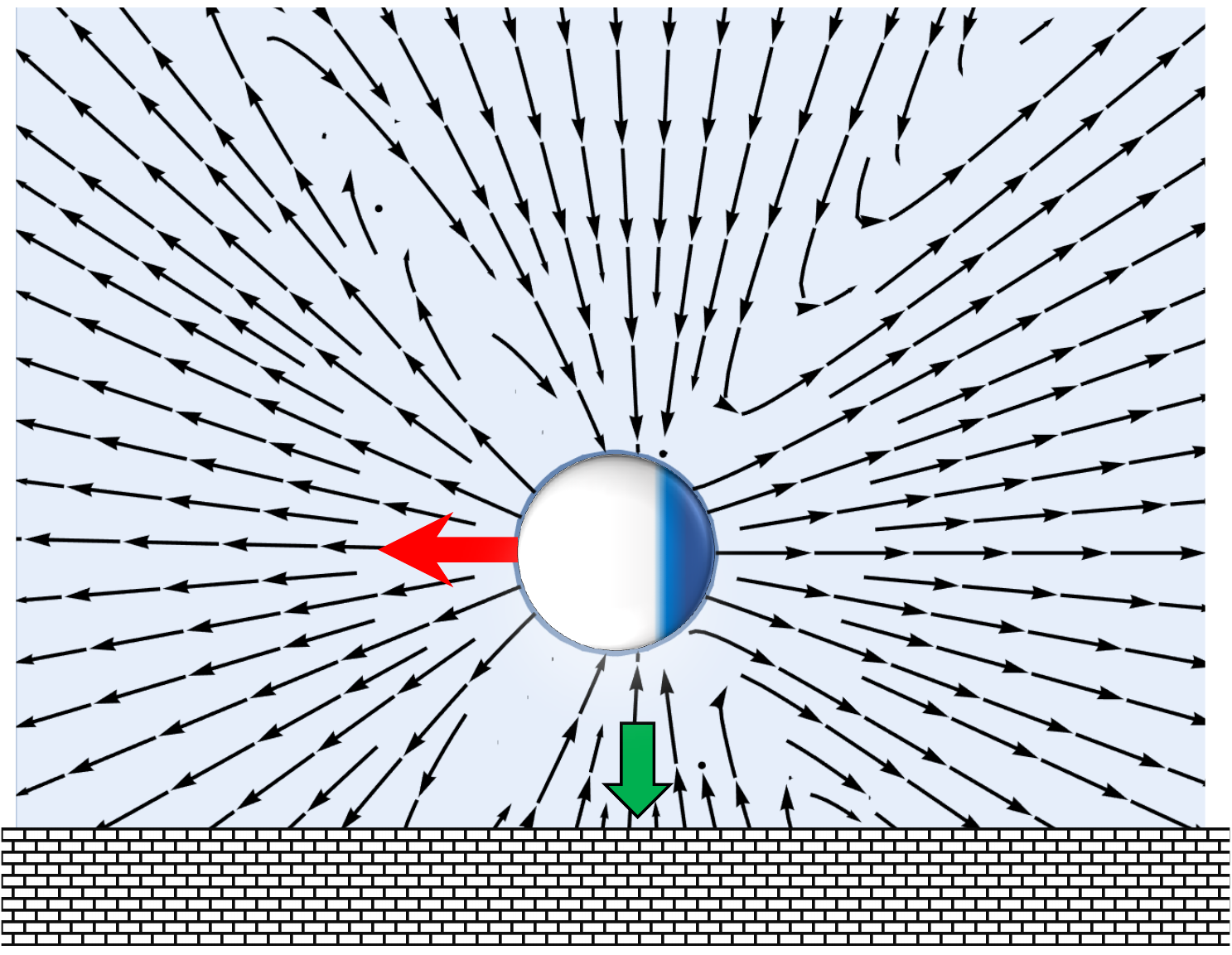} }}%
	\qquad \qquad \qquad \qquad
	\sidesubfloat[]{{\includegraphics[scale=0.26]{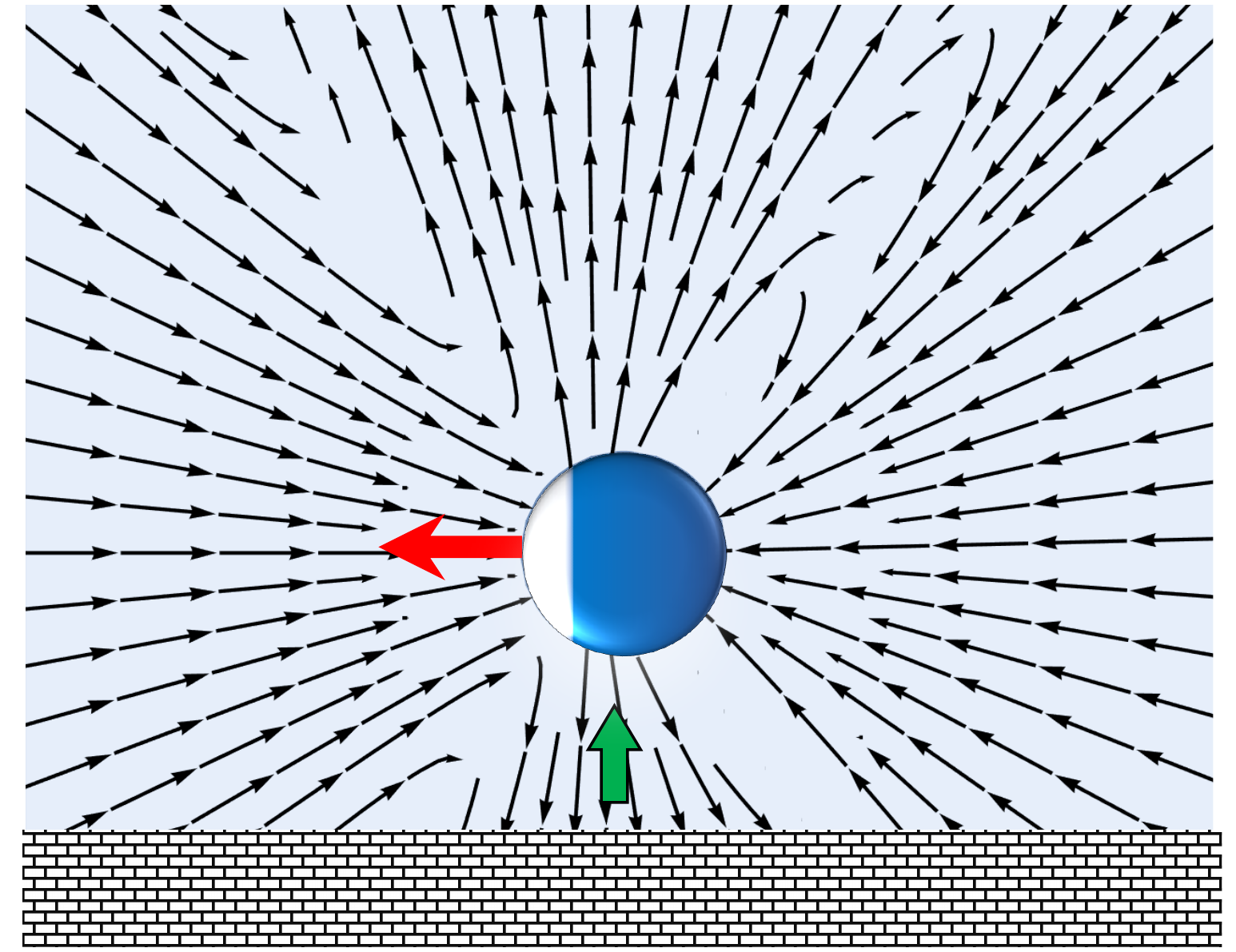} }}%
	\caption{{\small (a) Instantaneous perpendicular velocity for horizontally aligned $ (\theta_p=0) $ Janus-pusher $ (\theta_c=\pi/4, \, \beta=-3.5)  $ and Janus-puller $ (\theta_c=3\pi/4, \, \beta=3.5)  $, with attractive solute-particle interaction $ (M=-1) $ and $\kappa=0.05$. (b) Chemical and hydrodynamic components for a Janus-pusher in (a). $ O(\kappa^{3}) $ corrections are zero as they are proportional to $ \sin \theta_p $. The dashed line is drawn to show that the hydrodynamics effects are greater.
			Schematic showing the unbounded flow field around the particle; (c) Janus-pusher ($ \beta=-3.5 $) (d) Janus-puller ($ \beta=3.5 $). The green arrows depict the wall-induced lateral velocity.}}%
	\label{fig:non-half-inst-perp-horiz}%
\end{figure}

As noted in \S \ref{points} (point 2), there exists a finite wall-perpendicular velocity for a horizontal orientation of a Janus-pusher (see Fig. \ref{fig:non-half-inst-perp-horiz}(a)); this contribution is generated due to the $ O(\kappa^{2}) $ hydrodynamic and chemical interactions, as shown in Fig. \ref{fig:non-half-inst-perp-horiz}(b). 
The hydrodynamic effects are generated by the force-dipole disturbance, and as noted by earlier studies \cite{spagnolie2012hydrodynamics,berke2008hydrodynamic}, a flow field generated by a horizontally aligned pusher attracts it to the nearby wall.
The relatively weaker chemical effects (still $ O(\kappa^{2}) $), repel the inert facing Janus pusher (see \S \ref{points} point 1).
Since the hydrodynamic effects are dominant here, we plot the unbounded flow field around the particle in Fig. \ref{fig:non-half-inst-perp-horiz}(c-d) to understand the lateral motion. 
%The activity coverage corresponds to a Janus-pusher/puller of moderate strength: $ \beta=\mp 3.5 $. 
Similar to classical pushers \cite{berke2008hydrodynamic,spagnolie2012hydrodynamics}, for a Janus-pusher, the fluid is sucked from the sides and ejected from its front \& rear ends. The presence of wall gives rise to breaking of symmetry and lateral motion.
	%
%	Similar behavior is also observed for biological pushers such as \textit{E.coli} in the presence of boundaries \cite{berke2008hydrodynamic}.
	In a similar but reversed manner, a Janus-puller (having more than half active coverage) is hydrodynamically repelled from the walls.

\begin{figure}[H]
	\centering
	\sidesubfloat[]{{\includegraphics[scale=0.47]{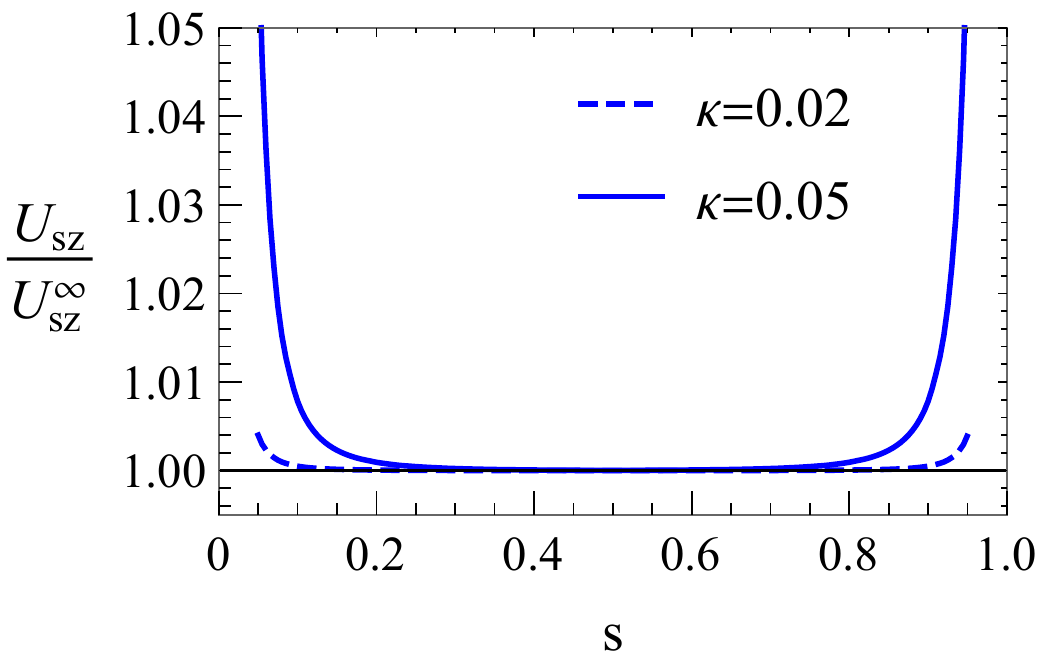} }}%
	\sidesubfloat[]{{\includegraphics[scale=0.46]{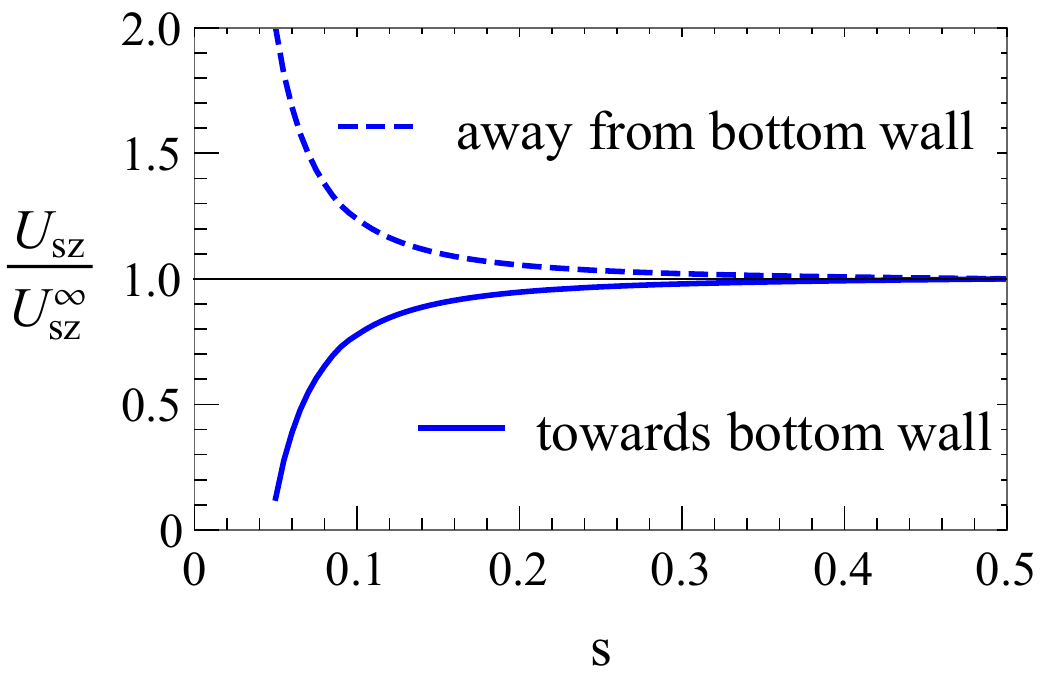} }}%
	\sidesubfloat[]{{\includegraphics[scale=0.49]{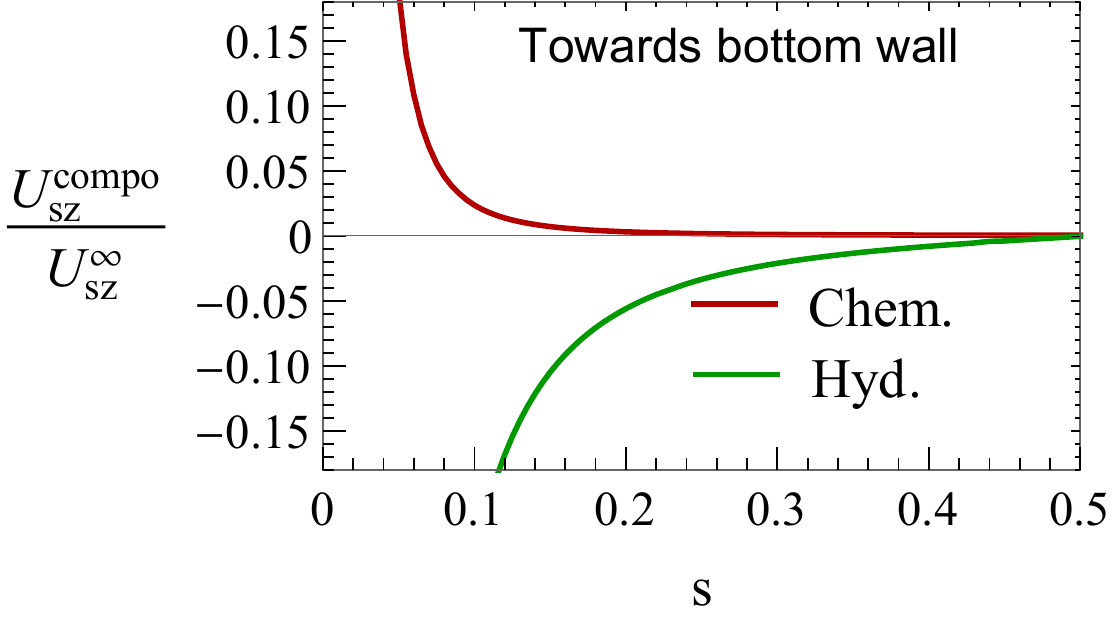} }}%
	\\	
	\sidesubfloat[]{{\includegraphics[scale=0.23]{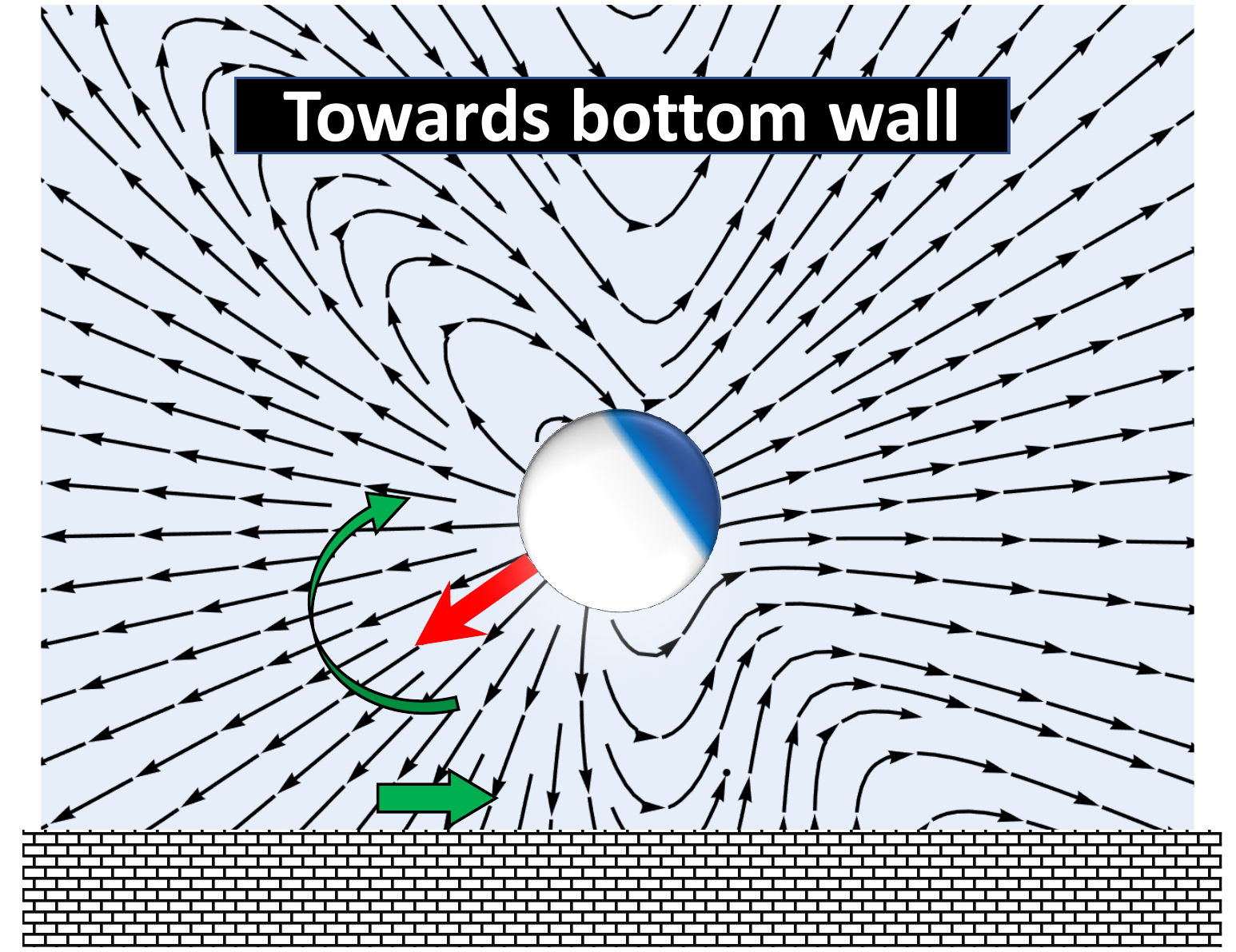} }}%
	\qquad \qquad
	\sidesubfloat[]{{\includegraphics[scale=0.23]{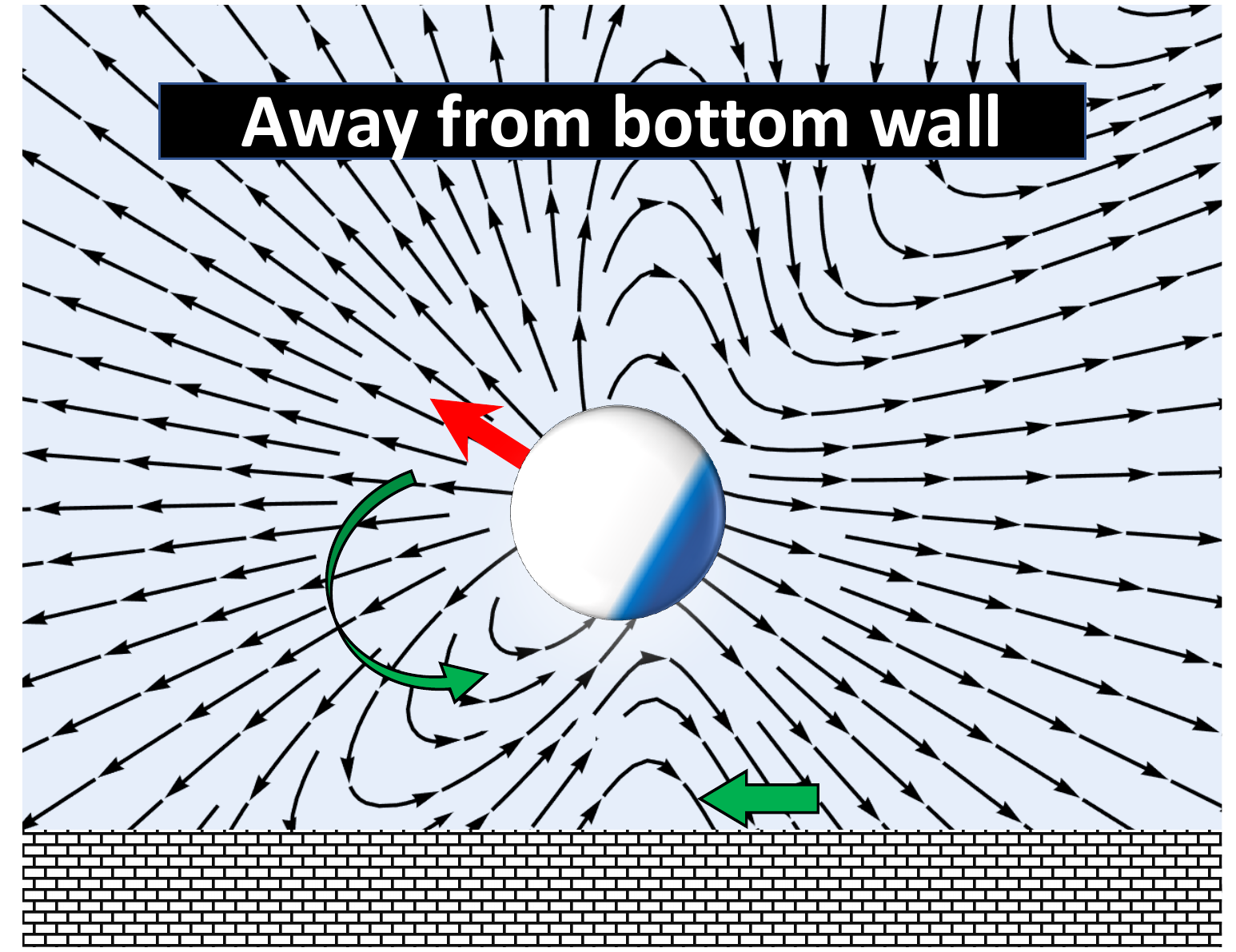} }}%
	
	\caption{{\small (a) Horizontal velocity of the Janus-pusher ($ \theta_{c}=\pi/4 $) aligned horizontally with the wall ($ \theta_{p}=0 $). (b) Horizontal velocity of Janus-pusher making an angle with the wall; towards and away correspond to $ \theta_{p}=\pm \pi/4 $, respectively for $ \kappa=0.05 $. (c) Chemical and hydrodynamic interactions for Janus-pusher: $ \theta_p = \pi/4 $ and $ \kappa=0.05 $. Schematics for Janus-pushers (d) and (e) correspond to (b)-towards and (b)-away, respectively. The red arrows indicate the axis of propulsion, and green arrows represent the wall-induced corrections to horizontal and rotational velocities.}}%
	\label{fig:non-half-inst-horz}%
\end{figure}

Fig.\ref{fig:non-half-inst-horz}(a) shows the horizontal velocity for a Janus-pusher with orientation $ \theta_{p}=0 $. The wall-induced correction is identical to that observed for half-coated particle (\ref{fig:half-inst-horiz}-a) because the hydrodynamic wall-effect associated with  force-dipole vanishes in horizontal orientation ($ \theta_{p}=0 $), as it is proportional to $ \sin 2\theta_{p} $ (see Eq. \ref{U_s_hyd}). 
Thus, here too, we observe a weak enhancement in horizontal velocity due to chemical interactions with walls.

We next consider the horizontal velocity of particle approaching and leaving the wall at an angle \textit{i.e.} $ \theta_{p}=\pi/4 $ \& $ \theta_{p}=-\pi/4 $, respectively.
In Fig. \ref{fig:non-half-inst-horz}(b), we show that the Janus-pusher approaching towards the walls experiences velocity reduction, and an enhancement if it is oriented away. Fig.\ref{fig:non-half-inst-horz}(c) shows that this enhancement is entirely from dominant hydrodynamic interactions. For this reason, we plot the first reflection of velocity field  in Fig. \ref{fig:non-half-inst-horz}(d-e) to develop an intuitive understanding of Fig. \ref{fig:non-half-inst-horz}(b). Since the particle is a moderate pusher ($ \beta= -3.5 $), fluid ejection from its front and rear end can be observed.
Such particle, when approaching walls at an angle ($ \pi/4 $), will find it difficult to push the fluid onto the walls, resulting in reduction of horizontal velocity (indicated by the green arrow). Whereas, a particle oriented away from the wall at an angle ($ -\pi/4 $) will push the fluid against the wall, resulting in increased horizontal velocity. For pullers, these results will be qualitatively reversed. Thus, the enhancement or reduction in horizontal velocity  depends on the angle of approach with the wall.

\floatsetup[figure]{style=plain,subcapbesideposition=top}

\floatsetup[figure]{style=plain,subcapbesideposition=top}
\begin{figure}[H]
	\centering
	\sidesubfloat[]{{\includegraphics[scale=0.5]{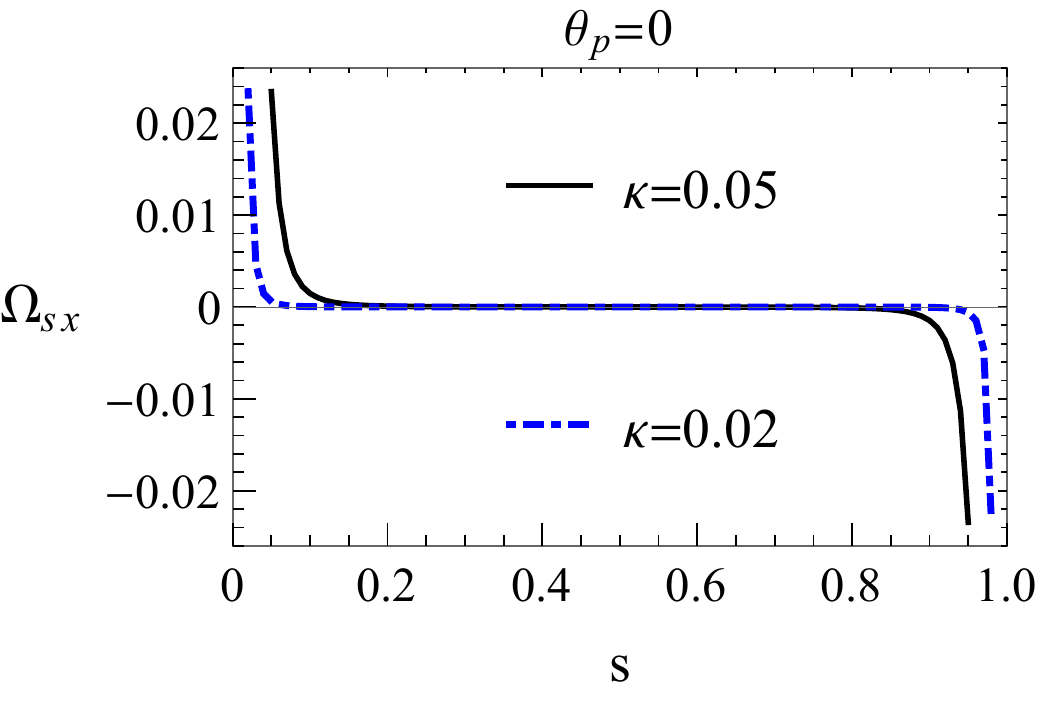} }}%
	\qquad \qquad
	\sidesubfloat[]{{\includegraphics[scale=0.5]{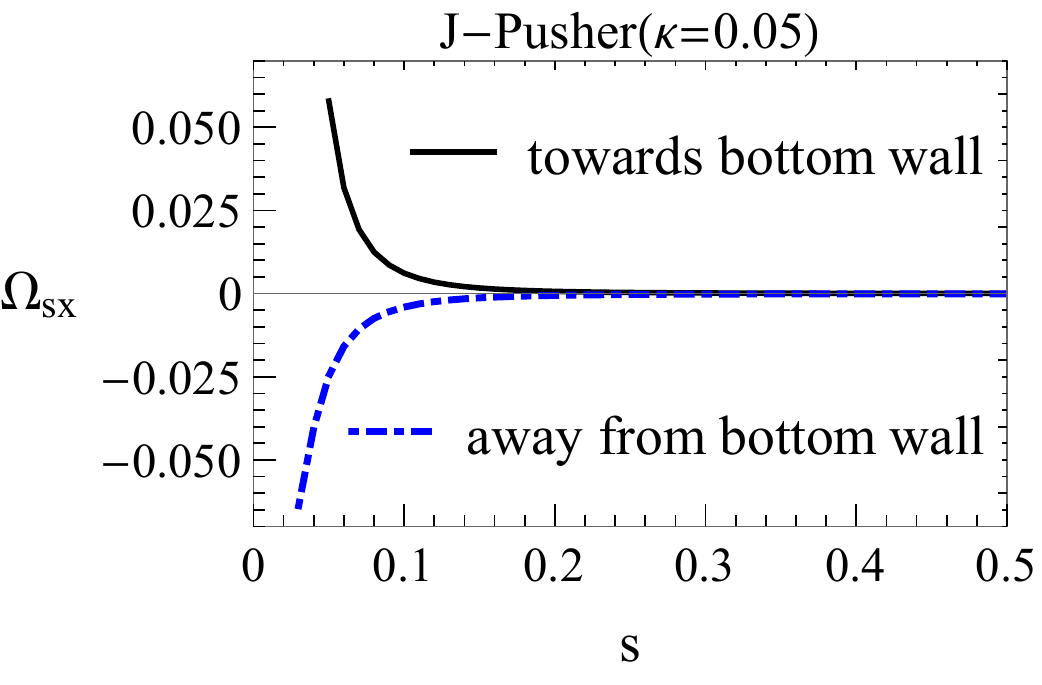} }}%
	
	\caption{{\small (a) Rotational velocity for horizontally aligned ($ \theta_{p}=0 $) Janus pusher. (b) Rotational velocity for particle making angle $ \theta_{p}=\pi/4 $ (towards) and $ -\pi/4 $ (away). }}%
	\label{fig:non-half-inst-rot}%
\end{figure}

Similar to the case of horizontal velocity, the rotational velocity of a horizontally aligned Janus-pusher/puller is identical to the half coated Janus particle as shown in Fig. \ref{fig:non-half-inst-rot}(a) (also see $ \sin 2\theta_{p} $ proportionality in Eq. \ref{Omega_s} \& Fig. \ref{fig:half-inst-horiz}(c); the intensity is reduced because $ \mathcal{K}_{1} $ is lower here).
Fig.\ref{fig:non-half-inst-rot}(b) shows that, the $ O(\kappa^{3}) $ interactions contribute and reorient the approaching (leaving) particle away (towards) from the walls. Fig.\ref{fig:non-half-inst-horz}(d-e) illustrate the direction of wall-induced reorientation.

\subsubsection{Phase diagrams}

\begin{figure}
	\centering
	\includegraphics[scale=0.35]{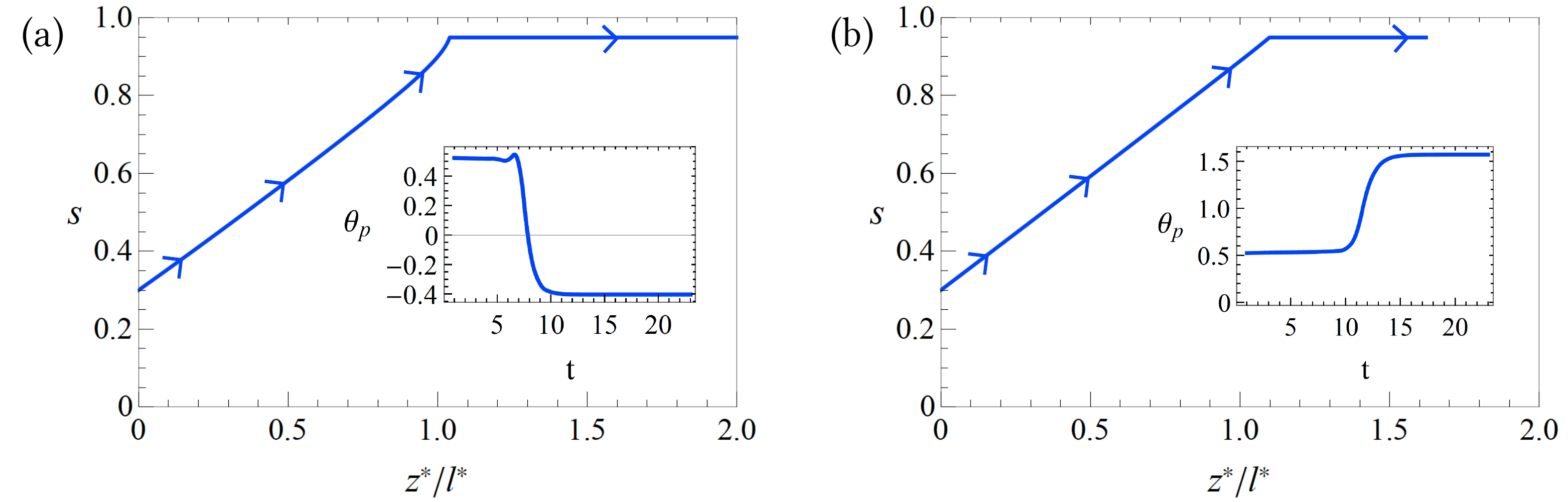}\\ 
	\vspace{2mm}
	\includegraphics[scale=0.35]{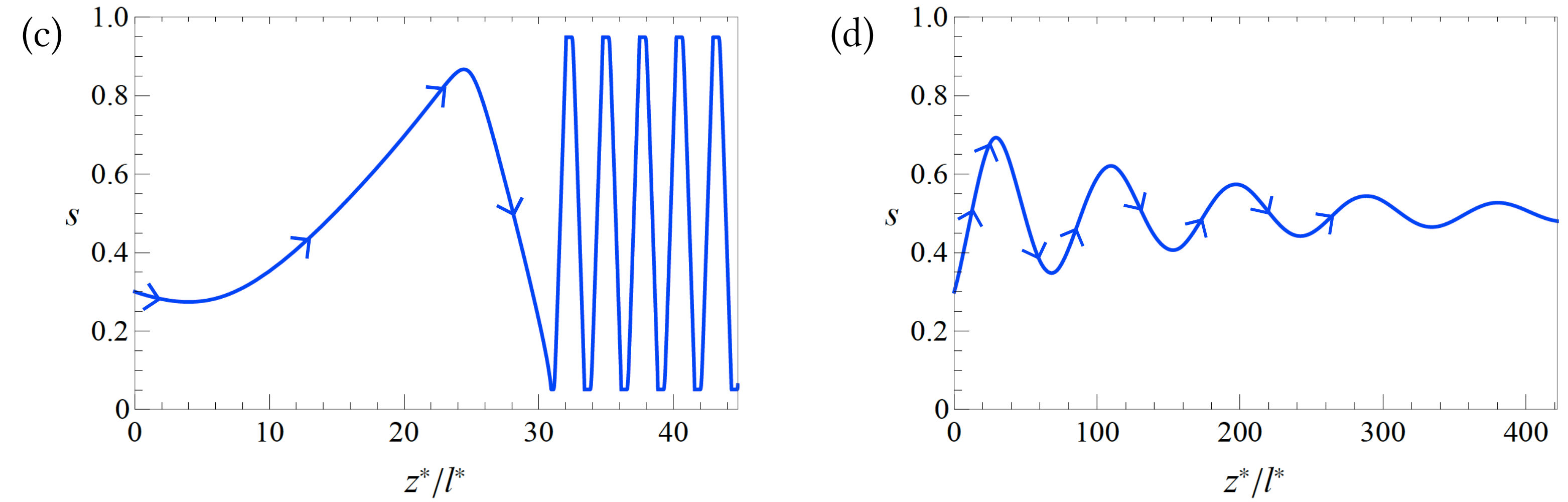}
	\caption{{\small Examples of various  states for $ M=-1, \, \kappa=0.02 $. (a) Sliding, (b) hovering, (c) oscillations, (d) damped oscillations. Inset in (a) and (b) shows the temporal evolution of the orientation. Hovering states attain $ \pm \pi/2 $ orientation, whereas sliding states can attain any non-perpendicular orientation.
			%			The first three figures in the top row correspond to sliding, hovering, and wall oscillatory states, respectively. The inset shows the evolution of orientation angle with time. The second row corresponds to undamped and damped channel oscillations.
	}}%
	\label{fig:state-example}%
\end{figure}

As the surface coverage of activity deviates from $ \pi/2 $, the contribution from second concentration mode $ \mathcal{K}_{2} $ becomes non-zero, which is responsible for the slowly decaying force-dipole disturbance ($ \sim 1/r^{2} $). In the presence of force-dipole, we observe four different dynamical states: two near wall-states and two channel-bulk states. (i) `Sliding' state is characterized near-wall horizontal motion (Fig. \ref{fig:state-example}-a), (ii) `Hovering' is a state of stagnation near wall with perpendicular orientation (Fig. \ref{fig:state-example}-b), (iii) `Osc.' refers to the undamped oscillatory state of particle where it periodically bounces off the opposite walls (Fig. \ref{fig:state-example}-c), (iv) `Damped Osc.' represents the damped oscillatory state where particle focuses at the channel center (Fig. \ref{fig:state-example}-d).
%As an illustration of various  states, Fig. \ref{fig:state-example} shows the trajectories for $ \kappa=0.02 $ and attractive particle-solute interaction ($ M=-1 $).
%Analyzing the temporal evolution of the orientation angle helps us differentiate between the near-wall dynamical states: (i) for sliding, the particle attains a steady non-perpendicular orientation (also see Fig. \ref{fig:intro_illustration}c); (ii) for hovering, the orientation is perpendicular to the walls (also see Fig. \ref{fig:intro_illustration}d).
We next discuss phase diagrams, summarizing the various  states across the parameter space: (i) orientation angle ($ \theta_{p} $), (ii) particle to channel size ratio ($ \kappa $), (iii) activity coverage ($ \theta_{c} $), and (iv) particle-solute interactions ($ M $) \textit{i.e.} inert/active -facing.

\floatsetup[figure]{style=plain,subcapbesideposition=top}

\vspace{2mm}

\begin{figure}
	\centering
	\includegraphics[scale=0.34]{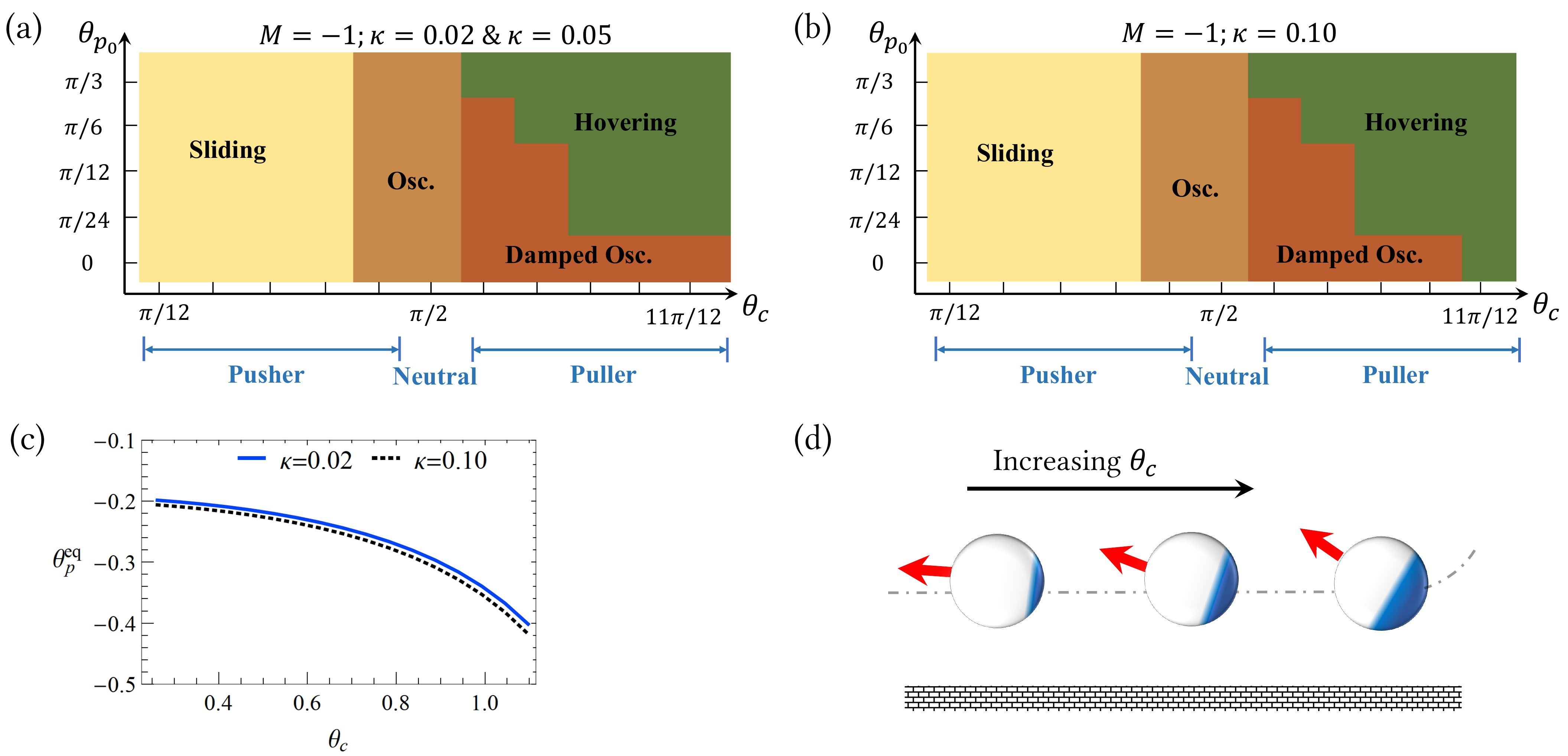}
	\caption{{\small (a-b) Phase diagrams for squirmer of three different size (\textit{i.e.} $ \kappa=0.02,0.05, 0.10 $) and $ M=-1 $. Phase diagram for $ \kappa =0.05 $ is identical to $ \kappa =0.02 $.
			(c) Variation of equilibrium angle with the activity coverage ($ \theta_{c} $) for sliding state in (a,b). (d) Schematic explaining (c): as the activity coverage increases, equilibrium orientation points further away from the wall. For each phase diagram, we chose eleven $ \theta_{c} $ and five $ \theta_{p_0} $ points.}}%
	\label{fig:phase-plots-SQ}%
\end{figure}

\textbf{The role of hydrodynamic interactions}:
To better understand the phase diagrams, we first explain the simpler case of squirmers (equivalent to unbounded Janus particle) because they interact only hydrodynamically with the walls. Thus, it helps us to differentiate between hydrodynamic and chemical interactions.
 Fig.\ref{fig:phase-plots-SQ} shows the phase diagrams for three squirmer sizes.
  We perform this by \textit{switching off} the chemical interactions in (\ref{UOmega}).
   Activity coverage ranges from $ \pi/12 $ to $ 11\pi/12 $; these extremes correspond to moderate pushers and pullers, with force-dipole strength ($ \beta $) ranging from $ -4.83 $ to $ +4.83 $.

\floatsetup[figure]{style=plain,subcapbesideposition=top}

\noindent
	1. 	Due to the hydrodynamic attraction to the wall, strong pushers ($\theta_c>5\pi/12$) exhibit a sliding state. This behavior is insensitive to the choice of $\kappa$ and initial orientation.
	During this state, the pushers move parallel to the wall and are oriented slightly away from the wall (shown in Fig. \ref{fig:phase-plots-SQ} (c)-(d)). 
	As the activity coverage increases the magnitude of equilibrium orientation increases, which, beyond a limit, results in a transition to oscillatory state (discussed next).
	\\
	2. As we increase the activity coverage $ \theta_{c} $, the force-dipole strength $ \beta $ weakens (weak pushers), and thus the hydrodynamic attraction reduces. 
	 In this case, the wall induced angular velocity reorients the swimmer away from the wall, which results in channel-wide oscillations.
	\\
	3. Further increase in $ \theta_c $ (in $ \beta $) increases the force-dipole strength of  pullers. They show two dynamical behaviors, namely (i) damped oscillations, and (ii) stationary hovering state, which depends on the initial orientation. When orientation of the puller is parallel or at a small angle with respect to walls, it gets repelled as shown in Fig. \ref{fig:non-half-inst-perp-horiz}(d).
	If the pullers makes an obtuse angle with the walls, it gets reoriented to attain hovering state: the perpendicularly oriented stationary state ($ \theta_{p}=\pi/2 $).

\floatsetup[figure]{style=plain,subcapbesideposition=top}
\begin{figure}
	\centering
	\includegraphics[scale=0.34]{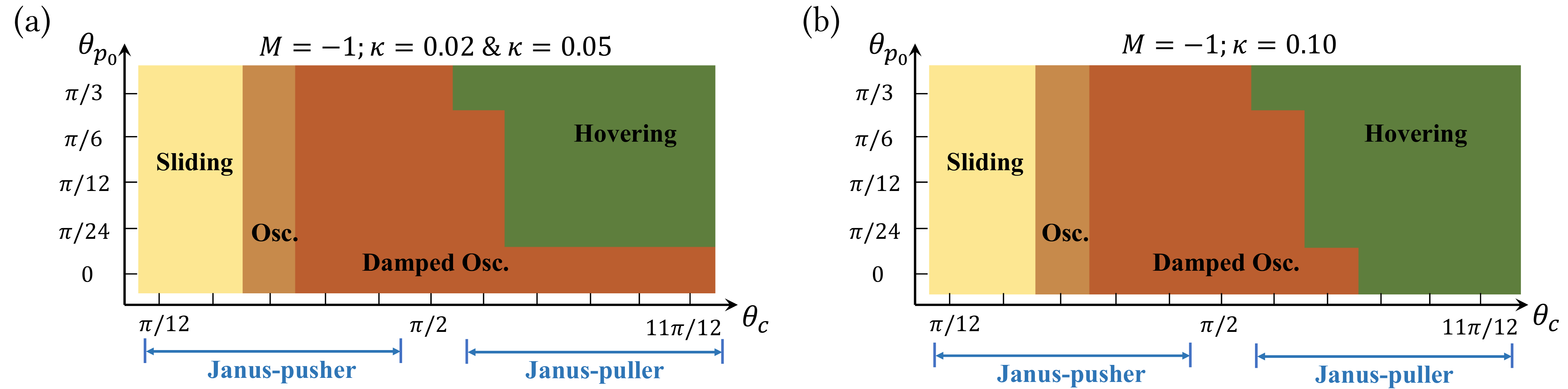}
	
	\includegraphics[scale=0.34]{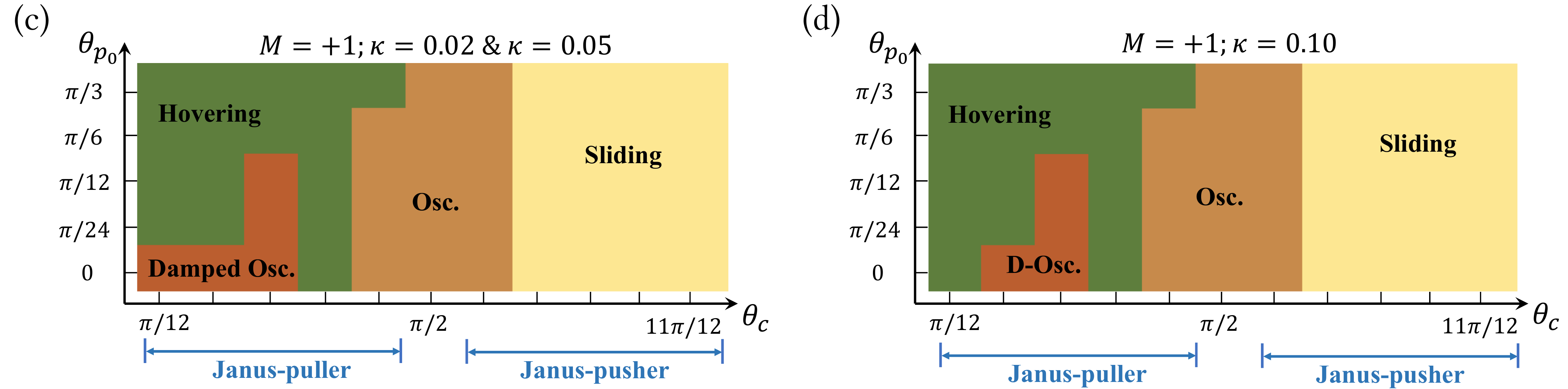}
	\caption{{\small Phase diagrams showing the various  states for Janus particle of three different size ratios with M=-1 (a-b) and M=+1 (c-d). 
			Phase diagram for $ \kappa =0.05 $ is identical to $ \kappa =0.02 $.
			For each phase diagram, we chose eleven $ \theta_{c} $ and five $ \theta_{p_0} $ points.}}%
	\label{fig:phase-plots}%
\end{figure}

\vspace{3mm}

\textbf{Janus particle}:
Fig.\ref{fig:phase-plots}(a-b) show the phase diagrams for particles propelling with inert-face forward (attractive particle-solute interaction+solute consumption) for three different particle sizes, where the left side of x-axis ($ \theta_{c} < \pi/2 $) corresponds to Janus-pushers and right side ($ \theta_{c}>\pi/2 $) to Janus-pullers. 

\noindent
1.  For strong Janus-pushers ($ \pi/12 \leq \theta_{c} \leq \pi/4 $), the sliding state resembles classical pushers because the hydrodynamic interactions of force-dipole dominate over the chemical ones. 

\noindent
2.  As the activity coverage increases, (i) chemical interactions increase, and (ii) relative force-dipole strength decreases. Consequently, the leading order chemical repulsion associated with $ M \mathcal{K}_{0} $ (see Eq. \ref{U_s_chem}) dictates the periodic oscillations and damped trajectories across the channel centerline.

\noindent
3. As the activity coverage further increases beyond $ \theta_{c}=\pi/2 $, the force-dipole strength of Janus-puller increases, which again competes with chemical effects to give rise to the hovering state (similar to classical pullers).

\floatsetup[figure]{style=plain,subcapbesideposition=top}
\begin{figure}
	\centering
	\includegraphics[scale=0.35]{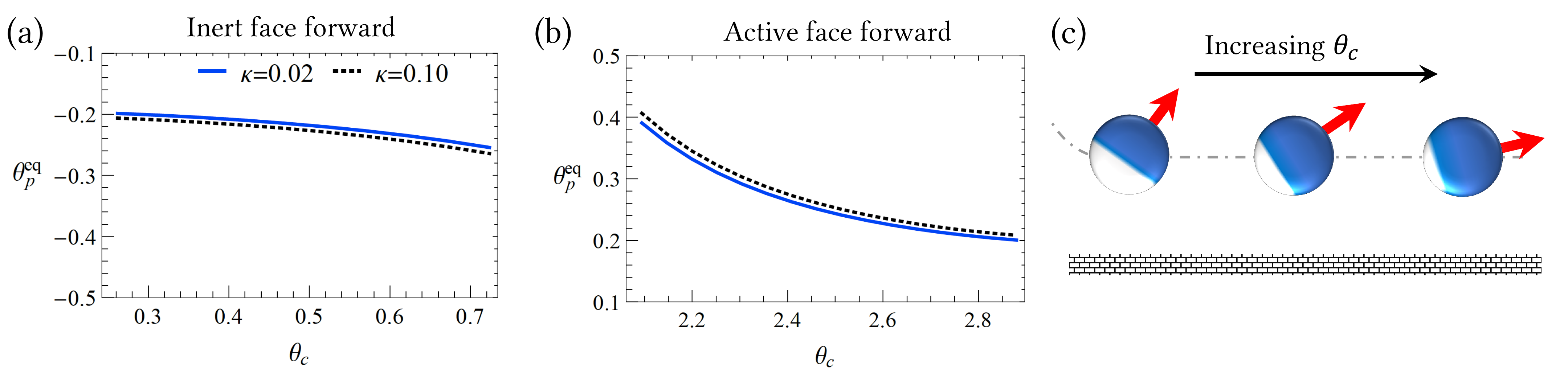}
	\caption{{\small Variation of equilibrium angle for sliding states of (a) inert face forward particle ($ M=-1 $), and  (b) active face forward particle ($ M=+1 $). (c) Schematic illustrating the effect of surface coverage on equilibrium orientation for (b). As activity coverage increases, the particle points more horizontally.}}%
	\label{fig:sliding_Janus}%
\end{figure}

For active-face forward Janus particles ($ M>0 $), in Fig. \ref{fig:phase-plots}(c-d) the left side of x-axis ($ \theta_{c} < \pi/2 $) corresponds to Janus-pullers and right side ($ \theta_{c}>\pi/2 $) to Janus-pushers.

\noindent
1.  We noted in \S\ref{points} (point 3) that inert-facing Janus-pullers/pushers are different from active-facing Janus-pullers/pushers. Thus, the phase diagrams Fig. \ref{fig:phase-plots}(c-d) cannot be expected to be the exact mirror image of Fig. \ref{fig:phase-plots}(a-b) because a transition from $ M=-1 $ to $ M=+1 $ changes both the chemical and hydrodynamic signature of particle\footnote{ If there were sole hydrodynamic interactions, as we observe for the case for squirmers, a change in sign of $ M $ would merely mirror the phase diagram about $ \theta_{c}=\pi/2 $.}.

\noindent
2.  The major deviation arises for the regime of weak force-dipole (\textit{i.e.} $ 5\pi/12 \leq \theta_{c} \leq 7\pi/12 $); we observe an oscillatory state across the channel. This is due to the leading order chemical attraction associated with net solute consumption/release ($ M \mathcal{K}_{0} $ proportionality in Eq. \ref{U_s_chem}) and subsequent reorientation.

\noindent
3.  Although not exact, but the qualitative nature of strong Janus-pullers and -pushers here is similar to that observed for inert-facing Janus particles; strong pullers demonstrating hovering or damped oscillatory state, and strong pushers sliding along the walls. Fig. \ref{fig:sliding_Janus} shows the variation in equilibrium orientation for sliding state. As the activity coverage increases, the equilibrium orientation of (i) inert-face-forward particles points further into the channel bulk, (ii) active-face-forward particles points further along the channel walls.

\noindent
4.   The current results take the first three concentration modes into account ($ \mathcal{K}_{0}, \, \mathcal{K}_{1}, \, \mathcal{K}_{2} $). Since near-wall dynamics is affected by short-ranged disturbances, in Appendix, we include the single wall-corrections (derived by \citet{ibrahim2016walls}) to the chemical and hydrodynamic field resulting from the fourth mode: $ \mathcal{K}_{3} $. 
We find that this inclusion does not qualitatively alter the conclusions of previously analyzed phase diagrams.

%\vspace{2mm}

\section{Conclusions}
Motivated by the fact that confinement (i)  is ubiquitous to biological and
artificial microswimmers alike \cite{nelson2014micro}, and (ii)  can affect microbial and colloidal assembly \cite{elgeti2016microswimmers,kanso2019phoretic}, we studied the dynamics of a channel confined Janus particle.
Inspired by the early theoretical works on particle-boundary interactions \cite{happel2012low}, we assume that the particle is much smaller than the confinement, and build a general framework based on the method of reflections in conjunction with the Faxen's transformations to capture two-wall effects. 
One of the key insight that this analysis delivers is the finding of two new channel-bulk states: damped and periodic oscillations around the centerline. These dynamical states depend on the surface characteristics of the particle and are in addition to sliding and hovering states explored in the earlier studies \cite{Uspal2014,ibrahim2016walls}. 
We capture and compare the hydrodynamic and chemical wall effects on the dynamical behavior separately, which is found to be sensitive to particle-solute interaction, activity coverage, and orientation.
%we analyze particle kinematics, trajectories  and phase diagrams over a wide range of parameter space.
%For near-half activity coverage, the wall-induced chemical effects are found to be primarily responsible for these two states.
%This leading order $ O(\kappa^{2}) $ chemical effect depends on particle mobility and net consumption (characterized by zeroth concentration mode $ \mathcal{K}_{0} $), where $ \kappa  $ is particle to channel size ratio.

In unbounded domains, an inert facing Janus particle behaves as a pusher (puller) if activity coverage is less (more) than half \cite{michelin2014phoretic}.
Since the activity coverage determines net solute consumption ($ \mathcal{K}_{0} $), in bounded domains, a Janus-pusher will experience weak chemical effects, whereas a Janus-puller is strongly affected. This effect would reverse if the Janus particle propels with active-face forward. Thus, the boundary interaction of inert facing Janus-puller (or pusher) is different than active facing Janus-puller (or -pusher). Previous studies have majorly focused on inert facing Janus particles \cite{Uspal2014,ibrahim2016walls,Ibrahim2015Sep}.
To address this gap in literature, we explored the parameter space (activity coverage, initial orientation, particle size) for both active and inert facing particles.
 Janus-pushers and -pullers with strong force-dipole strength overcome chemical effects, and exhibit sliding \& hovering states, respectively. On the other hand, weak Janus-pushers and -pullers are strongly affected by the chemical interactions with the walls: (i) inert-face-forward particles are chemically repelled from both walls, leading to damped oscillations, (ii) active-face-forward particles are attracted and reoriented away from both walls, yielding channel-wide periodic oscillations.
This insight suggests that Janus particles can separate themselves based on the mode of propulsion\footnote{Inert faced movement occurs for (i) attractive (or  repulsive) particle-solute interaction and (ii) solute consumption (or release) from the active site. If any one of the conditions reverse, the particle moves with active face forward.}.

The use of far-field approach has been cautioned by earlier studies \cite{Popescu2017Apr} because the truncation of chemical field (to first two modes) may result in deviation from the actual physics, as the associated rapidly decaying velocity disturbances can play an important role in determining near-wall dynamical states. 
Thus, we incorporate first three modes to attain a better prediction. \ac{However, the estimates on near-wall states may still be imprecise, also the nature of soft wall-repulsion may play a significant part \cite{bayati_dynamics_2019}.} For this, a numerical approach can be used to explore the parameter space for both active and inert facing Janus particles \cite{Shen2018Mar}.
Nevertheless, for channel-spanning states, it is unlikely that higher modes would amount to a qualitative change in phase-diagrams that are reported here. 
%For steady-states in the channel bulk need not require such high precision.

For the current problem, we find that the instantaneous velocity of Janus particle obtained by accounting for two walls
matches closely with those obtained via superposition of single-wall analysis. This suggests that either of the approaches (Faxen's transformation or superposition) can be used to analyze the trajectories and phase-diagrams of a confined Janus particle.
Nevertheless, our framework can be easily extended for a variety of interesting and open problems where disturbances are long-ranged.  To list a few, the framework can be used to analytically study (i) the confinement effect on bottom-heavy squirmers \& Janus particles \cite{Ruhle2020May}, and (ii) such particles in confined shear \& pressure-driven flows \cite{elgeti2016microswimmers,Katuri2018Jan,Zottl2012May}. 
The finite stokeslet resulting from bottom-heaviness of the swimmer decays very slowly ($ \sim 1/r $), and thus will be greatly influenced by the surrounding confinement.
We intend to explore these extensions in our future work.

\vspace{5mm}

\noindent
\textbf{\large{Acknowledgments}}\\
A.C. and K.V.S. thank Indian Ministry of Human Resource Development for the financial support. A.C. also thanks Alexander von Humboldt foundation for the partial financial support.
This work was also partly supported by the European Research Council (ERC) under the European Union's Horizon 2020 research and innovation program (Grant Agreement No. 714027 to S.M.).

\vspace{5mm}

%\vspace{8mm}

\setcounter{equation}{0}
\renewcommand\theequation{A.\arabic{equation}}

\noindent
\textbf{\large{Appendix A: $\mathcal{K}_{3}$  mode inclusion}}

\begin{figure}[H]
	\centering
	\includegraphics[scale=0.29]{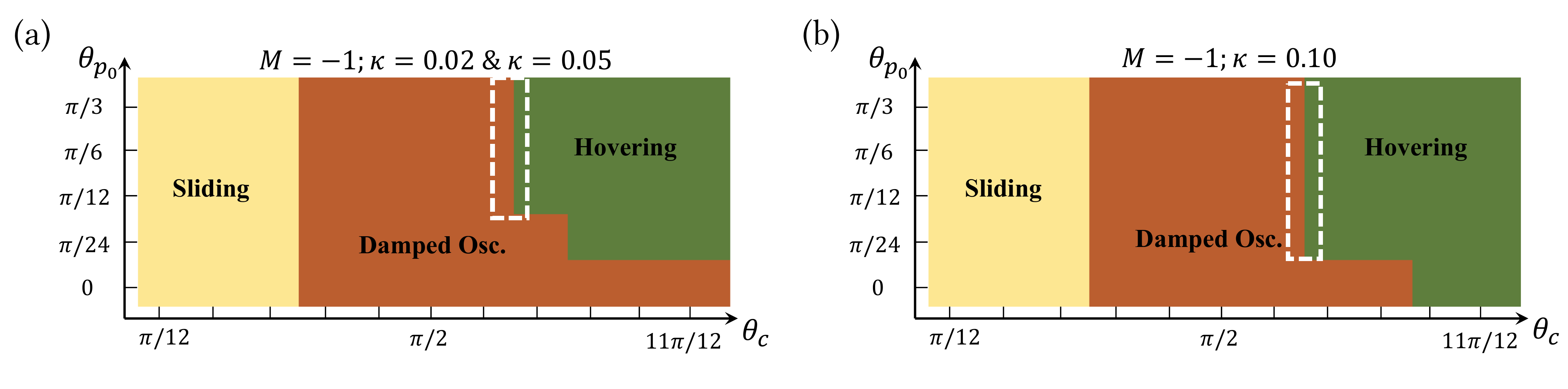}
	
	\includegraphics[scale=0.29]{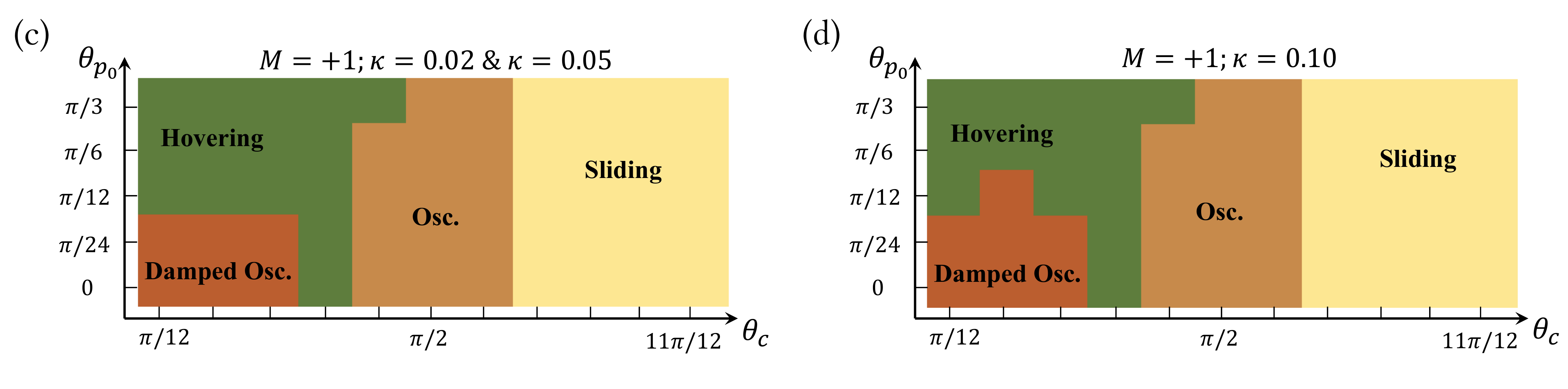}
	
	\includegraphics[scale=0.29]{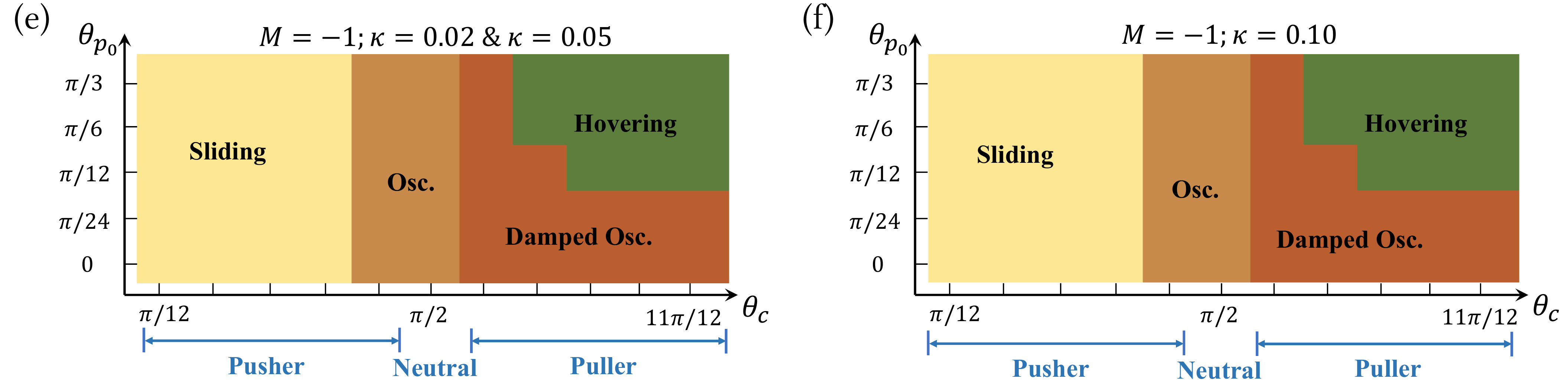}
	\caption{{\small Phase diagrams with inclusion of 3rd concentration mode ($ \mathcal{K}_{3} $). (a-b) Inert facing Janus particles, (c-d) Active facing Janus particles, (e-f) Squirmers.
			Other parameters are identical to Fig. \ref{fig:phase-plots}. The dotted portion consists of a sliding region, which is shown in Fig. \ref{fig:hidden}}.}%
	\label{fig:phase-plotsK3}%
\end{figure}

\begin{figure}[H]
	\centering
	\includegraphics[scale=0.55]{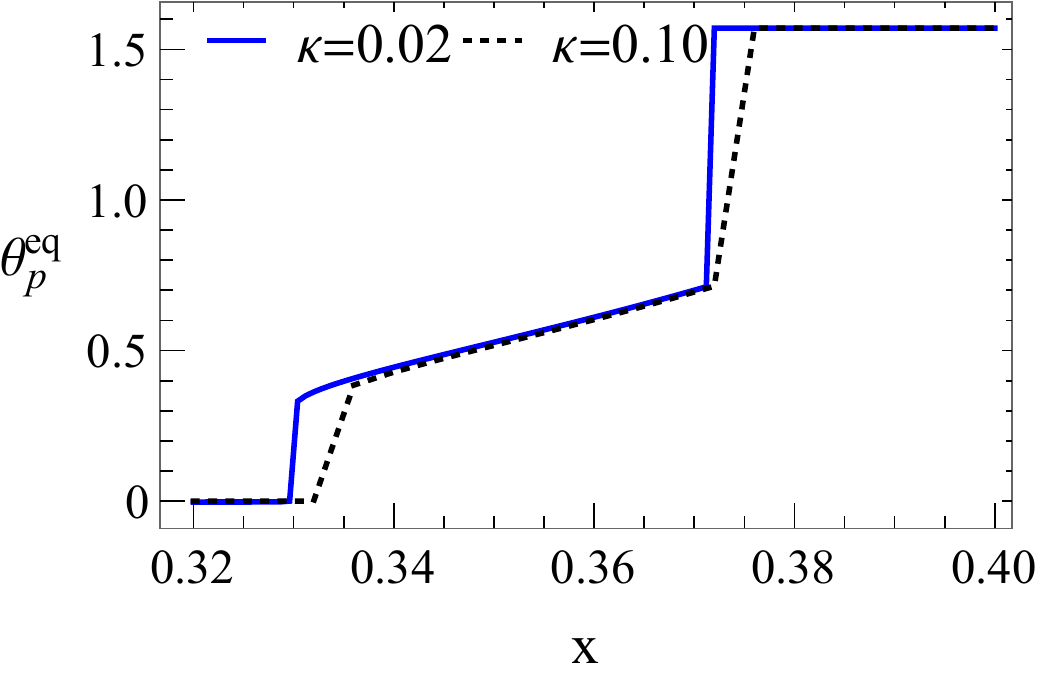}
	\caption{{\small Variation of the equilibrium orientation with the activity coverage: $ \cos^{-1}(-x) $. The plot corresponds to the zoomed portion in Fig. \ref{fig:phase-plotsK3}(a-b).}}%
	\label{fig:hidden}%
\end{figure}

%\subsection{}\label{K3_app}
We utilize the wall corrections corresponding to $ \mathcal{K}_{3} $ derived by \citet{ibrahim2016walls} and add them in (\ref{UOmega_detail}). Then we integrate (\ref{dynamic}) and explore the same parameter space (as in Fig. \ref{fig:phase-plots}) to generate the Fig. \ref{fig:phase-plotsK3} for Janus particles and squirmers. 
A comparison  of Fig. \ref{fig:phase-plotsK3} with Fig. \ref{fig:phase-plots-SQ} and Fig. \ref{fig:phase-plots}  would reveal that first three modes sufficiently capture the qualitative nature of various dynamical states.
However, there is an exception: for inert-facing Janus particle (Fig. \ref{fig:phase-plotsK3}a-b), a sliding state occurs between damped oscillatory and hovering state (indicated by the white dashed region). 
Fig. \ref{fig:hidden} shows that, as activity coverage increases, the equilibrium orientation transitions from 0 (corresponding to damped oscillations) to $ \pi/2 $ (hovering state) via a sliding state where the particle points into the walls. This is consistent with earlier studies \cite{ibrahim2016walls,Uspal2014}, and is qualitatively different from the sliding state that is found for other activity coverage in Fig. \ref{fig:phase-plots} and \ref{fig:phase-plotsK3}; there, the sliding particle points away from the walls.

\floatsetup[figure]{style=plain,subcapbesideposition=top}
\floatsetup[figure]{style=plain,subcapbesideposition=top}

%\pagebreak
%
%
%\setcounter{equation}{0}
%\renewcommand\theequation{A.\arabic{equation}}
%
%\begin{center}
%	{\large \textbf{Supplementary Information}}
%\end{center}

\noindent
\textbf{\large{Appendix B: Comparison with \citet{ibrahim2016walls} }}

Below we compare the instantaneous velocities derived in \S IV with those derived for single walls by Ibrahim and Liverpool \cite{ibrahim2016walls}. To extend their expressions for two-walls, we replace their parameter $ \epsilon $; it represents the inverse particle-wall distance normalized by the particle size i.e. $ \epsilon \equiv ({d/a})^{-1} = a/d$. Dividing both numerator and denominator with the channel size $ l $, we get: $ \epsilon = \kappa/s $ and $ \kappa/(1-s) $, for bottom and top wall, respectively. Using this, we convert the single-wall expression of Ibrahim and Liverpool for first three concentration modes:\\

\small{
$ \begin{aligned}
	\begin{array}{l}
		\Omega_{s\,x} = -\left[-\frac{1}{8} \cos (\text{$\theta $p}) (\mathcal{K}_{1} M) \left(\frac{\kappa ^4}{s^4}-\frac{\kappa ^4}{(1-s)^4}\right)- \frac{9}{16}  \mathcal{K}_{2} M \left(\frac{\kappa ^3}{16 s^3}+\frac{\kappa ^3}{ (1-s)^3}\right) \sin (\text{$\theta $p}) \cos (\text{$\theta $p})\right], \\[5mm]
		\IB{U}_{s}^{chem} = \left[ \frac{-1}{8} \left(\frac{\kappa ^3}{s^3}+\frac{\kappa ^3}{(1-s)^3}\right)  \cos (\text{$\theta $p}) ( \mathcal{K}_{1} M)  \right] \IB{e}_{z} \\[3mm]
		\qquad \qquad \qquad \qquad \qquad \qquad \qquad \qquad + \left[ \frac{1}{4} (\mathcal{K}_{0} M) \left(\frac{\kappa ^2}{s^2}-\frac{\kappa ^2}{(1-s)^2}\right)  - \frac{1}{4} \mathcal{K}_{1} M \sin (\text{$\theta $p}) \left(\frac{\kappa ^3}{s^3}+\frac{\kappa ^3}{(1-s)^3}\right)  \right] \IB{e}_{y} ,\\[5mm]
		\IB{U}_{s}^{hyd} = \left[  \frac{9}{8} \sin (\text{$\theta $p}) \cos (\text{$\theta $p}) ( \mathcal{K}_{2} M) \left(\frac{\kappa ^2}{s^2}-\frac{\kappa ^2}{(1-s)^2}\right)
		+ \frac{1}{12} \cos (\text{$\theta $p}) (\mathcal{K}_{1} M) \left(\frac{\kappa ^3}{s^3}+\frac{\kappa ^3}{(1-s)^3}\right) \right] \IB{e}_{z}\\[3mm]
		\qquad \qquad + \left[  -\frac{9}{16} \mathcal{K}_{2} M \left(1-3 \sin ^2(\text{$\theta $p})\right) \left(\frac{\kappa ^2}{s^2}-\frac{\kappa ^2}{(1-s)^2}\right)
		+ \frac{1}{3} \sin (\text{$\theta $p}) (\mathcal{K}_{1} M) \left(\frac{\kappa ^3}{s^3}+\frac{\kappa ^3}{(1-s)^3}\right)  \right] \IB{e}_{y}.
	\end{array}
\end{aligned} $
}
\vspace{3mm}

\noindent
The rotational velocity is multiplied by a negative sign because the x-axis (axis of rotation) in our study is opposite to that reported by Ibrahim and Liverpool \cite{ibrahim2016walls}.
\vspace{6mm}

\floatsetup[figure]{style=plain,subcapbesideposition=top}
\begin{figure}[H]
	\centering
	\includegraphics[scale=0.32]{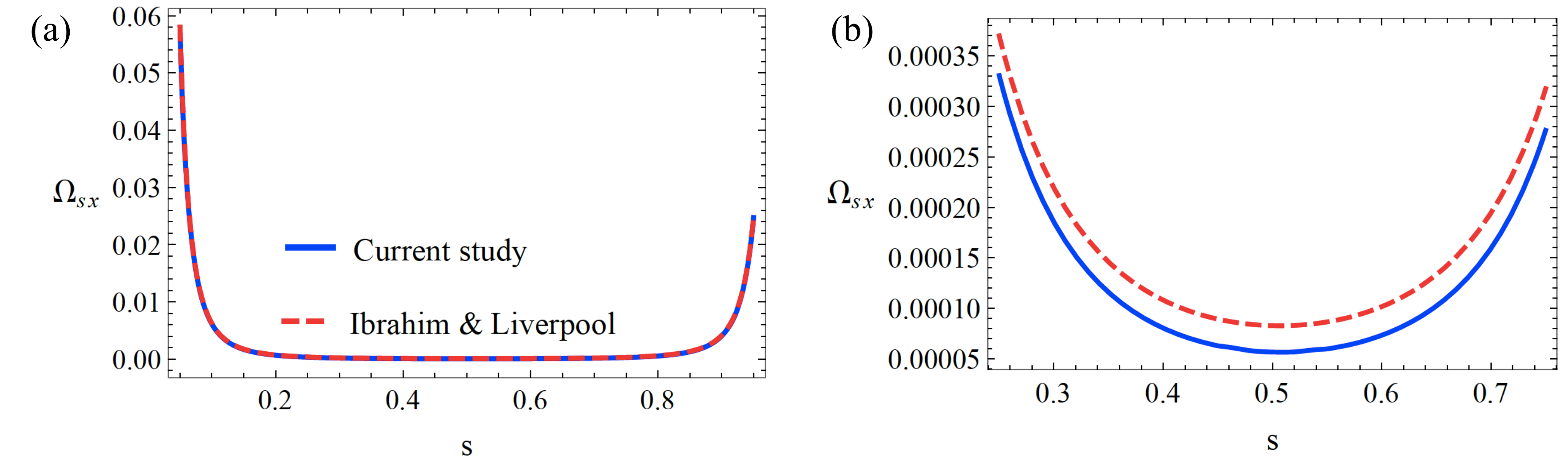}
	\caption{{\small Comparison of wall-induced rotational velocity for $ \theta_{c} = \pi/4 $, $ \theta_{p}=\pi/4 $, and  $\kappa=0.05$. (b) Zoomed-in version of (a).}}%
	\label{fig:comparison_Omega}%
\end{figure}

\floatsetup[figure]{style=plain,subcapbesideposition=top}
\begin{figure}[H]
	\centering
	
	\includegraphics[scale=0.32]{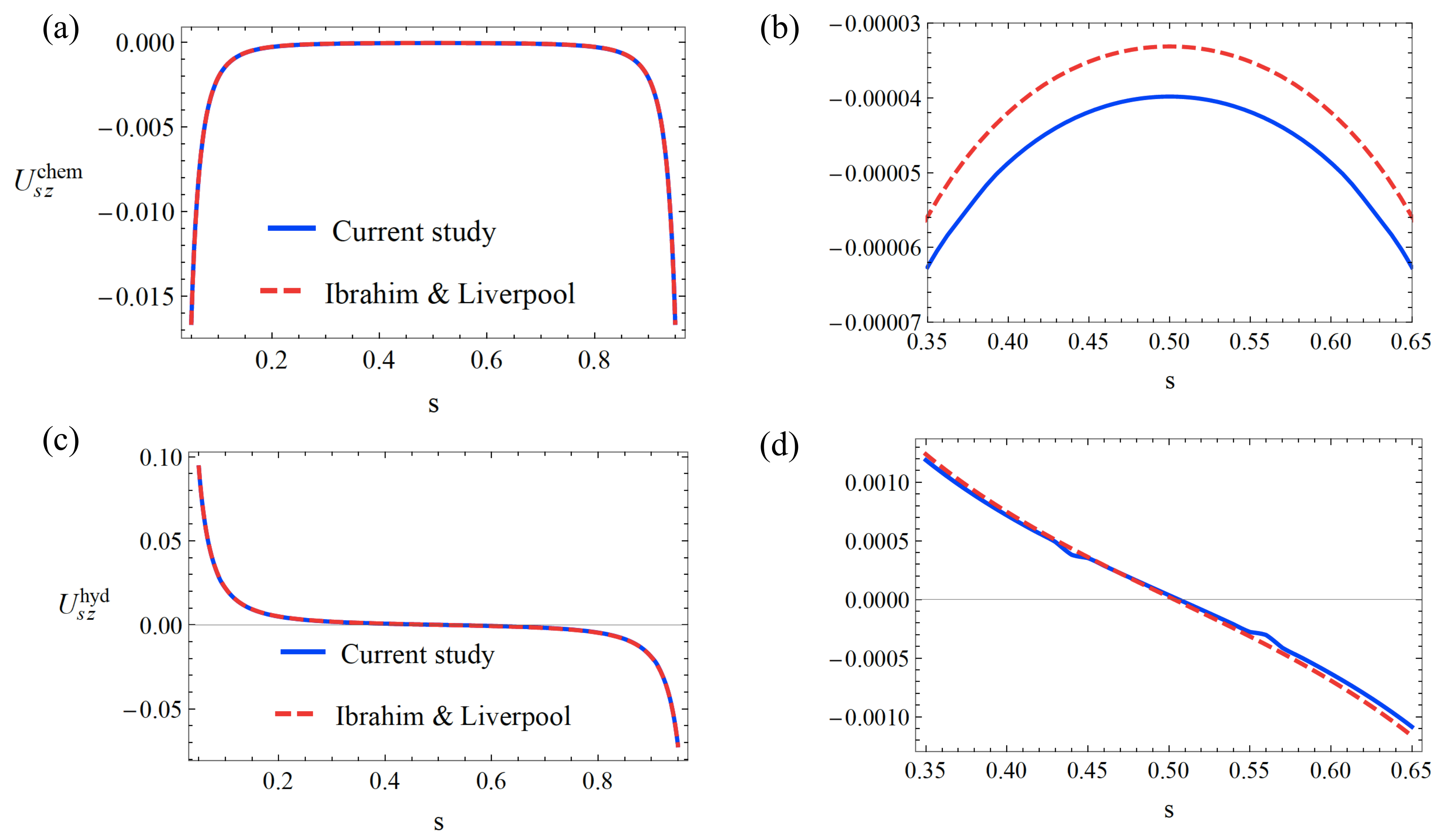}
	
	\caption{{\small Comparison of wall-induced horizontal velocity for $ \theta_{c} = \pi/4 $, $ \theta_{p}=\pi/4 $, and  $\kappa=0.05$: (a) wall-induced chemical correction (c) hydrodynamic correction. (b) and (d) are zoomed-in version of (a) and (c), respectively.}}%
	\label{fig:comparison_Usz}%
\end{figure}

\floatsetup[figure]{style=plain,subcapbesideposition=top}
\begin{figure}[H]
	\centering
	
	\includegraphics[scale=0.32]{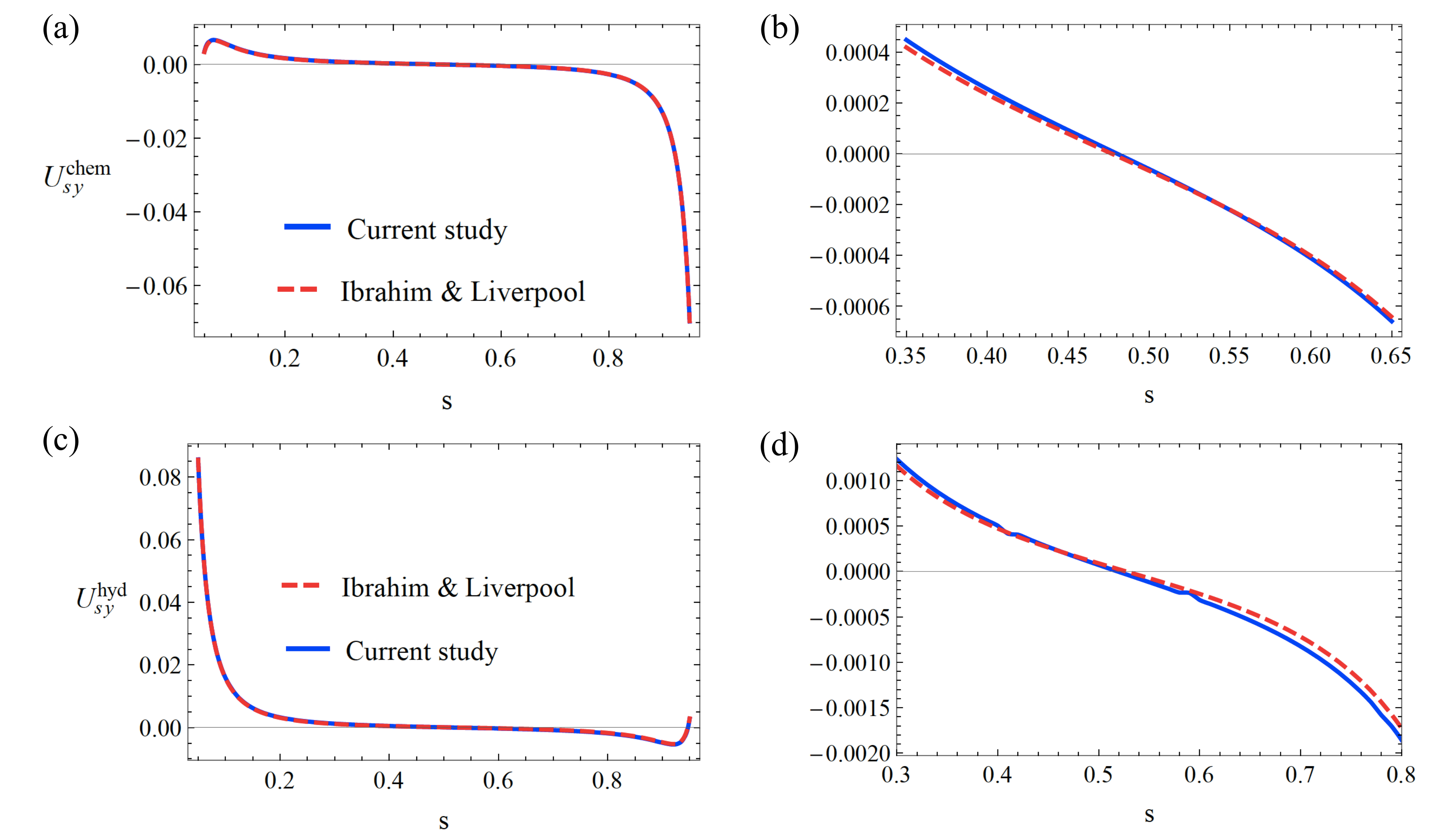}
	\caption{{\small Comparison of wall-induced vertical velocity for $ \theta_{c} = \pi/4 $, $ \theta_{p}=\pi/4 $, and  $\kappa=0.05$: (a) wall-induced chemical correction (c) hydrodynamic correction.. (b) and (d) are zoomed-in version of (a) and (c), respectively.}}%
	\label{fig:comparison_Usy}%
\end{figure}

\noindent
\textbf{\large{Appendix C: Expressions of wall corrections }}

{\footnotesize
	
	\centering
	
	\begin{equation}
		\mathbb{F}_{\mathcal{K}0}^{y} =   \int_{0}^{\infty}	 \frac{-1}{4} \lambda  \left(\frac{e^{\lambda  s}+1}{e^{\lambda }-1}-\frac{e^{\lambda  (1-s)}+1}{e^{\lambda }-1}\right)  {\rm d} \lambda , \nonumber
	\end{equation}

	\begin{equation}
		\mathbb{F}_{\mathcal{K}1}^{y} =  \int_{0}^{\infty}	\frac{-1}{4} \lambda  \left(-\frac{\lambda  \left(1-e^{\lambda  (1-s)}\right)}{2 \left(e^{\lambda }-1\right)}-\frac{\lambda  \left(1-e^{\lambda  s}\right)}{2 \left(e^{\lambda }-1\right)}\right) {\rm d} \lambda, \nonumber
	\end{equation}

	\begin{equation}
		\mathbb{F}_{\mathcal{K}1}^{z} =	\int_{0}^{\infty}  \frac{1}{8} \lambda ^2 \left(-\frac{e^{\lambda  (1-s)}+1}{2 \left(e^{\lambda }-1\right)}-\frac{e^{\lambda  s}+1}{2 \left(e^{\lambda }-1\right)}\right)  {\rm d} \lambda, \nonumber
	\end{equation}

	$ \begin{aligned}
		\mathbb{F}_{B}^{y} = -  \int_{0}^{\infty} \frac{ e^{-s \lambda}  \lambda ^2 }{24 \Lambda}  
		\left[ \begin{array}{l}
			e^{\lambda +2 \lambda  s} (\lambda  (s-1)-1)-2 e^{\lambda  s}+e^{\lambda } (\lambda  (s-1)+1)+e^{2 \lambda  s} (1-\lambda  s)-e^{2 \lambda } (\lambda  s+1)\\
			+ \left(\lambda ^2+2 \lambda +2\right) e^{\lambda (1+ s)}
		\end{array} \right]  {\rm d} \lambda  ,
	\end{aligned} $

	\vspace{2mm}

	$ \begin{aligned}
		\mathbb{F}_{B}^{z} = \int_{0}^{\infty} \frac{ e^{-s \lambda} \lambda^{2} }{48 \Lambda}  
		\left[ \begin{array}{l}
			\left(\lambda ^2-2 \lambda +2\right) e^{\lambda +\lambda  s}+e^{2 \lambda } (\lambda  s-1)-2 e^{\lambda  s}+e^{2 \lambda  s} (\lambda  s+1)+e^{\lambda +2 \lambda  s} (\lambda - \lambda  s -1)
			\\
			+e^{\lambda } (\lambda - \lambda s+1)
		\end{array} \right] {\rm d} \lambda  ,
	\end{aligned} $
	
	\vspace{2mm}

	$ \begin{aligned}
		\mathbb{F}_{D}^{i} = \int_{0}^{\infty} \frac{ 3e^{-s \lambda} \lambda^{3} }{32 \Lambda}  
		\left[ \begin{array}{l}
			e^{\lambda } (s-1)^2+(s-1)^2 e^{\lambda +2 \lambda  s} - e^{2 \lambda } s^2 - s^2 e^{2 \lambda  s} + (4 s-2) e^{\lambda  (1+s)}
		\end{array} \right] {\rm d} \lambda  ,
	\end{aligned} $
	
	\vspace{2mm}

	$ \begin{aligned}
		\mathbb{F}_{D}^{ii} = \int_{0}^{\infty} \frac{ -3e^{-s \lambda} \lambda^{3} }{16 \Lambda}  
		\left[ \begin{array}{l}
			e^{\lambda } (s-1)^2+(s-1)^2 e^{\lambda +2 \lambda  s} - e^{2 \lambda } s^2 - s^2 e^{2 \lambda  s} + (4 s-2) e^{\lambda  (1+s)}
		\end{array} \right] {\rm d} \lambda  ,
	\end{aligned} $
	
	\vspace{2mm}

	$ \begin{aligned}
		\mathbb{F}_{D}^{iii}= \int_{0}^{\infty} \frac{ -3e^{-s \lambda} \lambda }{32  \left(e^{\lambda }-1\right) \Lambda}  
		\left[ \begin{array}{l}
			-e^{2 \lambda  s} \left(\lambda ^2 s^2+2 \lambda  s+4\right)+e^{3 \lambda } \left(\lambda ^2 s^2-2 \lambda  s+4\right)+e^{\lambda +2 \lambda  s} \left(\lambda ^2 \left(2 s^2-2 s+3\right)+\lambda  (4 s-2)+8\right)
			\\
			-e^{2 \lambda } \left(\lambda ^2 \left(2 s^2-2 s+3\right)+\lambda  (2-4 s)+8\right) -e^{2 \lambda  (s+1)} \left(\lambda ^2 (s-1)^2+2 \lambda  (s-1)+4\right)
			\\
			+e^{\lambda } \left(\lambda ^2 (s-1)^2-2 \lambda  (s-1)+4\right)
		\end{array} \right]  {\rm d} \lambda ,
	\end{aligned} $
	
	\vspace{2mm}

	$ \begin{aligned}
		\mathbb{T}_{D}^{iii}=  \int_{0}^{\infty}  \frac{ 3e^{-s \lambda} \lambda^{2} }{64  \left(e^{\lambda }-1\right) \Lambda}  
		\left[ \begin{array}{l}
			4 \left(\lambda ^2-\lambda +3\right) e^{\lambda +\lambda  s}+e^{\lambda +2 \lambda  s} \left(\lambda ^2+\lambda  (4 s-2)+6\right)+e^{2 \lambda } \left(\lambda ^2+\lambda  (2-4 s)+6\right)
			\\
			-2 \left(\lambda ^2-2 \lambda +3\right) e^{\lambda  (s+2)}+e^{\lambda } (2 \lambda  (s-1)-3)-6 e^{\lambda  s}+e^{3 \lambda } (2 \lambda  s-3)-e^{2 \lambda  s} (2 \lambda  s+3)
			\\
			+e^{2 \lambda  (s+1)} (-2 \lambda  (s-1)-3)
		\end{array} \right]  {\rm d} \lambda ,
	\end{aligned} $
	
	\vspace{4mm}

	$ \begin{aligned}
		\mathbb{T}_{B}^{z}= \int_{0}^{\infty} \frac{ \left(e^{\lambda }-1\right)  e^{\lambda  (-s)} \left(e^{2 \lambda  s}-e^{\lambda }\right)}{48 \Lambda} \lambda ^3   {\rm d} \lambda  .
	\end{aligned} $

	\vspace{5mm}
	
}

\noindent
Here, $ \Lambda = 1-e^{\lambda } \left(\lambda ^2+2\right)+e^{2 \lambda } $.

%\vspace{5mm}

%\large{
%	\textbf{
%REFERENCES
%}}

%\bibliographystyle{abbrv}
%\bibliography{Akash}% Produces the bibliography via BibTeX.

\end{document}